\documentclass[12pt]{article}%
\usepackage{amsfonts}
\usepackage{amsmath}
\usepackage{amssymb}
\usepackage{graphicx}
\usepackage{natbib}
\usepackage{setspace}
\usepackage{geometry}
\usepackage{rotating}
\usepackage{subfigure}
\usepackage{url}
\usepackage{caption}
\usepackage{version}
\usepackage{titletoc}
\usepackage{afterpage}
\usepackage{placeins}
\usepackage{float}
\usepackage{accents}%
\graphicspath{ {/path/to/graphics/folder/} }

\allowdisplaybreaks
\makeatletter
\g@addto@macro\normalsize{
	\setlength\abovedisplayskip{7pt}
	\setlength\belowdisplayskip{7pt}
	\setlength\abovedisplayshortskip{7pt}
	\setlength\belowdisplayshortskip{7pt}
}
\makeatother
\ifx\pdfoutput\relax\let\pdfoutput=\undefined\fi
\newcount\msipdfoutput
\ifx\pdfoutput\undefined\else
\ifcase\pdfoutput\else
\ifx\paperwidth\undefined\else
\ifdim\paperheight=0pt\relax\else\pdfpageheight\paperheight\fi
\ifdim\paperwidth=0pt\relax\else\pdfpagewidth\paperwidth\fi
\fi\fi\fi
\begin{document}

\title{\textbf{Causal effects of the Fed's large-scale asset purchases on firms'
capital structure}\thanks{E-mail addresses: and.nocera@gmail.com (Nocera),
pesaran@usc.edu (Pesaran). We are grateful to Ron Smith, Ivan Petrella,
Alessandro Rebucci, Harry DeAngelo, Yunus Aksoy, and Rashad Ahmed, and the
participants at the Birkbeck Economics Department 50th Anniversary and NIESR
2022 conference on quantitative easing, for constructive comments and
suggestions. The views expressed in this paper are those of the authors and do
not necessarily reflect those of Norges Bank Investment Management. Andrea
Nocera was a postdoctoral scholar at USC Dornsife INET when he started working
on this paper.} \textbf{{\Large {} }}}
\author{Andrea Nocera\\{\small Norges Bank Investment Management}\\M. Hashem Pesaran\\{\small University of Southern California, USA, and Trinity College,
Cambridge, UK}}
\date{October 22, 2023}
\maketitle

\begin{abstract}
We investigate the short- and long-term impacts of the Federal Reserve's large-scale asset purchases (LSAPs) 
on non-financial firms' capital structure using a threshold panel ARDL model. To isolate the effects of LSAPs 
from other macroeconomic conditions, we interact firm- and industry-specific indicators of debt capacity with measures 
of LSAPs. We find that LSAPs facilitated firms’ access to external financing, with both Treasury and MBS purchases
having positive effects. Our model also allows us to estimate the time profile of the effects of LSAPs on 
firm leverage providing robust evidence that they are long-lasting. These effects have a half-life of 4-5 
quarters and a mean lag length of about six quarters. Nevertheless, the magnitudes are small, suggesting that LSAPs 
have contributed only marginally to the rise in U.S. corporate debt ratios of the past decade.

\vspace{0.5cm}



\end{abstract}

%

\bigskip

\pagenumbering{gobble}

\baselineskip0.245in

\newpage\pagenumbering{arabic} \setcounter{page}{1}

\section{Introduction}

The 2007-2009 financial crisis affected the U.S. corporate sector in a number
of important respects. Due to the reduction in the supply of external finance,
many non-financial firms found it difficult to roll over their debt
obligations, with consequent cuts in spending, investment, and employment
(e.g., \cite{almeida2012}, \cite{campello2010}, and \cite{duchin2010}). To
revitalize the economy, after cutting the policy rate close to zero, the
Federal Reserve resorted to large-scale asset purchases (LSAPs). The empirical
evidence so far suggests that these LSAPs have been successful at easing
financial conditions (\cite{bernanke2020}).\footnote{The empirical literature
on the effects of quantitative easing (QE) has grown very rapidly in the last
decade. \cite{bhattarai2020}, and \cite{kuttner2018} provide recent reviews.}
Yet the debate on the effectiveness of such policies is still far from being
settled. The vast majority of event studies show that LSAPs significantly
lowered long-term Treasury and corporate bond yields by reducing both
expected future short rates and the term premium (e.g., \cite{bauer2014},
\cite{damico2013}, \cite{gagnon2011}, and \cite{krishnamurthy2011}). At the
same time, \cite{greenlaw2018} find that the Fed's interventions only had
modest and uncertain impact on yields. They also note that these effects
tended to die out quickly. Other studies cast some doubt on the persistence of
such effects. Notably, \cite{wright2012} shows that the Fed's unconventional
monetary policies reduced both Treasury and corporate bond yields but these
effects were fairly short-lived. In contrast, \cite{ihrig2018} and
\cite{swanson2021} find that the effects of LSAPs on yields were quite persistent.

The literature so far has also provided contrasting evaluation of the efficacy
of LSAPs in stimulating corporate lending by financial institutions.
\cite{rodnyansky2017} show that banks more exposed to mortgage-backed
securities (MBS) significantly increased both their real estate and commercial
loans, whilst \cite{chakraborty2020} document a crowding out effect, whereby
banks benefiting from MBS purchases increased mortgage origination, largely at
the expense of reducing their commercial and industrial lending.

In this paper we focus on non-financial companies' leverage responses, and ask
whether LSAPs facilitated non-financial firms' access to external financing.
By characterizing the time profile of the effects of LSAPs, we also
investigate whether the Fed's purchases systematically affected the way firms
finance their operations beyond the transitory responses around policy
announcements. Thus, we extend the evidence on the persistence of the effects
of LSAPs on interest rates to quantities, namely firm leverage. To this end,
we estimate dynamic panel data models with threshold effects using quarterly
firm-level data covering the period of the Great Recession, to investigate the
impact of LSAPs on non-financial firms' capital structure, distinguishing
between short-term and long-term effects.

The key challenge is to isolate the effects of LSAPs on firms' financing
decisions from that of concurrent general macroeconomic conditions typically
represented in panel data models by unobserved time effects. As it is well
recognized in the literature, the effects of macro policy interventions cannot
be identified when using standard panel regressions with time effects, since
any attempt at eliminating the unobserved time effects will also end up
eliminating the observed macro variables. Isolating the impact of LSAPs from
the general business cycle conditions is all the more important in light of
the strong link between macroeconomic conditions and firms' ability to raise
capital, as documented in \cite{begenau2019}, \cite{bhamra2010},
\cite{erel2011}, and \cite{halling2016}, among others.

We address the identification problem by exploiting the heterogeneity that
exists in firms' debt capacity constraints both before and after LSAPs.
Specifically, we interact measures of LSAPs, generically denoted by
$q_{t}$, with indicators of firms' spare debt capacity to be
defined below. In line with \cite{myers1984}, we say that a firm has exhausted
its debt capacity if its debt to asset ratio reaches a level where further
debt issuance could result in substantial additional costs or increased
default risk. In practice, this threshold debt level is unknown.
\cite{leary2010} define debt capacities in terms of the leverage ratios of
investment-grade rated firms in the same industry-year combination. We propose
a new indicator of debt capacity which does not require to specify an
\textit{a priori} given threshold value.

We start by considering firm-specific indicator variables, $d_{is,t}(\gamma)$,
that take the value of one if firm $i$ in industry $s$ at time $t$ has a debt
to asset ratio ($y_{is,t}$) below a given threshold quantile,
$\gamma$. At the same time, because firms' financing decisions are not made in
isolation but are dependent on the financing choices of other firms in the
same industry (e.g., \cite{grieser2021}, \cite{leary2014}, and
\cite{mackay2005}), we average $d_{is,t}(\gamma)$ across firms within a given
industry to obtain an industry-specific indicator of debt capacity, which we
denote by $\pi_{st}(\gamma)$. This gives the proportion of firms within
industry $s$, whose $y_{is,t}$ lie below the threshold quantile, $\gamma$. We
investigate the relevance of this measure of debt capacity empirically. To
avoid simultaneity bias we interact $q_{t}$ with one-quarter lagged values of
this proportion. This allows us to use cross-industry
variations in $q_{t}\times\pi_{s,t-1}(\gamma)$ to separate the effects of
$q_{t}$ from other factors that are common across all industries.\footnote{In
the paper we focus on the industry-specific debt capacity indicator as it
provides an even stronger differentiation between the effects of dynamics
(past firm-leverage) from the effects of debt capacity. This choice is also in
line with the findings in the literature that industry variables are powerful
predictors of firms' leverage. Results based on firm-specific indicators,
$d_{is,t}(\gamma)$, are reported in Section \ref{Appx:firmdebtcpcty} of the
online supplement, and are generally in line with our conclusions obtained
using industry-specific measures of debt capacity.}

We estimate the quantile threshold parameter, $\gamma$, by grid search
together with other unknown parameters. Thresholding has been widely used in
the time series literature and more recently in panel data regressions to
capture differential impacts of macroeconomic shocks or policy interventions
across groups or categories.\footnote{See, for example, \cite{tong1990},
\cite{hansen1999}, \cite{dang2012} \cite{seo2016}, and \cite{chudik2017},
among others. \cite{hansen2011} provides a review of econometric applications
of threshold models.} Similar ideas are used in corporate finance, but often
using threshold values that are fixed \textit{a priori}. For example, firms
are classified based on lowest/highest tercile or quartile of the empirical
distribution of some particular firm or industry characteristic of
interest.\footnote{See, for example, \cite{flannery2006}, and
\cite{greenwood2010}.} In our empirical strategy, the unknown quantile
threshold values are estimated, allowing for possible changes in such
threshold values due to the policy intervention under consideration.

The main hypothesis behind our identification strategy is that the effects of
LSAPs are heterogeneous and depend on the ability of firms in an industry to
raise\ debt (as indicated by our debt capacity indicator). This assumption is
motivated by the most frequently discussed channels through which LSAPs may
reduce interest rates and ease financial conditions.\footnote{See
\cite{bernanke2020}, \cite{krishnamurthy2011}, and \cite{kuttner2018} for
detailed discussions on the transmission mechanisms of quantitative easing.}
Here, we highlight three main channels. According to the \textquotedblleft
portfolio balance channel\textquotedblright, by purchasing a large quantity of
assets held by the private sector, central banks increase their prices. In
order to rebalance their portfolios, the sellers of these financial assets may
use the proceeds to purchase other assets that have similar characteristics to
the assets sold, thus pushing up prices of other ``safe'' substitute assets. A
second channel is the so called \textquotedblleft bank lending
channel\textquotedblright\ via which the Fed's LSAPs increase the value of
existing assets on banks' balance sheets. This raises banks' capital ratios
making them more willing to lend. A third mechanism is the \textquotedblleft signalling channel\textquotedblright,
whereby purchases of assets by the Fed reinforce its commitment to maintain
interest rates low for long. Our hypothesis is that for each of these
channels, firms with adequate debt capacity and more financial flexibility
ought to benefit more from the Fed's asset purchases, whilst over-leveraged
firms may find it difficult to take full advantage of the reduction in the
cost of credit or the additional credit supply generated by LSAPs without the
risk of becoming financially distressed.\footnote{See \cite{flannery2006}, and
\cite{leary2010}, among others, for a discussion on the inability of raising
further debt for highly leveraged firms. \cite{greenwood2010} show that bond
issuance of firms that are relatively unconstrained is more elastic to changes
in the supply of government debt. \cite{ottonello2020} find that firms with
low default risk were the most responsive to changes in (conventional)
monetary policy during the period preceding the global financial crisis.}

We find that existing firms' debt burdens play an important role in the
transmission of LSAPs. In our main specifications, the threshold parameter,
$\gamma$, is estimated to be $0.77$, just above the upper quartile of the
cross-section distribution of firms' leverage at a given point in time,
indicating that firms with high debt burdens tended to benefit the least from
LSAPs. Our estimation results clearly show that industries with higher
proportion of firms with debt to assets ratio below the $77^{th}$ quantile
experienced, on average, a larger increase in external debt financing in
response to LSAPs.

At the same time, by considering a dynamic panel data model we are able to
estimate the time profile of the effects of LSAPs on firms' capital structure,
providing a clear and strong evidence that such effects are 
long-lasting. We find that the effects of LSAPs have a half-life of 4-5
quarters and a mean lag length of approximately 6 quarters. Nevertheless, the
magnitudes of these effects are relatively small, suggesting that LSAPs have
contributed only marginally to the rise in U.S. corporate debt ratios of the
last decade (as documented for instance by IMF (\citeyear{IMF2019})).

In one additional exercise, we separate the effects of MBS from Treasury
purchases to show that both programs facilitated non-financial firms' access
to external financing. Also in this case, we find that both type of purchases
had long-lasting effects on firm debt to asset ratios but the magnitudes are
rather small. Finally, we also replace our quantitative measures of LSAPs with
four qualitative variables equal to one during policy on periods and zero
otherwise. Consistently with the literature that studies the effects of LSAPs
on yields, we find that the first LSAP program (typically referred to as QE1)
had the strongest impact on firm leverage. This corroborates the view that
LSAPs can be particularly effective during periods of dysfunctions in
financial markets (e.g., \cite{damico2013}). Among the other programs, we find
that both the so called QE2 and QE3 programs had positive and statistically
significant effects on firm leverage. This suggests that LSAPs can also be an
effective tool outside periods of market stress. In contrast, the maturity
extension program (MEP) of 2011, where the Fed purchased long-term Treasuries
offset by the sale of short-term government bonds, didn't have a statistically
significant impact on firms' debt to asset ratios.

Our empirical results are robust to a number of specification choices. The
main paper reports short-term and long-term estimates obtained using a
relatively general panel autoregressive distributed lag (PanARDL) model of
order two. To show the robustness of our results to the choice of dynamic
specification, in the online supplement we also report results for the
standard partial adjustment model and the PanARDL(1) specification. Regarding
the control variables, in addition to firm-specific fixed effects we also
control for several time-varying industry-specific covariates to account for
differential growth opportunities and to further reduce possible omitted
variables bias due to the fact that firms in a given industry face common
factors that may drive their financing choices. In addition to time effects,
we also allow for industry-specific trend differentials and hence allow firms'
leverage to follow different time trends across industries. We also check the
robustness of our results to another popular measure of common effects whereby
real output growth is interacted with industry-specific dummies. Finally, our
results continue to hold after correcting for potential small-sample bias
arising from the fact that we employ a dynamic panel model with fixed effects
where the number of time series observations could be small for some of the
firms included in the panel, due to its unbalanced nature.

In summary, we find statistically highly significant effects of LSAPs on
corporate debt financing, but at the same time we find the magnitude of such
effects to be rather small in the short run (on impact) as well as in a longer
run when the business cycle effects are allowed to iron out.

\paragraph{Related literature.}

Our paper relates to a number of different strands in the literature. One
recent strand investigates the relationship between corporate debt issuance
and government debt supply. \cite{greenwood2010} document that firms tend to
issue more long-term (short-term) debt when the maturity of government debt
decreases (increases). This gap filling is more pronounced for firms with more
financial flexibility. \cite{badoer2016} argue that this gap filling behaviour
is more prominent in the issuance of long-term (LT) corporate bonds and that
the supply of LT government bonds affect both the maturity choice and the
level of corporate borrowing. \cite{graham2014} find that government debt is
negatively correlated with corporate debt, especially for larger and less
risky firms. Although these studies mostly cover the period before the
introduction of LSAPs, they provide some insight on how LSAPs may impact
firms' financing choices by affecting the overall supply of Treasuries. The
current paper provides direct evidence on the effects of LSAPs on firms'
capital structure.

There is also a growing literature that looks at the impact of LSAPs using
micro-level evidence. \cite{foley2016}, FRY henceforth, show that firms with
greater dependence on longer-term debt issued more long-term debt
as a result of the Fed's MEP.\footnote{\cite{giambona2020} also use firm-level
data at annual frequency (2004-2011) and identify the effects of QE on firm
investment by exploiting differences in firms' access to the bond market. The
same arguments that differentiate our paper from \cite{foley2016} apply to this study as well.} Our analysis differs from this study in at least two respects.
First, we quantify the effects of both MBS and Treasury purchases
on firms' capital structure. Second, we characterize the time profile of these
effects, evaluating whether they persist or dissipate immediately after the
implementation of one particular program. Assessing the overall long-term
effects of LSAPs and their persistence is particularly important from a policy
perspective given that quantitative easing (QE) is now part of the standard central
bank toolkit in the U.S..

When evaluating the first major four Fed's programs separately using
qualitative policy indicators, we find that the effects of the MEP are
positive but not statistically significant. Thus, while FRY document a
significant impact on long-term debt growth, we find that the MEP didn't lead
to higher debt to asset ratios. We focus on debt to assets instead of debt
growth consistently with the fact that asset and liability side of a firm's
balance sheet are jointly determined.

Our paper also contributes to the methodological discussion on the
identification of macro policy effects. First, while FRY's research question
only requires a static specification, our empirical model is dynamic and thus
accounts for the highly persistence nature of firm leverage (e.g.,
\cite{flannery2006} and \cite{lemmon2008}). Second, we use quarterly
observations which are better suited to distinguish the effects of LSAPs from
other macroeconomic conditions represented in our model by unobserved time
effects. More importantly, our empirical strategy doesn't require to specify a
single treatment date.

Our study is also related to the literature which studies the link between QE
and bank lending. \cite{rodnyansky2017} use a difference-in-difference
approach which exploits the fact that banks differ in their relative exposure
to MBS. They demonstrate that banks with a relatively large fraction of MBS on
their balance sheets expanded both real estate lending, and commercial and
industrial loans as a results of QE. \cite{chakraborty2020} also exploit the
fact that banks differ in their exposure to MBS purchase to find that banks
benefiting from MBS purchases increased mortgage origination. They also document a crowding out effect: QE encouraged exposed
banks to lend more to the housing markets while reducing their commercial and
industrial lending. Compared to these two studies, we focus on non-financial
firms' capital structure, distinguishing between short- and long-term effects.
We find that firms' debt to asset ratios increased as a results of both
Treasuries and MBS purchases.

Our paper also partly relates to the literature that tries to understand the
role of financial frictions in the transmission of monetary policy. For
example, focusing on the period preceding the global financial crisis,
\cite{ottonello2020} find that firms with low default risk were the most
responsive to changes in monetary policy. Our paper highlights the important
role of pre-existing firms' debt capacity within an industry in the transmission
of LSAPs.

More generally, our paper relates to the vast literature which studies the
relative importance of various factors in non-financial firms' capital
structure decisions. Excellent reviews are provided by \cite{deangelo2022},
\cite{frank2022}, and \cite{graham2011}. In line with the findings of
\cite{mackay2005}, \cite{frank2009}, and \cite{leary2014}, amongst others,
we find that industry factors are powerful predictors of firms' leverage. Our
study is also connected to the research that advocates that capital market
segmentation and supply conditions play an important role in observed
financial structures (see \cite{baker2009} for a comprehensive review).%

\vspace{5mm}%

\section{Panel data and sources \label{Sec:Data}}

We use an unbalanced panel of U.S. publicly traded non-financial firms
observed at quarterly frequencies over the period $2007$-Q1 to $2018$-Q3. We
employ Compustat database to obtain selected measures of firm size,
tangibility, cash holdings, leverage, and other firm characteristics which are
commonly used in the corporate finance literature.

As a proxy for capital structure we use firm leverage, defined as the ratio of
debt to assets, both measured at book values. We prefer book leverage to
market leverage to reduce concerns over the possibility that the effects of
LSAPs on firms' debt ratios are anticipated. This is because, as noted by
\cite{frank2009}, contrary to market measures which are typically forward
looking, book leverage is a backward looking variable.

In addition to firm-specific data, we also consider several variables at the
industry level. To construct such industry-specific variables, we group firms
in our sample into various industries, based on the three-digit Standard
Industrial Classification (SIC). Specifically, firms are grouped into $67$
three-digit SIC industries, such that each industry group contains at least
$20$ firms.\footnote{In line with the existing literature, we employ the
three-digit SIC industry classification instead of the two-digit SIC industry
classification which would also result in fewer industry groups, namely 41.}

To align our analysis with previous studies on firms' capital structure, we
focus on non-financial firms and exclude firms in the regulated utilities (SIC
$4900$-$4999$) and those that belong to the non-classifiable sector (SIC codes
above or equal to $9900$).\footnote{The SIC codes of excluded financial firms
are $6000$-$6999$.} In total, our data consists of $95,489$ firm-quarter
observations, comprised of $3,647$ distinct firms observed on average over
$26$ quarters. Firms in our sample have at least $5$ time observations (T)
while the maximum T is $47$. For brevity, a detailed description of both the
variables under consideration and the sample selection screens, as well as the
classification of firms by industry are provided in Section
\ref{Appendix_DataAnalysis} of the online supplement, where we also provide a
number of descriptive and summary statistics at both firm- and industry-level.

\subsection{Large-scale asset purchases\label{Sec:qt}}

To estimate the effects of the Fed's asset purchases on firms' debt to asset
ratios, we employ a quantitative measure of LSAPs obtained from the New York
Fed's website. Our primary policy variable of interest is the total gross
amount of U.S. Treasuries and agency mortgage-backed securities (MBS) purchased by
the Fed, denoted by $q_{t}$. The use of gross instead of net amount is in line with
\cite{chakraborty2020} who focus on gross purchases to capture the Maturity
Extension Program through which the Fed used the proceeds of its sales of
shorter-term Treasuries to purchase longer-term Treasury securities.

We scale our policy variable so that its average value is unity over the
policy sample. This scaling facilitates the interpretations of the estimation
results, and makes our estimates based on the quantitative measure directly
comparable to the estimates obtained using qualitative (0,1) policy variables.
While we report results for both the quantitative and qualitative measure of
LSAPs, our main focus is on the quantitative measure which is better suited to
capture the magnitude of the Fed's purchases.\footnote{Further information on
both the quantitative and qualitative measures of LSAPs are provided in
Section \ref{Appendix_DataAnalysis} of the online supplement.}

\section{Identification of macro policy effects with heterogeneous outcomes
\label{sec: Ident_Strategy}}

\subsection{Firm-specific and industry-average debt capacity measures}

The rationale for our identification strategy is based on the \textit{a
priori} belief that firms with higher debt capacity and financial flexibility
are likely to be more responsive to the Fed's LSAPs.\footnote{See for example,
\cite{graham2014}, and \cite{greenwood2010} on the heterogeneous responses of
firms to government debt issuance. \cite{bolton2021}, \cite{leary2010}, and
\cite{lemmon2010} provide discussions on the ability of firms to issue debt
according to their debt capacity.} Our hypothesis is that in order to take
advantage of the reduction in the cost of debt and/or increase in the supply
of external finance resulting from LSAPs, firms should have enough spare debt
capacity. The basis for this argument is twofold. On the one hand, firms with
lower levels of leverage are better able to borrow and deviate from the
long-run target to meet their funding needs (e.g., \cite{flannery2006},
\cite{leary2005}, \cite{lemmon2010}). On the other hand, over-leveraged firms
are less able to fill the gap of safe assets' supply created by the Fed's
asset purchases because issuing further public debt or resorting to additional
bank borrowing could lead to financial distress (e.g., \cite{bolton2021},
\cite{leary2010}). It is in fact well recognized that higher debt burdens are
powerful predictors of future default probabilities and, as such, constitute
an important measure of credit risk (e.g., \cite{bhamra2010},
\cite{ottonello2020}). Debt ratios have also been found to be a significant
predictor of firms' financial constraints (e.g., \cite{kaplan1997},
\cite{hadlock2010}).

To account for differences in debt exposure we consider both firm-specific and
industry-average measures. We measure firm-specific debt capacity by the
indicator, $d_{is,t}(\gamma)$, defined by
\begin{equation}
d_{is,t}(\gamma)=\mathcal{I}\left[  y_{is,t}<g_{st}(\gamma)\right]  ,
\label{Firm_DebtCap}%
\end{equation}
where $y_{is,t}$ is the ratio of debt to assets (DA) of firm $i$ in industry
$s$ for quarter $t$, $g_{st}(\gamma)$ is the $\gamma^{th}$ quantile of the
cross-sectional distribution of DA over all firms in industry $s$ at time $t$,
and $\mathcal{I}\left(  \mathcal{A}\right)  $ is an indicator variable that
takes the value of one if $\mathcal{A}$ is true and zero otherwise. The
industry-average measure of debt capacity is defined by%
\begin{equation}
\pi_{st}(\gamma)=\frac{1}{N_{st}}\sum_{i=1}^{N_{st}}\mathcal{I}\left[
y_{is,t}<g_{t}(\gamma)\right]  , \label{Frac_debt}%
\end{equation}
where $g_{t}(\gamma)$ is the $\gamma^{th}$ quantile of the cross-sectional
distribution of DA of all firms at time $t$, and $N_{st}$ denotes the number
of firms in industry $s$ during quarter $t$. In effect, $\pi_{st}(\gamma)$ is
the proportion of firms in industry $s$ in quarter $t$ with DA below
$g_{t}(\gamma)$, an economy-wide time-varying threshold. The industry-average
measure, $\pi_{st}(\gamma)$, recognizes that firms in a given industry tend to
closely align their own financing decisions with the financial choices made by
firms from the same industry.\footnote{See, for example, \cite{frank2009},
\cite{grieser2021}, and \cite{leary2014}.} The quantile threshold parameter
$\gamma$ ($0<\gamma<1$) is estimated using a grid search procedure to be
explained in Subsection \ref{QuantileEst}.

Both firm-specific and industry-average measures of debt capacity are important 
in classifying firms with respect to their debt exposure relative to the 
existing debt levels within an industry or in the economy.
In the main paper we focus on $\pi_{st}(\gamma)$ which yields results that are
more readily interpretable and in some respects more convincing. However, for
completeness we provide estimation results based on $d_{is,t}(\gamma)$ in
Section \ref{Appx:firmdebtcpcty} of the online supplement.

\subsection{Identification strategy \label{subsec: Ident_Strategy}}

As with all macro policy changes, identification of the effects of LSAPs on
firms' debt to asset ratios is complicated by the concurrent effects of other
macroeconomic developments. A number of recent papers try to exploit
differences in banks' holdings of MBS to identify the effects of QE on banks'
lending (e.g., \cite{chakraborty2020}, and \cite{rodnyansky2017}). To this end,
banks' MBS exposure is interacted with a measure of Fed's purchases, and
identification of the policy effect is achieved from the differential effects
of the policy on bank lending. Interactions are also employed by
\cite{foley2016} who utilize differences in firms' long-term debt dependence
to study the effects of MEP on firms' long-term debt growth and other
characteristics. In this paper, in line with this literature, we employ
interactive terms to exploit differences in firms' debt capacity across industries.

The basic idea behind our identification strategy is best described in the
context of a static model without dynamics or control variables. Consider the
panel regression model
\begin{equation}
y_{is,t}=\mu_{is}+\phi_{st}+\beta_{0}\pi_{s,t-1}(\gamma)+\beta_{1}q_{t}%
\times\pi_{s,t-1}(\gamma)+u_{is,t}, \label{basic panel 1}%
\end{equation}
where as before $y_{is,t}$, is the DA ratio of firm $i$ in industry
$s=1,2,...,S,$ for quarter $t$, while $q_{t}$ is the quantitative policy
variable measuring the size of the Fed's U.S. Treasury and agency MBS
purchases, which we interact with our industry-specific debt capacity
proportion, $\pi_{s,t-1}(\gamma)$. Note that equation (\ref{basic panel 1})
only includes a one-quarter lag of $\pi_{st}(\gamma)$ to avoid simultaneous
determination of this proportion and the dependent variable, that could occur
when the number of firms in a given industry is rather small.

We use firm-specific effects, $\mu_{is}$, to remove systematic differences
across firms in different industries, and consider industry-time fixed
effects, $\phi_{st}$, to remove differences in time effects across industries.
Allowing for time effects is critical if we are to avoid confounding the
policy effects with other unrelated factors that are likely to have pervasive
effects on the outcome variable, $y_{is,t}$. Within the above framework,
$\phi_{st}$ is included to capture such time-industry effects that fully take
account of non-policy macro factors with differential industry effects. It is
clear that at this level of generality it is not possible to identify
$\beta_{1}$, which is the policy effectiveness coefficient of interest. Some
restrictions on $\phi_{st}$ must be entertained. One possible option is to
consider an interactive time effect by specifying
\begin{equation}
\phi_{st}=\delta_{t}+\phi_{s}f_{t}, \label{phi_st}%
\end{equation}
where $\delta_{t}$ is the common component of $\phi_{st}$, the so-called fixed
time effects, and $\phi_{s}f_{t}$ is the industry-specific component which is
intended to capture non-policy macro variables that have differential outcomes
across industries. To identify $\phi_{s}$ we first note that
\[
S^{-1}\sum_{s=1}^{S}\phi_{st}=\delta_{t}+\left(  S^{-1}\sum_{s=1}^{S}\phi
_{s}\right)  f_{t},
\]
and to identify the homogenous effects of non-policy variables from the
industry-specific ones we need to set%
\begin{equation}
\bar{\phi}_{\circ}=S^{-1}\sum_{s=1}^{S}\phi_{s}=0\text{.} \label{phi_res}%
\end{equation}
Under this (normalization) restriction, $\delta_{t}$ is identified as the
common component of non-policy macro variables. But to identify $\phi_{s}$,
and hence $\beta_{1}$, further restrictions are required. One possibility is
to assume $\phi_{s}$ are distributed independently across $s$ with mean zero
and a constant variance, and then estimate $f_{t}$ for $t=1,2,..,T$, along
with other parameters. See, for example, \cite{ahn2001}, \cite{bai2013}, and
\cite{hayakawa2020}. In this paper we consider an alternative estimation
strategy which allows $\phi_{s}$ to be treated as free parameters to be
estimated subject to (\ref{phi_res}) and for alternative specifications of
$f_{t}$. Using (\ref{phi_st}) in (\ref{basic panel 1}) we have
\begin{equation}
y_{is,t}=\mu_{is}+\delta_{t}+\phi_{s}f_{t}+\beta_{0}\pi_{s,t-1}(\gamma
)+\beta_{1}q_{t}\times\pi_{s,t-1}(\gamma)+u_{is,t}. \label{panel 1}%
\end{equation}

The fixed and time effects, $\mu_{is}$ and $\delta_{t}$, can now be eliminated
using standard de-meaning techniques.\footnote{In our empirical applications,
where the panel is unbalanced, we use \cite{wansbeek1989} transformations to
eliminate $\mu_{is}$ and $\delta_{t}$. Wansbeek and Kapteyn procedure is
equivalent to including both time and fixed effect dummies in the panel
regressions, but it is less computationally cumbersome when $sup_{t}\sum
_{s=1}^{S}N_{st}$ is large.} In standard panel regressions with fixed and time
effects identification is achieved by setting $\phi_{s}=0$ for all $s$. Here
we place the restrictions on $f_{t}$ and consider identification of $\beta
_{1}$ for arbitrary choices of $\phi_{s}$ but conditional on alternative
specification of $f_{t}$. In the empirical applications we consider linear
trends and set $f_{t}=t/T$. The panel estimates of $\beta_{1}$ do not depend
on the scales of $f_{t},$ and it is therefore convenient to set $f_{t}=f(t/T)$
where $f(x)$ is a general function of $x=t/T$. Changing the scale of $f_{t}$
only affects the estimates of $\phi_{s}$, with no consequence for the policy
effectiveness coefficient, $\beta_{1}$. In view of the uncertainty surrounding
the choice of $f_{t}$, the robustness of the estimates of $\beta_{0}$ and
$\beta_{1}$ are further investigated by also including U.S. real GDP growth as
a proxy for $f_{t}$, which is a commonly used indicator of macroeconomic
conditions in the corporate finance literature (e.g., \cite{erel2011},
\cite{frank2009}).\footnote{In Section \ref{MacroVar} of the online supplement
we also experiment with alternative observed macro-variables as proxies for
$f_{t}$. When using the firm-specific debt capacity indicator, $d_{is,t}%
(\gamma)$, we also allow for industry-time fixed effects (without imposing
restrictions on $\phi_{st}$). We find that using industry-time fixed effects
or the restriction described in equation (\ref{phi_st}) leads to very similar
results.}

Identification of $\beta_{1}$ also requires a sufficient degree of variations
in $q_{t}$ over time and $\pi_{s,t-1}(\gamma)$ over $s$, such that there is a
unique solution for $\beta_{0}$ and $\beta_{1}$ to our estimation problem.
This is indeed the case in our application, as shown in Section
\ref{Appendix_Threshold_Optim} of the online supplement.

\subsection{Average policy effect at industry and national levels
\label{AveEffect_IndustryNational}}

For clarity of exposition, suppose the policy is introduced at time $t=T_{0}$,
and the full sample period $t=1,2,...,T,$ is split into policy on ($t>T_{0}$)
and policy off ($t\leq T_{0}$) sub-periods. It is clear that post $t=T_{0}$ we
only observe the policy on outcomes, which we denote by $y_{is,t}^{1}%
=y_{is,t}$, for $t=T_{0}+1,T_{0}+2,...,T$. The policy off outcomes over the
policy on sample, denoted by $y_{i,st}^{0}$ are not observed but can be
estimated using (\ref{panel 1}). Specifically, assuming that the proportions,
$\pi_{s,t-1}(\gamma)$, are not materially affected by the policy change, we
have
\[
y_{is,t}^{0}=E\left(  y_{is,t}\left\vert q_{t}=0,\pi_{s,t-1}(\gamma)\right.
\right)  =\mu_{is}+\delta_{t}+\phi_{s}f_{t}+\beta_{0}\pi_{s,t-1}(\gamma),
\]
for $t=T_{0}+1,T_{0}+2,...,T$. The predicted policy effects are given by
\[
y_{is,t}^{1}-y_{is,t}^{0}=\beta_{1}q_{t}\times\pi_{s,t-1}(\gamma)+u_{is,t}.
\]
Using this result, we can now compute the average policy effect over the
policy on sample at the industry or national level. At the industry level, the
average policy effect (per quarter) is
\[%
\begin{array}
[c]{ccl}%
\overline{PE}_{s} & = & \frac{1}{T-T_{0}}\sum_{t=T_{0}+1}^{T}\left[  \frac
{1}{N_{st}}\sum_{i=1}^{N_{st}}\left(  y_{is,t}^{1}-y_{is,t}^{0}\right)
\right] \\
&  & \\
& = & \beta_{1}\left[  \frac{1}{T-T_{0}}\sum_{t=T_{0}+1}^{T}q_{t}\times
\pi_{s,t-1}(\gamma)\right]  +\frac{1}{T-T_{0}}\sum_{t=T_{0}+1}^{T}\left(
\frac{1}{N_{st}}\sum_{i=1}^{N_{st}}u_{is,t}\right)  .
\end{array}
\]
The random component of the last term is likely to be small and will tend to
zero with $N_{st}$ and $T-T_{0}+1$ sufficiently large, and the industry level
policy effect is well approximated by%
\begin{equation}
\overline{PE}_{s}=\beta_{1}\left[  \frac{1}{T-T_{0}}\sum_{t=T_{0}+1}^{T}%
q_{t}\times\pi_{s,t-1}(\gamma)\right]  +o_{p}(1). \label{PE_s}%
\end{equation}
At the national level the average per quarter policy effect is given by
\begin{equation}
\overline{PE}=\beta_{1}\left[  \frac{1}{T-T_{0}}\sum_{t=T_{0}+1}^{T}%
q_{t}\times\sum_{s=1}^{S}w_{s}\pi_{s,t-1}(\gamma)\right]  , \label{PE}%
\end{equation}
where $w_{s}$ is the share of industry $s$ in the economy, which can be
measured for example by employment shares.

Although the above expressions apply irrespective of whether the strength of
the policy varies over the policy period or not, our preferred measure of
$q_{t}$ is the size of the Fed MBS and U.S. Treasuries' purchases because of
its greater degree of variability over time as compared to when $q_{t}$ is a
qualitative measure equal to $1$ over the policy on period and $0$ otherwise.
We scale our quantitative measure so that its average value over the policy
sample is unity. Specifically, let $Q_{t}>0$ for some $t$, denote the size of
Fed's MBS and Treasuries purchases (TY), namely $Q_{t}=MBS_{t}+TY_{t}$. Then
$q_{t}$ ought to be scaled as
\[%
\begin{array}
[c]{clll}%
q_{t} & = & 0, & \text{policy off period }(t=1,2,...,T_{0}),\\
&  &  & \\
q_{t} & = & \frac{Q_{t}}{\frac{1}{T-T_{0}}\sum_{\tau=T_{0}+1}^{T}Q_{\tau}}, &
\text{policy on period }(t=T_{0}+1,T_{0}+2,...,T).
\end{array}
\]
This normalization, besides removing the unit of measurement of the variable,
also makes the policy outcomes directly comparable under both qualitative and
quantitative policy measures.

\subsection{Possible confounding effects of policy changes on threshold
parameters \label{Motivation_Different_Thresholds}}

The above analysis assumes the threshold parameter, $\gamma$, used to compute
the industry proportion, $\pi_{st}(\gamma)$ described in equation
(\ref{Frac_debt}), is the same under the policy on and policy off periods.
Denoting the threshold values during the policy off and policy on periods by
$g_{t}(\gamma_{0})$ and $g_{t}(\gamma_{1})$, respectively, and using a similar
line of reasoning as above we have
\[
y_{is,t}^{1}-y_{is,t}^{0}=\beta_{0}\left[  \pi_{s,t-1}(\gamma_{1})-\pi
_{s,t-1}(\gamma_{0})\right]  +\beta_{1}q_{t}\times\pi_{s,t-1}(\gamma
_{1})+u_{is,t},
\]
for $t=T_{0}+1,T_{0}+2,...,T$. The first term can be viewed as an indirect
effect of the policy change, which needs to be taken into account. To allow
for such a possibility, in our empirical application we consider a more
general formulation of (\ref{panel 1}) and distinguish between the threshold
parameter for the construction of the industry-specific proportions before and
after the policy change, namely we consider the two-threshold panel
regression
\begin{equation}
y_{is,t}=\mu_{is}+\delta_{t}+\phi_{s}f_{t}+\beta_{0}\pi_{s,t-1}(\gamma
_{pre})+\beta_{1}q_{t}\times\pi_{s,t-1}(\gamma_{post})+u_{is,t},
\label{panel 1_2thresh}%
\end{equation}
which then ensures that
\[
y_{is,t}^{1}-y_{is,t}^{0}=\beta_{1}q_{t}\times\pi_{s,t-1}(\gamma
_{post})+u_{is,t}.
\]
The separate threshold parameters $\gamma_{pre}$ and $\gamma_{post}$ can be
estimated using grid search techniques.

\pagebreak

\section{Panel ARDL regressions with debt capacity thresholds\label{ARDL}}

In our empirical analysis, we extend the simple static models described in
equation (\ref{panel 1_2thresh}) by adding dynamics as well as firm- and
industry-specific control variables. We consider the following general
$p^{th}$ order threshold panel autoregressive distributed lag, PanARDL($p$),
model:%
\begin{align}
\lambda(L,p)y_{is,t}  &  =\mu_{is}+\delta_{t}+\boldsymbol{\phi}_{s}^{\prime
}\mathbf{f}_{t}+\beta_{0}(L,p)\pi_{s,t-1}(\gamma_{pre})+\beta_{1}(L,p)\left[
q_{t}\times\pi_{s,t-1}(\gamma_{post})\right] \label{ARDL_model_lsap}\\
&  +\boldsymbol{\varphi}^{\prime}(L,p)\mathbf{w}_{st}+\boldsymbol{\psi
}^{\prime}(L,p)\mathbf{x}_{is,t}+u_{is,t},\nonumber
\end{align}
where $\lambda(L,p)$, $\beta_{0}(L,p)$, $\beta_{1}(L,p)$, $\boldsymbol{\varphi
}(L,p)$ and $\boldsymbol{\psi}(L,p)$ are $p^{th}$-order polynomials in the lag
operator, $Ly_{is,t}=y_{is,t-1}$.\footnote{Specifically, $\lambda
(L,p)=1-\lambda_{1}L-...-\lambda_{p}L^{p}$; $\beta_{j}(L,p)=\beta_{j,0}%
+\beta_{j,1}L+...+\beta_{j,p}L,$ for $j=0,1$, $\boldsymbol{\varphi
}(L,p)=\boldsymbol{\varphi}_{0}+\boldsymbol{\varphi}_{1}%
L+...+\boldsymbol{\varphi}_{p}L^{p}$, and $\boldsymbol{\psi}%
(L,p)=\boldsymbol{\psi}_{0}+\boldsymbol{\psi}_{1}L+...+\boldsymbol{\psi}%
_{p}L^{p}$.} As before, $y_{is,t}$ is the DA ratio of firm $i$ in industry $s$
for quarter $t,$ $\mu_{is}$ and $\delta_{t}$ are firm and time fixed effects,
$\boldsymbol{\phi}_{s}^{\prime}\mathbf{f}_{t}$ is industry interactive
effects, with $\mathbf{f}_{t}$ proxied by a linear time trend, real GDP
growth, or both. $q_{t}$ is the scaled measure of the Fed's asset purchases as
described in Subsection \ref{Sec:qt}, and $\pi_{st}(\gamma_{post})$ is the
proportion of firms in industry $s$ with DA below the $\gamma_{post}$-th
quantile, as defined by equation (\ref{Frac_debt}).

In addition, we control for time-varying industry-specific covariates to
further reduce possible omitted variables bias due to the fact that firms in
an industry face common forces that may drive their financing decisions. The
vector $\mathbf{w}_{st}$ includes the median (three-digit SIC) industry
leverage, and the median industry growth (computed as the median of the
changes in the logarithm of firm total assets). We also control for
firm-specific covariates, $\mathbf{x}_{is,t}$, that include the ratio of cash
to total assets (TA), property, plant, and equipment (PPE) scaled by TA (as a
proxy for tangibility), and a measure of firm size (the natural logarithm of
TA). The choice of control variables is motivated by the findings of
\cite{frank2009}, and is in line with the corporate finance
literature.\footnote{\cite{frank2009} document that the most relevant
variables for explaining firm leverage are firm size, market to book ratio,
measures of tangibility and profitability, the median industry leverage, and
expected inflation. We do not include expected inflation (or other observed
macroeconomic variables) as our model is more general as it allows for time
effects. We exclude the market to book ratio from our main model because the
associated coefficients were often insignificant, both in statistical and
economic terms. At the same time, as we shall see, our estimation results are
robust to the inclusion of additional explanatory variables, such as market to
book ratio, median industry Tobin's Q, and other industry-specific variables.}

Regarding the model choice, the PanARDL($p$) approach is particularly
attractive for our empirical analysis since, among its advantages, it can be
used for the analysis of long-run relations, and it is robust to
bi-directional feedback effects between firm leverage and its determinants
(\cite{pesaran1998}). In other words, unlike the partial adjustment
specification, the PanARDL model takes into account the effects of lagged
explanatory variables onto the dependent variable, and it allows for feedback
effects from the dependent variable onto the regressors.

The policy parameters of interest are $\beta_{1,\ell}$, for $\ell=$
$0,1,...,p$, namely the coefficients of $\beta_{1}(L,p)=\beta_{1,0}%
+\beta_{1,1}L+...+\beta_{1,p}L$, and the lagged dependent variable
coefficients, $\lambda_{\ell},$ for $\ell=$ $1,2,...,p$. The policy impact is
given by $\beta_{1,0}$ while the policy long-run effects are defined by
\begin{equation}
\theta=\frac{\sum_{\ell=0}^{p}\beta_{1,\ell}}{\left(  1-\sum_{\ell=1}%
^{p}\lambda_{\ell}\right)  }.\label{LR_definition}%
\end{equation}
The numerator of $\theta$, $\beta_{1}=\sum_{\ell=0}^{p}\beta_{1,\ell}$, is
often referred to as the net short-run effect, and will be reported as a
summary measure for short-run effects. Due to the highly persistent nature of
debt-to-asset ratios, the net short-run effects will be smaller (in absolute
value) than the long-run effects.

To estimate $\beta_{1}$ and $\theta$ we need to choose the lag order, $p$, and
the threshold parameter, $\boldsymbol{\gamma}=\mathbf{(}\gamma_{pre}%
,\gamma_{post})^{\prime}$. A simultaneous estimation of $p$ and
$\boldsymbol{\gamma}$ is computationally demanding and could involve a
considerable degree of data mining. Here we follow the literature and estimate
$\boldsymbol{\gamma}$ for $p=1$ and $2$ as well as for the partial adjustment
model, a commonly used specification in the empirical capital structure
research (\cite{graham2011}).\footnote{The partial adjustment model is given
by
\begin{align*}
y_{is,t} &  =\mu_{is}+\delta_{t}+\boldsymbol{\phi}_{s}^{\prime}\mathbf{f}%
_{t}+\lambda_{1}y_{is,t-1}+\beta_{0}\pi_{s,t-1}(\gamma_{pre})\\
&  +\beta_{1}q_{t}\times\pi_{s,t-1}(\gamma_{post})+\boldsymbol{\varphi
}^{\prime}\mathbf{w}_{st}+\boldsymbol{\psi}^{\prime}\mathbf{x}_{is,t}%
+u_{is,t},
\end{align*}
which is a special case of the PanARDL model, where $p=0$, except for the lag
operator applied to $y_{is,t}$ whose order is set to $p=1$.} Also, since
allowing for different lag orders for policy and control variables involves
many permutations with a large number of dynamic specifications to choose
from, we use the same lag order across the regressors which seems a reasonable
empirical strategy. For the sake of brevity, in the remainder of the paper, we
focus on reporting the estimates for the PanARDL(2) model.\footnote{Results
for both the panel partial adjustment model and PanARDL(1), which are special
cases of the PanARDL(2) specification, are reported in Section
\ref{Apx_Dyn_Specif} of the online supplement.}

As to the estimation of $\boldsymbol{\gamma}$, we follow the threshold
literature and estimate $\boldsymbol{\gamma}$ by grid search, and treat the
resultant estimate as given when it comes to estimating the policy parameters
of interest. This two-step strategy is justified since the estimates of the
threshold parameters are super consistent in the sense that they converge to
their true values much faster than the estimate of the policy parameters. This
result is shown formally in the context of static threshold panel data models
by \cite{hansen1999}, and investigated further for panel threshold ARDL models
by \cite{chudik2017}. In view of these theoretical results in what follows we
do not provide standard errors for threshold estimates and compute the
standard errors of the policy effects taking the estimated value of the
threshold parameter as given.

\section{Estimation and empirical findings\label{sec:empirical}}

\subsection{Quantile threshold parameter estimates\label{QuantileEst}}

The quantile threshold parameter $\boldsymbol{\gamma}$ in (\ref{Frac_debt}), is
estimated by minimizing the sum of squared residuals (SSR) for different
values of $\boldsymbol{\gamma}$ in the range $0.25\leq\gamma_{pre}%
,\gamma_{post}\leq0.9$ in increments of $0.01$.\footnote{We start our grid
search for $\gamma$ from $0.25$ instead of $0.1$ because the $q$-th quantile
of DA is equal to zero for all values of $q$ below $0.21$. Further details are
provided in Section \ref{Appendix_Threshold_Optim} of the online supplement.}
Specifically, for a given choice of $p$ and for each value of
$\boldsymbol{\gamma}$ within the grid, we run the panel regressions described
in equation (\ref{ARDL_model_lsap}) by both fixed and time effects (FE--TE)
over the sample period $2007$-Q1 to $2018$-Q3.%

\begin{table}[H]
\caption{\textbf{Estimated quantile threshold parameters}}\label{tab: thresholds_LSAPs}
\vspace{-0.2cm}
\footnotesize
Estimates of the quantile threshold parameters from a grid search procedure
for the PanARDL(2) model described in equation (\ref{ARDL_model_lsap}). Panel
A shows the estimated threshold parameters for the single-threshold panel
regression model, where $\gamma_{pre}=\gamma_{post}$. Panel B displays results
for the two-threshold model, where $\gamma_{pre}\neq\gamma_{post}$. In column
(1) and (2), we use linear time trends or real GDP growth as a proxy for
$f_{t}$, respectively. Column (3) reports results when including both linear
trends and real GDP growth at the same time. The estimation sample consists of
an unbalanced panel of $3,647$ U.S. publicly traded non-financial firms
observed at a quarterly frequency over the period $2007$:Q1 - $2018$:Q3.
\normalsize

\begin{center}%
\begin{tabular}
[c]{cccc}\cline{2-4}\cline{3-4}\cline{4-4}
& (1) & (2) & (3)\\\hline
\textit{Panel A}: & \multicolumn{3}{c}{$\gamma_{pre}=\gamma_{post}=\gamma$}\\
$\hat{\gamma}$ & 0.76 & 0.56 & 0.56\\[0.2cm]%
\textit{Panel B}: & \multicolumn{3}{c}{$\gamma_{pre}\neq\gamma_{post}$}\\
$\hat{\gamma}_{pre}$ & 0.56 & 0.56 & 0.56\\
$\hat{\gamma}_{post}$ & 0.77 & 0.77 & 0.77\\\hline
linear trends & Yes & No & Yes\\
RGDP growth & No & Yes & Yes\\\hline
\end{tabular}

\end{center}

%

\end{table}%

Panel A of Table \ref{tab: thresholds_LSAPs} reports the estimated threshold
parameters for the single-threshold PanARDL(2) model (where $\gamma
_{pre}=\gamma_{post}=\gamma$) across different choices of $f_{t}$. The
estimated quantile threshold parameter, $\hat{\gamma}$, is equal to $0.76$
when using linear trends as a proxy for $f_{t}$, and it is smaller at $0.56$
when proxying $f_{t}$ by either real GDP growth or both linear trends and GDP
at the same time.

The difference in the estimates obtained for $\gamma$, depending on the choice
of $f_{t}$, only applies to the single-threshold case. Following the more
general model discussed in Subsection \ref{Motivation_Different_Thresholds},
we re-estimate the threshold parameters allowing these parameters to differ
over the periods pre- and post-introduction of LSAPs. The grid search
procedure is now carried out over values of $\gamma_{pre}$ and $\gamma_{post}$
in the grid formed by $0.25\leq\gamma_{pre}\leq0.9$ and $0.25\leq\gamma
_{post}\leq0.9$, in $0.01$ increments for both $\gamma_{pre}$ and
$\gamma_{post}$. The estimation results for this case are reported in panel B
of Table \ref{tab: thresholds_LSAPs}. It can be seen that we obtain the same
estimates $\hat{\gamma}_{pre}=$ $0.56$ and $\hat{\gamma}_{post}=0.77$, across
all the three choices of $f_{t}$. Both threshold estimates lie well within the
grid, with the estimate for the post LSAPs period being higher.

The estimates of $\gamma_{pre}$ suggest that the higher the proportions of
firms in an industry with relatively low levels of leverage (below median
levels), the more likely it is that firms in that industry can take advantage
of their lower debt burdens to increase their DA ratios, as compared to firms
in industries with higher proportions of more leveraged firms. In addition,
the estimates of $\gamma_{post}$ suggest that the Fed's purchases may have
also benefited firms with somewhat high debt levels conditional on not being
over-leveraged (i.e. with leverage below the upper quartile), with the effects
of LSAPs being stronger when the proportions of firms in an industry without
high debt burdens are higher. This may be due to the fact that these firms,
being less constrained by concerns over debt capacity, can act most
aggressively in response to LSAPs and increase their leverage ratios.

We shall see in the next subsection that the estimated policy coefficients
associated with the interaction of our measure of LSAPs and the
industry-specific threshold leverage variable, $\pi_{st}$, corroborate these hypotheses.

\subsection{Short-run effects of LSAPs and other drivers of firms' capital
structure}

Given the estimated threshold values, we now present the estimates of some of
the key parameters of the panel regressions in equation (\ref{ARDL_model_lsap}%
) using both fixed and time effects (FE--TE) over the period $2007$-Q1 to
$2018$-Q3. The results are summarized in Table \ref{tab:SR_benchmark} where we
report the estimates of the net short-run effects defined as the sum of
estimated coefficients of current and the $p$ lagged values of the regressor
under consideration. In this way we allow for possible over-shooting of the
estimates whereby a large positive initial impact may be reversed subsequently
with some negative lagged effects. For example, as seen in Section \ref{ARDL},
the policy net short-run (SR) effect is defined by $\beta_{1}=\sum_{\ell=0}^{p}%
\beta_{1,\ell}$, where $\beta_{1,\ell}$ is the coefficient of $q_{t-\ell
}\times\pi_{s,t-\ell-1}(\hat{\gamma}_{post})$ in the threshold-panel
regression defined by (\ref{ARDL_model_lsap}), with $p=2$ for the PanARDL(2) model.

The first three columns of Table \ref{tab:SR_benchmark} show results for the
single-threshold panel regression model across different choices of $f_{t}$,
while the last three columns report the estimates for the two-threshold model,
which is our preferred (more general) specification. Full panel regression
estimation results are provided in Section \ref{OS_Main_Results} of the online supplement.

The estimates of policy SR effects are positive and highly statistically
significant under all specifications while the magnitudes differ across them.
We find that the higher the \textit{ex ante} proportion of not over-leveraged
firms in an industry, the more effective the LSAPs in facilitating firms'
access to external financing. This corroborates our hypothesis that firms with
adequate debt capacity are the most responsive to the introduction of LSAPs.
Nevertheless, even the largest estimate of the policy SR effect obtained for
the two-threshold PanARDL(2) model at $0.0088$ $(0.0017)$ is rather
small in economic importance.%

\begin{table}[H]
\caption{\textbf{FE--TE estimates of the net short-run effects of LSAPs on debt to asset ratios of non-financial firms}}\label{tab:SR_benchmark}
\vspace{-0.2cm}
\footnotesize
Estimates of net short-run effects of LSAPs on firms' debt to asset ratios
(DA) as well as the effects of both firm- and industry-specific variables on
DA, for the PanARDL(2) model described in equation (\ref{ARDL_model_lsap}).
Net short-run effects are defined as the sum of the estimated coefficients of
current and lagged values of the regressor under consideration. The first
three columns report results for the single-threshold panel regression model,
where $\gamma_{pre}=\gamma_{post}$. The last three columns report results for
the two-threshold panel regression, where $\gamma_{pre}\neq\gamma_{post}$. The
estimated quantile threshold parameters are shown in Table
\ref{tab: thresholds_LSAPs}. All regressions include both firm-specific
effects and time effects. Columns (1) and (4) include industry-specific linear
time trends, columns (2) and (5) include the interaction of industry dummies
and real GDP growth, while columns (3) and (6) include both. $LSAP$ is the
(scaled) amount of U.S. Treasuries and agency MBS purchased by the Fed;
$\pi_{-1}(\gamma)$ denotes the one-quarter lagged proportion of firms in an
industry with DA below the $\gamma$-th quantile. The sample consists of an
unbalanced panel of $3,647$ U.S. publicly traded non-financial firms observed
at a quarterly frequency over the period $2007$:Q1 - $2018$:Q3. Robust
standard errors (in parentheses) are computed using the delta method (***
$p<0.01$, ** $p<0.05$, * $p<0.1$).
\normalsize

\begin{center}
\scalebox{0.85}{
			\begin{tabular}{l|ccc|ccc}
				\multicolumn{1}{c}{} & \multicolumn{6}{c}{Dependent variable: debt to assets (DA)} \\
				\cline{2-7} \cline{3-7} \cline{4-7} \cline{5-7} \cline{6-7} \cline{7-7}
				\multicolumn{1}{c}{} & \multicolumn{3}{c|}{$\gamma_{pre}=\gamma_{post}=\gamma$} & \multicolumn{3}{c}{$\gamma_{pre}\neq\gamma_{post}$} \\
				\multicolumn{1}{c}{} & (1) & (2) & (3) & (4) & (5) & (6) \\
				\hline
				$\pi_{-1}(\hat{\gamma}_{pre})$ & 0.0156{*}{*}{*} & 0.0101{*}{*} & 0.0167{*}{*}{*} & 0.0186{*}{*}{*} & 0.0123{*}{*}{*} & 0.0194{*}{*}{*} \\
				& (0.0049) & (0.0047) & (0.0054) & (0.0050) & (0.0046) & (0.0051) \\
				$LSAP\times\pi_{-1}(\hat{\gamma}_{post})$ & 0.0068{*}{*}{*} & 0.0035{*}{*} & 0.0041{*}{*}{*} & 0.0088{*}{*}{*} & 0.0060{*}{*}{*} & 0.0077{*}{*}{*} \\
				& (0.0019) & (0.0014) & (0.0015) & (0.0017) & (0.0017) & (0.0018) \\
				Lagged DA & 0.8386{*}{*}{*} & 0.8408{*}{*}{*} & 0.8386{*}{*}{*} & 0.8386{*}{*}{*} & 0.8409{*}{*}{*} & 0.8387{*}{*}{*} \\
				& (0.0050) & (0.0050) & (0.0050) & (0.0050) & (0.0050) & (0.0050) \\
				Cash to assets & -0.0365{*}{*}{*} & -0.0368{*}{*}{*} & -0.0365{*}{*}{*} & -0.0364{*}{*}{*} & -0.0368{*}{*}{*} & -0.0364{*}{*}{*} \\
				& (0.0030) & (0.0029) & (0.0030) & (0.0030) & (0.0029) & (0.0030) \\
				PPE to assets & 0.0219{*}{*}{*} & 0.0208{*}{*}{*} & 0.0218{*}{*}{*} & 0.0220{*}{*}{*} & 0.0207{*}{*}{*} & 0.0219{*}{*}{*} \\
				& (0.0047) & (0.0046) & (0.0047) & (0.0047) & (0.0046) & (0.0047) \\
				Size & 0.0034{*}{*}{*} & 0.0037{*}{*}{*} & 0.0034{*}{*}{*} & 0.0034{*}{*}{*} & 0.0037{*}{*}{*} & 0.0034{*}{*}{*} \\
				& (0.0008) & (0.0007) & (0.0008) & (0.0008) & (0.0007) & (0.0008) \\
				Industry leverage & 0.0626{*}{*}{*} & 0.0519{*}{*}{*} & 0.0645{*}{*}{*} & 0.0710{*}{*}{*} & 0.0548{*}{*}{*} & 0.0688{*}{*}{*} \\
				& (0.0075) & (0.0079) & (0.0092) & (0.0090) & (0.0080) & (0.0092) \\
				Industry growth & -0.1004{*}{*}{*} & -0.1334{*}{*}{*} & -0.1053{*}{*}{*} & -0.1064{*}{*}{*} & -0.1366{*}{*}{*} & -0.1093{*}{*}{*} \\
				& (0.0209) & (0.0206) & (0.0218) & (0.0210) & (0.0206) & (0.0219) \\
				\hline
				Fixed effects & Yes & Yes & Yes & Yes & Yes & Yes \\
				Time effects & Yes & Yes & Yes & Yes & Yes & Yes \\
				Industry linear trends & Yes & No & Yes & Yes & No & Yes \\
				Ind. dummy$\times$RGDP & No & Yes & Yes & No & Yes & Yes \\
				Observations & 84548 & 84548 & 84548 & 84548 & 84548 & 84548 \\
				$N$ & 3647 & 3647 & 3647 & 3647 & 3647 & 3647 \\
				$max(T_{i})$ & 44 & 44 & 44 & 44 & 44 & 44 \\
				$avg(T_{i})$ & 23.2 & 23.2 & 23.2 & 23.2 & 23.2 & 23.2 \\
				$med(T_{i})$ & 19 & 19 & 19 & 19 & 19 & 19 \\
				$min(T_{i})$ & 2 & 2 & 2 & 2 & 2 & 2 \\
				\hline
		\end{tabular}	}
\end{center}

%

\end{table}%

The effects of the industry-specific debt capacity indicator, $\pi_{st}$
(without the interaction with the LSAPs variable) on firms' leverage are also
positive and statistically significant. This indicates that the proportion of
firms without high debt burdens within an industry helps predicting firms'
financing decisions. This is in line with the findings of \cite{flannery2006}
and \cite{lemmon2010}, among others, who show that concerns over debt capacity
influence firm financing behaviour.

With respect to the other control variables, our findings are in line with the
existing literature on firms' capital structure. First, leverage appears to be
highly persistent, an aspect which has been widely documented (e.g.,
\cite{lemmon2008}). Second, firms with more tangible assets and larger size
tend to have higher leverage. Third, firms with higher cash holdings tend to
operate with lower leverage. This finding is in line with the results of
\cite{hadlock2010}, who document that more financially constrained firms hold
cash for precautionary reasons. Finally, as in previous empirical studies, we
find that industry median leverage is one of the key drivers of capital
structure. The associated coefficient is one of the most important in
magnitude. We also find that higher industry median growth results in lower
leverage in line with the trade-off theory's prediction (\cite{frank2009}).

\paragraph{Additional control variables.}

In addition to the variables included in our main model, we also estimate the
panel regressions with additional firm-specific regressors (such as market to
book ratio (MB), and research and development (R\&D) expense scaled by total
assets) and industry-specific controls, such as the industry median MB and
Tobin's Q ratio. Together with industry growth, industry MB and Q ratio are
used to control for differential growth opportunities across industries. To
further reduce possible omitted bias concerns, we also include the industry
medians of the firm-specific regressors contained in $\mathbf{x}_{is,t}$,
which together with industry leverage and industry growth help to control for
differences in industry conditions. On top of this, our regression always
include fixed and time effects, as well as the interactions of $\phi_{s}$ with
$f_{t}$. The results reported in Section \ref{AddVar} of the online
supplement, show that the estimates of the policy effectiveness coefficients
and their statistical significance are not affected by the inclusion of these
additional control variables.

\paragraph{On the choice of $f_{t}$.}

We have seen that our estimation results hold across the three choices of
$f_{t}$, with the strongest estimation results obtained when simply allowing
for firm leverage to follow different time trends across industries. Let $M1$,
$M2$, and $M3$ denote the model for each choice of $f_{t}$, namely linear
trends, real GDP growth, and both linear trends and GDP at the same time,
respectively. To compare these three specifications, we test the joint
significance of the associated industry coefficients, $\phi_{s}$, using simple
Chow-type tests. Focusing on the more general two-threshold specification, we
obtain a F-test equal to $2.26$ and $1.32$ in $M1$ and $M2$,
respectively.\footnote{For a $0.05$ level of significance, the critical
F-value with $66$ degrees of freedom in the numerator and more than $120$
degrees of freedom in the denominator is equal to $1.3$.} This means we are
able to reject the null of $\phi_{s}=0$ for all $s$ at the $0.01$ and $0.05$
significance level, respectively. Thus, we find stronger statistical support
for the case of simple linear trends. Finally, we test the joint significance
of the $\phi_{s}$ in $M3$. Also in this case we are able to reject the null at
the $0.05$ significance level. When comparing $M3$ directly with $M1$ (which
is a restricted version of $M3$), the F-test is equal to $1.13$ which is below
the $10\%$ critical F-value, meaning we are not able to reject the null of all
$\phi_{s}$ associated with real GDP growth being zero when the model also
includes linear trends. Overall, we find that estimation results hold across
the three models but with stronger statistical support when simply using
(scaled) linear trends as a proxy for $f_{t}$, possibly because the model
already includes fixed time effects. Similar reasoning apply to the
single-threshold panel regressions.

\subsection{Half-life, mean lag, and long-run effects of LSAPs}

Another important question is whether the Fed's asset purchases had
long-lasting effects on firms' capital structure. While there is some evidence
on the persistence of the effects of LSAPs on corporate and Treasury yields,
albeit with some contrasting results (e.g., \cite{greenlaw2018},
\cite{swanson2021}, and \cite{wright2012}), less attention has been paid, to
the best of our knowledge, on how this translated into firms' preference about
their leverage ratios. Our dynamic panel model provides a suitable setting to
answer this question. To this end, we estimate the long-run effects of LSAPs
on firms' leverage ratios to measure the magnitude of the total impact of such purchases.

Results are shown in panel A of Table \ref{tab:LR_benchmark} where as before,
the first three columns report results for the single-threshold panel
regression model while the last three columns display results for the
two-threshold panel regression, for each choice of $f_{t}$. For brevity, we
focus on the long-run effects of LSAPs, defined by equation
(\ref{LR_definition}), with $p=2$ for the PanARDL(2) mode.\footnote{In Section
\ref{Apx_LR_bench} of the online supplement we also report the long-run
effects of the other regressors.} We find more economically meaningful effects
of LSAPs on firms' capital structure in the long-run. Our results are in line
with the findings of \cite{ihrig2018} on the persistent effects of the Fed's
asset purchases on yields, extending this evidence to firm leverage. We show
that LSAPs significantly contributed to higher debt to asset ratios in the
long-run, although the magnitude of the effects suggests that concerns over
firms' excessive risk-taking (in the forms of higher debt ratios) due to LSAPs
were at least in part overstated.

Taking advantage of our empirical approach, we can also compute the mean lag
of the effects of LSAPs, i.e. the average number of quarters it takes for
firms' leverage to return to the long-run equilibrium. Another important
measure of persistence is the half-life, which we define as the number of
periods required for the peak response of firm debt to assets to LSAPs to
dissipate by one half.\footnote{See Section \ref{Appx_HalfL_MeanL} of the
online supplement for calculations of half-life and mean lag.}%

\begin{table}[H]
\caption{\textbf{Half-life, mean lag, and long-run effects of LSAPs on debt to asset ratios of non-financial firms}}\label{tab:LR_benchmark}
\vspace{-0.2cm}
\footnotesize
Panel A reports estimates of the long-run effects of LSAPs, defined in
equation (\ref{LR_definition}), on firms' debt to asset ratios (DA) for the
PanARDL(2) model described in equation (\ref{ARDL_model_lsap}). $LSAP$ denotes
the (scaled) amount of U.S. Treasuries and agency MBS purchased by the Fed;
$\pi_{-1}(\gamma)$ denotes the one-quarter lagged proportion of firms in an industry with DA below the
$\gamma$-th quantile. Panel B displays the estimated mean lag and half-life of
LSAPs. The first three columns report results for the single-threshold panel
regression model, where $\gamma_{pre}=\gamma_{post}$. The last three columns
report results for the two-threshold panel regression, where $\gamma_{pre}%
\neq\gamma_{post}$. The estimated quantile threshold parameters are shown in
Table \ref{tab: thresholds_LSAPs}. All regressions include both firm-specific
effects and time effects. Columns (1) and (4) include industry-specific linear
time trends, columns (2) and (5) include the interaction of industry dummies
and real GDP growth, while columns (3) and (6) include both. Further
information on the sample used can be found in Table \ref{tab:SR_benchmark}.
Robust standard errors (in parentheses) are computed using the delta method
(*** $p<0.01$, ** $p<0.05$, * $p<0.1$).
\normalsize

\begin{center}
\scalebox{0.85}{
			\begin{tabular}{lcccccc}
				\cline{2-7} \cline{3-7} \cline{4-7} \cline{5-7} \cline{6-7} \cline{7-7}
				\multicolumn{1}{c}{} & \multicolumn{3}{c|}{$\gamma_{pre}=\gamma_{post}=\gamma$} & \multicolumn{3}{c}{$\gamma_{pre}\neq\gamma_{post}$} \\
				\multicolumn{1}{c}{} & (1) & (2) & \multicolumn{1}{c|}{(3)} & (4) & (5) & (6) \\
				\hline
				\multicolumn{1}{c}{} & \multicolumn{6}{c}{\textit{Panel A: Long-run effects of LSAPs}} \\
				$\pi_{-1}(\hat{\gamma}_{pre})$ & 0.0966{*}{*}{*} & 0.0637{*}{*} & 0.1033{*}{*}{*} & 0.1152{*}{*}{*} & 0.0772{*}{*}{*} & 0.1203{*}{*}{*} \\
				& (0.0306) & (0.0300) & (0.0335) & (0.0316) & (0.0291) & (0.0321) \\
				$LSAP\times\pi_{-1}(\hat{\gamma}_{post})$ & 0.0424{*}{*}{*} & 0.0220{*}{*} & 0.0254{*}{*}{*} & 0.0546{*}{*}{*} & 0.0379{*}{*}{*} & 0.0475{*}{*}{*} \\
				& (0.0116) & (0.0087) & (0.0092) & (0.0108) & (0.0106) & (0.0112) \\ [0.2cm]
				\multicolumn{1}{c}{} & \multicolumn{6}{c}{\textit{Panel B: Mean lag and half-life of LSAPs}} \\
				Mean lag & 6.1 & 5.6 & 5.6 & 6.0 & 5.7 & 5.8 \\
				Half-life & 5.0 & 4.0 & 5.0 & 5.0 & 4.0 & 5.0 \\
				\hline
				linear trends & Yes & No & Yes & Yes & No & Yes \\
				RGDP growth & No & Yes & Yes & No & Yes & Yes \\
				\hline
		\end{tabular}	}
\end{center}

%

\end{table}%

As shown in Panel B of Table \ref{tab:LR_benchmark}, the mean lags vary
between $5.6$ and $6.1$ quarters, while our estimates of half-life vary
between $4$ to $5$ quarters depending on the model specification used. These
results show that the effects of LSAPs do not dissipate immediately. It may
take a few quarters for these effects to play out, supporting the view that
the impact of LSAPs on firm leverage can be quite persistent. Our results
align more closely with the findings of \cite{swanson2021} showing that the
effects of LSAPs on yields tended to be very persistent as opposed to
\cite{wright2012} who document that the effects of the Fed's unconventional
monetary policy announcements on yields have a half-life of less than three
months.\footnote{It should be noted that event studies, by construction, tend
to capture asset market reactions over only a short period
(\cite{bernanke2020}).}

Overall, our results suggest that LSAPs facilitated firms' access to credit,
and that their effectiveness depends on the ability of firms to issue new debt
safely. The higher the proportion of firms without high leverage ratios in an
industry, the stronger the response of firms to LSAPs in the same industry. We
also document that the effects of LSAPs are long lasting.


\subsection{The effects of LSAPs at industry and national levels}

\begin{figure}[H]
	\caption{\textbf{Average policy effects ($\overline{PE}_{s}\times
100$) at the industry level ordered by industry median leverage}}
	\renewcommand\thefigure{1} \centering{\includegraphics
[scale=0.4]{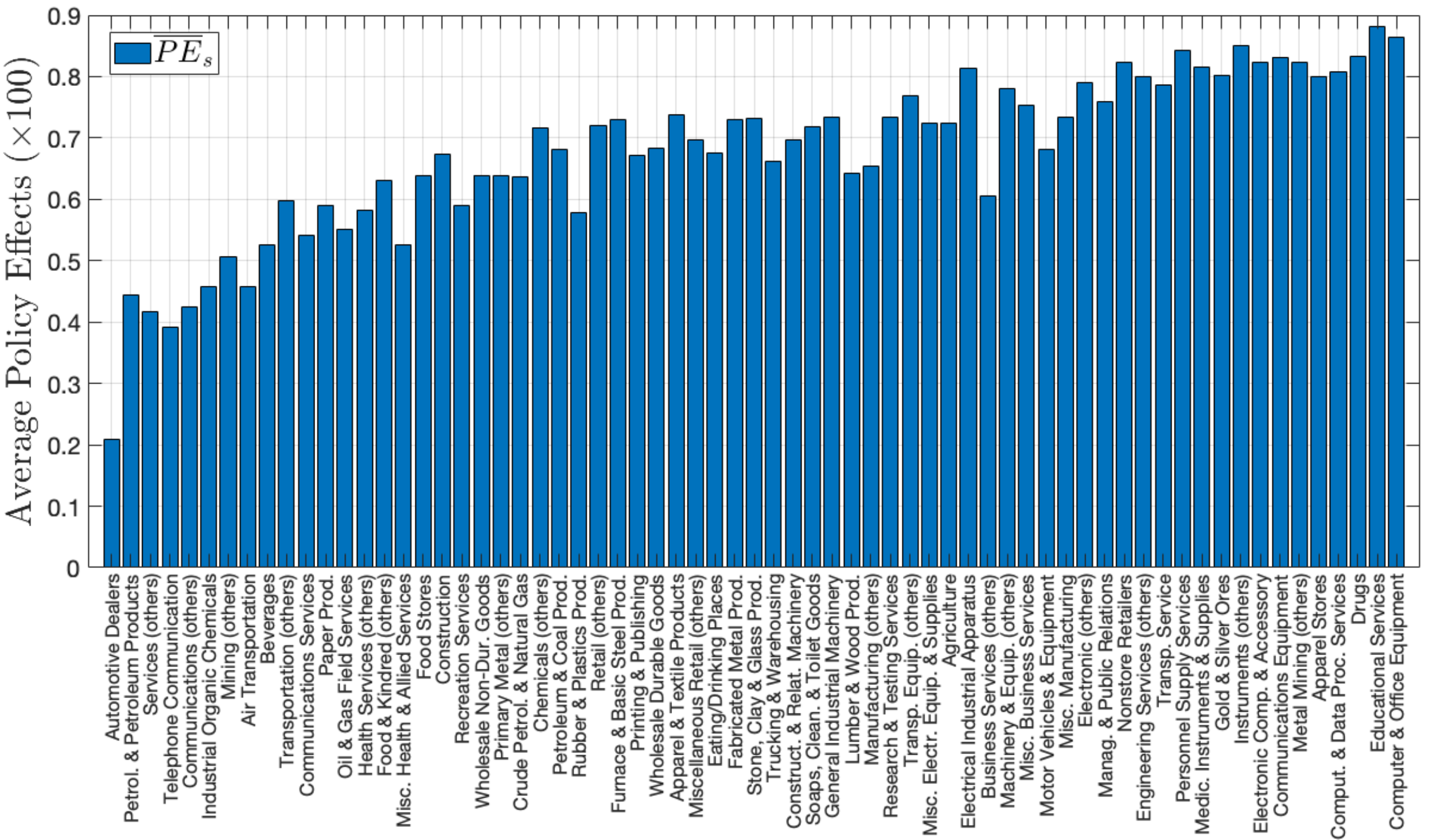}}
	\parbox{0.9\textwidth}{\footnotesize
{The blue bars display the average policy effects at the industry level
			described in equation (\ref{PE_s}), based on the interaction of our quantitative measure of LSAPs and one-quarter lagged values of $\pi
(\hat{\gamma}_{post})$.
			The x-axis reports the three-digit SIC industries sorted from largest to smallest
			industry median leverage, averaged over time. }} \label
{fig:APE_DA_ARDL2_two-thresh}
\end{figure}

We now discuss the estimates of the average policy effects (APE) at the
industry and national levels as set out in equations (\ref{PE_s}) and
(\ref{PE}), respectively. For brevity, we focus on the results for our
preferred specification, namely the two-threshold PanARDL(2) model with
$f_{t}$ denoting industry linear trends. The estimates are displayed in Figure
\ref{fig:APE_DA_ARDL2_two-thresh}. The blue bars report the estimated APE by
industry based on the interaction of our quantitative measure of LSAPs and
lagged the leverage threshold variable ($\pi$). Three-digit SIC industries are
sorted from largest to smallest industry median leverage (averaged over time).%

The estimates show a relatively high degree of heterogeneity in the effects of
LSAPs on firms' debt to asset ratios across industries, driven by
cross-industry variation in the proportions of firms without high debt burdens
($\pi(\hat{\gamma}_{post})$). The APE vary from $0.0021$ for the automotive
dealers' industry, which is one of the industries in our sample with largest
median leverage, to $0.0088$ for educational services' industry, one of the
least leveraged industry in our sample. As another example, we note that
airlines which typically rely more on debt financing than software companies
(see \cite{baker2009}) also tended to benefit less from LSAPs.

The policy effects at the national level are computed as averages using
industry-specific weights. As weights we consider both employment and firm
size shares over the full sample.\footnote{To compute the employment shares we
use annual data at the firm-level from the Compustat annual database. See
Section \ref{PE_appx} of the online supplement for more details on the weights
used.} The estimated APE at the national level is equal to $0.0065$ and
$0.0066$ when using average employment and firm size as share of an industry
in the economy, respectively. Due to the relatively large number of industries
in our sample, the weights do not seem to have a big impact on the estimated
national effects, and in fact using equal constant weights across industries
leads to a similar estimate, namely $0.0068$. We have also experimented with
using average sectoral employment or firm size over a three-year period
(instead of over the entire sample) to compute the weights, obtaining very
similar results.

These estimates once again highlight the rather small magnitude of the LSAPs'
effects despite the statistical significance of the underlying estimates.


\subsection{Robustness of the results to small--T bias}

It is well known that standard within-group estimators for linear dynamic
panel data models with fixed effects suffer from small--$T$ bias. In our
application, after using the first $3$ observations to generate the lagged
values as regressors (recall that $p_{max}=2$), we end up with a highly
unbalanced panel with the number of time series observations in panel
regressions ($T_{i}$ for firm $i$) ranging from $2$ to $44$, that correspond
to $5$ and $47$ available quarterly observations. The main reason for
including firms with $T_{i}=2$ observations in the panel regressions was to
avoid sample selection bias that could result from dropping newly founded
firms with a short history. However, the inclusion of such firms could lead to
small $T$ bias which we address here.

We approach the problem from two perspectives. First we consider the
implications of dropping firms with very few time series observations and see
if this makes that much of a difference to the estimates of the policy
effects. Accordingly, we re-estimate equation (\ref{ARDL_model_lsap})
including firms with at least $8$ or $10$ time series observations. As
documented in Section \ref{small-T_bias_OS} of the online supplement, the
streamlining of the data set to reduce the small--$T$ bias does not seem to
have meaningful effects on the estimates or their statistical significance.
The estimates of the net short-term policy effects are hardly affected by
dropping firms with very few time series observations. This is partly due to
the rather low proportion of firms in our sample with fewer than $8$ or $10$ observations.

Whilst this is reassuring, the FE-TE estimates could still be subject to the
small $T$ bias, since there is a large number of firms in our sample with
$T<20$, as documented in \cite{chudik2018} (CPY henceforth). Therefore, as a
second robustness check, we examine the extent to which our estimation results
hold after correcting for the small--$T$ bias by applying the half-panel
jackknife method also proposed by CPY.\footnote{See Section
\ref{small-T_bias_OS} of the online supplement for more details.} This
estimation procedure is well suited for our empirical analysis as it allows
for fixed and time effects, and it is appropriate for both balanced and
unbalanced panels with large cross-section dimension and moderate T. In
addition, it yields more accurate inference in the presence of weakly
exogenous regressors.\footnote{In particular, the half-panel jackknife method
is applicable even when the error terms are correlated with future values of
the regressors without requiring to specify the particular nature of weak
exogeneity of the regressors. We refer the interested reader to CPY for
further details.}

The implementation of the half-panel jackknife bias correction requires
splitting the time series observations on each firm into equal sub-samples,
with each sub-sample having at least $2$ observations. With this in mind, we
include firms with at least $8$ time series observations, and in the case of
firms with odd numbers of observations, we follow CPY and drop the first
observation before dividing the sample into two sub-samples. We then apply
\cite{wansbeek1989} transformation to remove the fixed and time effects from
each of the two sub-samples separately, before computing the half-panel
jackknife estimators.\footnote{Because of the super consistency property of
the threshold estimators, to compute the jackknife estimator we use the
threshold parameters estimated in the main specification as reported in Table
\ref{tab: thresholds_LSAPs}.}

The first notable implication of this new estimation strategy is the larger
estimates obtained for the coefficients of the lagged dependent variables,
which is in line with the known downward bias of the corresponding FE
estimates (\cite{nickell1981}). We also find that amongst the control
variables, cash to assets, industry leverage and industry growth continue to
be highly statistically significant, while the estimates for the PPE to asset
ratio and firm size become statistically insignificant. This is due to the
fact that jackknife standard errors are generally larger than the standard
FE--TE estimates which tend to be under-estimated.

More importantly, the estimates of net short-run policy effects continue to be
highly statistically significant even after applying the jackknife bias
correction. The jackknife estimates of the $SR$ for the two-threshold
PanARDL(2) model vary between $0.0061$ and $0.0080$ depending on the choice of
$f_{t}$, in line with standard FE-TE estimates shown in column (4) to (6) of
Table \ref{tab:SR_benchmark}. However, due to the larger estimates obtained
for the coefficients of lagged dependent variables, the estimated long-run
effects of LSAPs are much larger after the jackknife bias correction. Based on
the standard FE--TE estimates, the long-run policy effects for the
two-threshold PanARDL(2) model are estimated to vary between $0.0379$ and
$0.0546$ across the various specifications considered (as shown in Table
\ref{tab:LR_benchmark}). By comparison, the jackknife estimates vary between
$0.1224$ and $0.1584$, depending on the choice of $f_{t}$.

\section{Heterogeneous effects of various Fed's asset purchase programs}

\subsection{Short- and long-run effects across LSAP programs}

In the previous section, we have quantified the overall effects of the Fed's
total Treasuries and MBS purchases on firm leverage. We now consider the
effects of each asset class purchases separately. Thus, our regression model
now includes two distinct quantitative measures of LSAPs interacted with our
one-quarter lagged industry debt capacity indicator. These two policy measures
are defined as follows:
\begin{equation}%
\begin{array}
[c]{ccc}%
ty_{t}=\frac{TY_{t}}{\frac{1}{T-T_{0}}\sum_{\tau=T_{0}+1}^{T}Q_{\tau}}, &  &
mbs_{t}=\frac{MBS_{t}}{\frac{1}{T-T_{0}}\sum_{\tau=T_{0}+1}^{T}Q_{\tau}%
},\label{quanti_measures}%
\end{array}
\end{equation}
where $Q_{t}=MBS_{t}+TY_{t}$, with $TY$ and $MBS$ denoting the gross amount of
U.S. Treasuries and agency MBS purchased by the Fed, respectively.

For brevity, we focus on the more general two-threshold PanARDL(2)
specification which includes both fixed and time effects while allowing
industries to follow different time-trends.\footnote{Additional results are
provided in Section \ref{sec:TY_vs_MBS} of the online supplement.} We
re-estimate the quantile threshold parameters associated with the debt
capacity indicator, $\pi_{st}(\gamma)$, by grid search obtaining the same
results to those described in Table \ref{tab: thresholds_LSAPs}, namely
$\hat{\gamma}_{pre}$ is equal to $0.56$ while $\hat{\gamma}_{post}$ is equal
to $0.77$.

Estimation results are provided in Panel A of Table \ref{tab:heter_effects}
where we report both the net short-run (SR) and long-run (LR) effects of both
MBS and Treasury purchases. We find that both type of purchases facilitated
firm credit access. MBS purchases have a slightly higher impact on firm debt
to assets relative to Treasuries purchases, which in turn results into
marginally stronger long-run effects. The estimated coefficients associated
with MBS purchases have also a higher degree of statistical significance.
These results seem in line with the argument of \cite{krishnamurthy2013} that
MBS purchases tended to be more beneficial for the economy than Treasury purchases.

A slightly different picture emerges when using the firm-specific debt
capacity indicator, $d_{is,t}(\gamma)$. As shown in Section
\ref{Appx:firmdebtcpcty} of the online supplement, both Treasury and MBS
purchases have significant impacts on firm leverage but the effects are now
stronger for Treasuries.\footnote{The identification strategy based on
$d_{is,t}(\gamma)$ exploits cross-firm variation within an industry, implying
that firms which are not over-leveraged should benefit more from LSAPs
relative to peers in the same industry. However, estimation based on $\pi
_{st}(\gamma)$ exploits cross-industry variation, suggesting that firms in
less leveraged industries should benefit more, thus allowing for spillover
effects within an industry. This may explain some of the differences in the
estimation results based on the two approaches.} Taken together, these results
suggest that both MBS and Treasury purchases can facilitate firms' access to
credit, although the magnitude of the effects is rather small.

Our estimates, whether based on industry-average or firm-specific measures of
debt capacity, provide strong empirical evidence in support of the hypothesis
that non-financial firms with spare debt capacity benefited from MBS
purchases. This finding is in line with the results obtained by
\cite{rodnyansky2017} and corroborates the evidence of \cite{gagnon2011} which
suggest that LSAPs had wide ranging effects on borrowing costs not limited
exclusively to the type of asset being purchased by the Fed.\footnote{There is
not a consensus on this in the literature. For example,
\cite{krishnamurthy2013} emphasize the so called narrow channel, whereby asset
purchases have a stronger impact on the yields of the asset being purchased.
At the same time, it is widely accepted that asset purchases may have broader
effects on financial markets (beyond those on the asset being purchased)
through the reduction of investors' expectations on the path of the federal
funds rate (e.g., \cite{woodford2012}), and by easing financial conditions. Our
empirical findings demonstrate that these financial market impacts have
translated into greater reliance on external financing for non-financial
firms.} Because our identification strategy does not confine the effects of
LSAPs to firm-bank relationships, we do not find a \textquotedblleft crowding
out\textquotedblright\ effect as in \cite{chakraborty2020}, and instead
document that MBS purchases had positive effects on firm
leverage.\footnote{According to the \textquotedblleft crowding
out\textquotedblright\ behaviour documented by \cite{chakraborty2020},
following the Fed's MBS purchases, banks increased their mortgage origination
at the expense of commercial and industrial lending.}%

\begin{table}[H]
\caption{\textbf{FE--TE estimates of the net short-run (SR) and long-run (LR) effects of various LSAPs on debt to asset ratios of non-financial firms}}\label{tab:heter_effects}
\vspace{-0.2cm}
\footnotesize
Estimates of the net short-run (SR) and long-run (LR) effects, 
defined in Section \ref{ARDL}, of various Fed's
asset purchase programs on firms' debt to asset ratios (DA) for the
two-threshold PanARDL(2) model described in equation (\ref{ARDL_model_lsap}).
Panel A focuses on two quantitative measure of LSAPs, whereby $ty$ and $mbs$
denote the (scaled) amount of U.S. Treasuries and agency MBS purchased by the
Fed, respectively. Panel B displays results for the qualitative measures of
LSAPs, a set of dummy variables which take the value of one during policy on
periods and zero otherwise. $\pi_{-1}(\gamma)$ denotes the one-quarter lagged
proportion of firms in an industry with DA below the $\gamma$-th quantile. All
regressions include both firm-specific effects and time effects as well as
industry-specific linear time trends. Further information on the sample used
can be found in Table \ref{tab:SR_benchmark}. Robust standard errors (in
parentheses) are computed using the delta method (*** $p<0.01$, ** $p<0.05$, *
$p<0.1$).
\normalsize

\begin{center}%
\begin{tabular}
[c]{lccclcc}%
\multicolumn{3}{c}{\textit{Panel A: Treasuries versus MBS purchases}} &  &
\multicolumn{3}{c}{\textit{Panel B: Major QE programs}}\\\cline{1-3}%
\cline{2-3}\cline{3-3}\cline{5-7}\cline{6-7}\cline{7-7}
& SR & LR &  &  & SR & LR\\\cline{1-3}\cline{2-3}\cline{3-3}\cline{5-7}%
\cline{6-7}\cline{7-7}%
$\pi_{-1}(\hat{\gamma}_{pre})$ & 0.0188{*}{*}{*} & 0.1166{*}{*}{*} &  &
$\pi_{-1}(\hat{\gamma}_{pre})$ & 0.0199{*}{*}{*} & 0.1233{*}{*}{*}\\
& (0.0050) & (0.0317) &  &  & (0.0051) & (0.0321)\\
$ty\times\pi_{-1}(\hat{\gamma}_{post})$ & 0.0078{*}{*} & 0.0482{*}{*} &  &
$QE_{1}\times\pi_{-1}(\hat{\gamma}_{post})$ & 0.0189{*}{*}{*} & 0.1169{*}%
{*}{*}\\
& (0.0031) & (0.0191) &  &  & (0.0044) & (0.0278)\\
$mbs\times\pi_{-1}(\hat{\gamma}_{post})$ & 0.0092{*}{*}{*} & 0.0569{*}{*}{*} &
& $QE_{2}\times\pi_{-1}(\hat{\gamma}_{post})$ & 0.0114{*}{*} & 0.0705{*}{*}\\
& (0.0021) & (0.0134) &  &  & (0.0053) & (0.0332)\\\cline{1-3}\cline{2-3}%
\cline{3-3}
&  &  &  & $MEP\times\pi_{-1}(\hat{\gamma}_{post})$ & 0.0019 & 0.0120\\
&  &  &  &  & (0.0038) & (0.0234)\\
&  &  &  & $QE_{3}\times\pi_{-1}(\hat{\gamma}_{post})$ & 0.0069{*}{*} &
0.0426{*}{*}\\
&  &  &  &  & (0.0033) & (0.0204)\\\cline{5-7}\cline{6-7}\cline{7-7}%
\end{tabular}

\end{center}

%

\end{table}%

It is also interesting to compare the effects of each Fed's program separately
by replacing the two aforementioned quantitative measures of LSAPs with four
qualitative variables which take the value of one during policy on periods and
zero otherwise. Following the literature, we label these policy indicators as
QE1 (covering the period 2008Q4 to 2010Q1), QE2 (2010Q4 - 2011Q2), MEP (the
maturity extension program of 2011Q3 - 2012Q4), and QE3 (2012Q3 -
2014Q4).\footnote{Additional information on these four large-scale asset
purchase programs can be found in Section \ref{Appx: Var_Construct} of the
online supplement.} Also in this case, we re-estimate the quantile threshold
parameters by grid search obtaining results in line with previous findings,
namely $\hat{\gamma}_{pre}$ is equal to $0.56$, while $\hat{\gamma}_{post}$ is
equal to $0.73$.\footnote{Here, we focus on the two-threshold PanARDL(2)
specification allowing industries to follow different time-trends. Additional
results are provided in Section \ref{Apx_QE_sep} of the online supplement.}

Estimation results for the $(0,1)$ policy indicators are shown in Panel B of
Table \ref{tab:heter_effects}, where we report both the SR and LR effects of
each LSAP episode. Our results show some degree of variation in the
effectiveness of the four Fed's programs covered in our sample. In particular,
QE1 had the largest impact of firm debt to assets. This result is consistent
with previous findings in the literature where QE1 is typically found to have
the largest impact on MBS and Treasury yields, as well as corporate bond
yields (e.g., \cite{kuttner2018}). Our results also corroborate the view that
LSAPs can be particularly effective during periods of dysfunctions in
financial markets (e.g., \cite{damico2013}).

In line with the arguments of \cite{bernanke2020}, we also find that LSAPs can
also be effective outside periods of market stress. In particular, we find
that both QE2 and QE3 had statistically significant effects on firm leverage
albeit smaller in magnitude than QE1. Instead, the effects of MEP are much
lower in magnitude and not statically significant.\footnote{Differences across
the QE1, QE2, and QE3 are less pronounced when employing the firm-specific
debt capacity indicator. MEP continues to have the lowest impact, and is
generally non significant at the 5 per cent level.}

\subsection{Half-life and mean lag lengths of different Fed's QE programs}

We now discuss the time profile of the effects of the various Fed's asset
purchase programs separately. In particular, we report the mean lags and
half-life of the effects of each program in Table \ref{tab:HL_heterog}. The
first two columns focus on our quantitative measures. We find that Treasuries
and MBS purchases have the same mean lag length of $4$ quarters although the
half-life of MBS purchases is higher. In particular, the effects of Treasury
purchase dissipate by one half (from peak) after four quarters as opposed to
six quarters for MBS purchases, while the average number of periods after
which the effects of both purchases dissipate is about six quarters. These
results further corroborate the hypothesis that the effects of LSAPs do not
die out immediately but are instead quite persistent, in line with the
arguments of \cite{bernanke2020}.%

\begin{table}[H]
\caption{\textbf{Half-life and mean lag of various LSAP programs}}\label{tab:HL_heterog}
\vspace{-0.2cm}
\footnotesize
Estimates of mean-lag and half-life for the policy effectiveness coefficient,
$\beta_{1}$, associated with the interaction of our measures of LSAPs and
$\pi_{-1}(\gamma)$, the one-quarter lagged proportion of firms in an industry
with debt to assets below the $\gamma$-th quantile. Estimates are based on
two-threshold PanARDL(2) model described in equation (\ref{ARDL_model_lsap}),
which includes both fixed and time effects as well as industry-specific linear
trends.
\normalsize

\begin{center}%
\begin{tabular}
[c]{c|cc|cccc}\cline{2-7}\cline{3-7}\cline{4-7}\cline{5-7}\cline{6-7}%
\cline{7-7}%
\multicolumn{1}{c}{} & $ty$ & $mbs$ & $QE_{1}$ & $QE_{2}$ & $MEP$ & $QE_{3}%
$\\\hline
Mean lag & 6.1 & 6.1 & 6.4 & 6.0 & 4.5 & 5.5\\
Half-life & 4.0 & 6.0 & 4.0 & 5.0 & 0.0 & 0.0\\\hline
\end{tabular}

\end{center}

%

\end{table}%

Turning on each individual LSAP episode, using the four qualitative policy
indicators, we find that the mean lag lengths are higher for the first two
programs, with the magnitude being in line with those obtained under the
quantitative measures.%

\begin{figure}[H]
	\caption{\textbf
{Distributed effects of LSAPs on non-financial firms' debt to asset ratios}}
	\renewcommand\thefigure{1} \centering{\includegraphics
[scale=0.5]{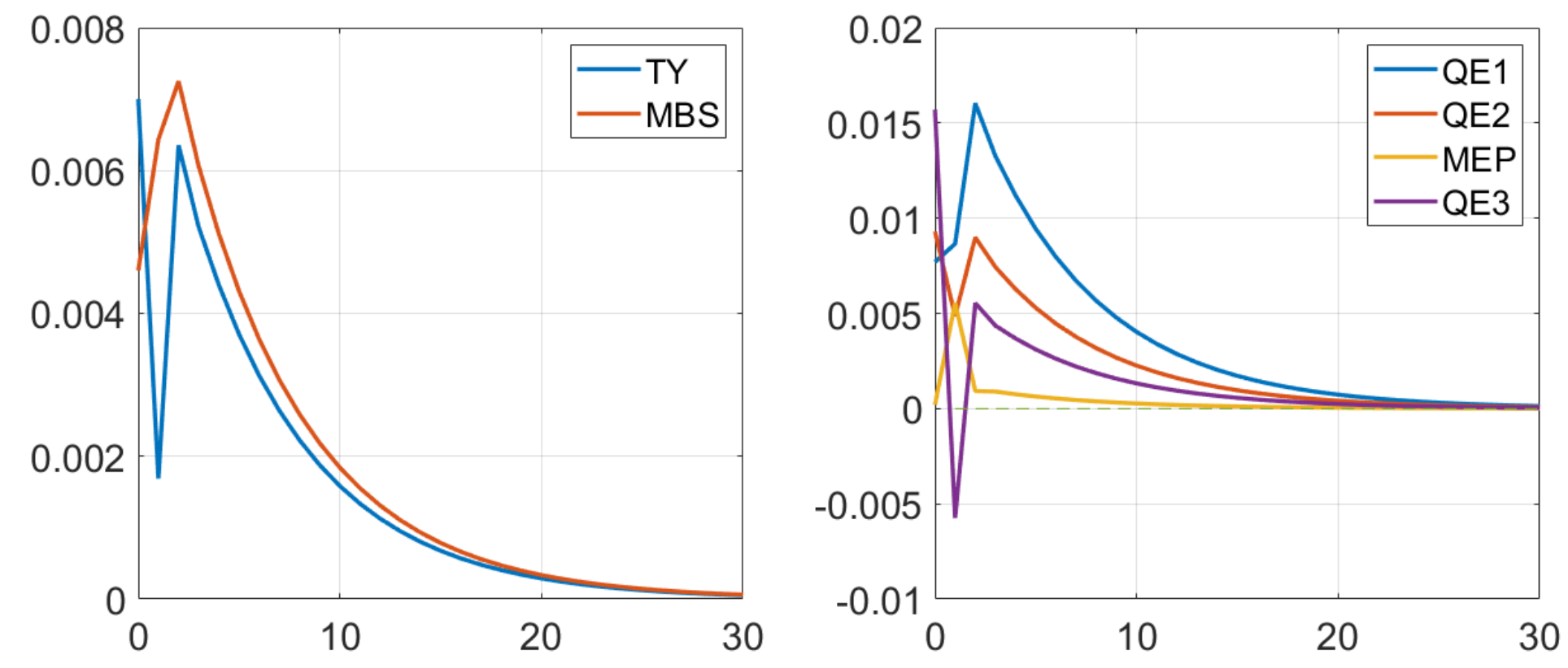}}
	\parbox{0.9\textwidth}{\footnotesize
{Plots of the distributed effects of LSAPs on firm debt to asset ratios. The left-hand side panel displays results for the quantitative measure of LSAPs described in (\ref
{quanti_measures}%
), namely $ty$ and $mbs$, denoting the scaled gross amount of U.S. Treasuries and agency MBS purchased by the Fed, respectively. The right-hand side panel shows results for the four qualitative policy indicators. The distributed effects are computed after rewriting the two-threshold ARDL(2) model described in equation (\ref
{ARDL_model_lsap}) in its distributed lag form. }}
	\label{fig:time_profile}
\end{figure}%

One notable difference emerges in the estimated half-lives. In particular, the
half-lives of both MEP and QE3 are zero, meaning that their effects on firm
leverage decrease by more than half immediately after having reached their
peak. This can be clearly seen in Figure \ref{fig:time_profile}, where we plot
the lag distribution of the effects of each LSAP program on firm debt to asset
ratios, computed after re-writing the PanARDL(2) model in its distributed lag
form.\footnote{Details on the derivation of the distributed lag form, and the
computation of mean lag and half-life are provided in Section
\ref{Appx_HalfL_MeanL} of the online supplement.} While the initial impact is
closer to zero, the effects of MEP reach its peak the subsequent quarter to
then fall close to zero immediately after. QE3 has a substantial immediate
impact on firm leverage which is partly offset the next quarter.

\subsection{Robustness of the results to small--T bias}

Results continue to hold after correcting for potential small-sample
bias.\footnote{See Section \ref{Appx:MBS_TY_SmallT} and
\ref{Apx_QE_sep_smallT} of the online supplement for more details on the
estimation results for the quantitative and qualitative measures of LSAPs,
respectively.} In particular, the effects of our quantitative measure of MBS
purchases are similar when only including firms with at least $8$ or $10$ time
series observations. The magnitude of the impact of MBS purchases increases to
$0.0107$ (0.0025) when applying the half-panel jackknife method, resulting in
a much more meaningful long-term impact ($0.2124$ (0.0580)). Turning to
Treasury purchases, their effects continue to hold when selecting firms with
at least $8$ or $10$ observations but they become insignificant when
implementing the jackknife bias correction.

Regarding the four qualitative measures of LSAPs, we note that the findings on
QE1 continue to hold even after applying the aforementioned small-T bias
corrections. Results are less clear-cut for QE2 and QE3. The effects of QE3
remain statistically significant when applying the half-panel jackknife method
but the same cannot be said for QE2. The effects of MEP remain statistically insignificant.

\section{Concluding remarks \label{Con}}

We estimate panel ARDL models with threshold effects to
quantify both the short- and long-term impacts of the Fed's LSAPs on firms'
capital structure. To disentangle the effects of LSAPs from that of concurrent
macroeconomic conditions, we exploit variations, within and across industries, in the ability of firms to raise additional external funds without exhausting their
debt capacity. To this end, we construct firm- and industry-specific measures of spare
debt capacity, which we then interact with our measures of LSAPs. 

The industry-specific measure of debt capacity is defined as the proportion of firms in an industry with debt to assets below a given threshold, and has the merit of taking into account interdependencies in financing decisions across firms within each industry. We treat the threshold quantile as an unknown parameter, and find that the quantile value that gives the best fit in our preferred specification is equal
to 0.77. We then test whether a higher proportion of firms in an industry with
leverage below the $77th$ quantile predicts a stronger impact of LSAPs on
firms' capital structure. We find robust evidence in support of this
hypothesis. Our results demonstrate that existing debt burdens within an
industry are a good predictor of a firm's ability to increase its debt
financing in response to the Fed's asset purchases.

When separating the effects of MBS from Treasury purchases, we find that both
programs facilitated firm credit access and that their effects do not
dissipate immediately. We also examine the effects of the first four episodes
of LSAPs separately, to find that the first round of purchases (QE1) had the
largest positive impact on firms' external financing. At the same time, our
results show that LSAPs can also be effective outside periods of market stress.

Finally, our dynamic panel data models enable us to identify the time profile
of the effects of LSAPs on firms' capital structure. Our analysis provides a
clear and strong evidence that such effects are long-lasting, extending 
the existing findings on the persistence of the effects
of LSAPs on interest rates to firm leverage. To the best of
our knowledge, this aspect has not received enough attention so far.

To conclude, our results suggest that LSAPs facilitated firms' access to
external debt financing, and that their effectiveness depends on the ability
of firms within an industry to raise new debt ``safely''. At the same time,
albeit highly statistically significant, the relatively small magnitude of the
estimated long-run effects indicates that LSAPs have contributed only
marginally to the rise in U.S. corporate debt ratios of the last decade or so.

\pagebreak

%

\nocite{*}%

\bibliographystyle{chicago}
\bibliography{Bibliography_QE}


\pagebreak

\begin{center}
{\Large Online Supplement for}

{\Large \textbf{Causal effects of the Fed's large-scale asset purchases on
firms' capital structure}}

\vspace{1cm}

Andrea Nocera

Norges Bank Investment Management \vspace{0.5cm}

M. Hashem Pesaran

University of Southern California, USA, and Trinity College, Cambridge, UK
\vspace{0.5cm}

October 2023
\end{center}

\section*{Introduction}

This online supplement is organized in six sections. Section
\ref{Appendix_DataAnalysis} provides detailed information on the data used in
the empirical analysis, and the various filters used in the sample selection
process. It also discusses the classification of firms by industries while
providing several summary statistics at both firm and industry levels.

Section \ref{Appendix_Threshold_Optim} provides additional information on both
the identification and estimation strategy, while Section
\ref{Appx_HalfL_MeanL} describes the concept of half-life and mean lag, and
their calculations.

Section \ref{OS_Main_Results} reports the estimation results for the threshold
panel autoregressive distributed lag, PanARDL(2), specifications where the
macro policy intervention variable, $q_{t}$, is the scaled gross amount of
U.S. Treasuries and mortgage-backed securities (MBS) purchased by the Fed. We
also report several robustness exercises. Subsection \ref{AddVar} presents the
estimates of the policy effectiveness coefficients when including several
additional control variables to the main regressions. Subsection
\ref{MacroVar} shows estimation results using a number of alternative
macro-variables interacted with industry-specific dummies. Subsection
\ref{small-T_bias_OS} shows estimation results after correcting for potential
small-sample bias. In Subsection \ref{Apx_Dyn_Specif} we also report results
for the standard partial adjustment model and the PanARDL(1) specification.

In Section \ref{sec:TY_vs_MBS}, we show estimation results when separating the
effects of MBS from Treasury purchases. In Section \ref{Apx_QE_sep}, we
compare the effects of each Fed's asset purchase program by replacing the two
aforementioned quantitative measures of LSAPs with four qualitative variables
which take the value of one during policy on periods and zero otherwise.

In Section \ref{Appx:firmdebtcpcty}, we report results using a firm-specific
measure of debt capacity. \thispagestyle{empty}\pagebreak

\appendix
%

\setcounter{equation}{0}%
\renewcommand{\theequation}{\Alph{section}.\arabic{equation}}%
%

\setcounter{table}{0}%
\renewcommand\thetable{\thesection.\arabic{table}}%
%

\setcounter{page}{1}%
\renewcommand{\thepage}{S\arabic{page}}%

\section{Data sources, data filters and summary statistics
\label{Appendix_DataAnalysis}}

This section provides detailed information on the data used in our empirical
analysis. In subsection \ref{Appx: Var_Construct}, we describe the main
variables of our dataset. In subsection \ref{Sample_Sel_Appdx}, we discuss the
sample selection screens. Summary statistics are reported in subsection
\ref{Stats}. In subsection \ref{Industry_Stats}, we describe our
classification of firms by industries. Subsection \ref{[Stat_Industry]}
provides some summary statistics at the industry level.

\subsection{Construction of the dependent and explanatory
variables\label{Appx: Var_Construct}}

Table \ref{table:Definition_Variable} describes the main firm- and
industry-specific variables used in our empirical analysis, which are obtained
from Compustat (quarterly) database.%

\begin{table}[H]
\caption{\textbf{List of variables and definitions}}\label{table:Definition_Variable}
\vspace{-0.2cm}
\footnotesize
This table describes the main variables considered in our empirical analysis.
The market to book ratio is based on \cite{badoer2016}. To calculate the
Tobin's Q we use the definition of \cite{duchin2010} which is the ratio of the
market value of assets (MVA) to a weighted average of MVA and total assets
(TA). When data on deferred taxes (\textit{txdbq}), used in the construction
of MVA, are missing we set them equal to zero. This is consistent with the
numerator used in the definition of Tobin's Q in \cite{foley2016}. By
construction, our measure of Tobin's Q is bounded above at $10$. Following
\cite{badoer2016}, when computing research and development expense
(\textit{xrdq}) scaled by total assets, we set \textit{xrdq} to zero if missing.

\begin{center}
\scalebox{0.75}{
			\begin{tabular}
				[c]{lp{10cm}p{6cm}}\hline
				\textbf{Variable} & \textbf{Definition} & \textbf{Compustat}\\\hline
				Total debt to total assets  & Sum of short- and long-term debt
				scaled by total assets & (dlttq+dlcq)/atq\\[0.1cm]
				Long-term debt to TA  & Long-term debt scaled by total assets &
				dlttq/atq\\[0.1cm]
				Short-term debt to TA  & Debt in current liabilities scaled by total
				assets & dlcq/atq\\[0.1cm]
				Debt to equity  & Ratio of total debt to book value of equity &
				(dlttq+dlcq) / ceqq\\[0.1cm]
				Market to book & Market capitalization divided by total book value &
				(ltq-txditcq+prccq{*}cshoq+pstkq)/atq.\\[0.1cm]
				Market value of assets (MVA) & The sum of total assets and market value of common equity minus common equity and deferred taxes & (atq + (cshoq{*}prccq)
				- ceqq - txdbq)\\[0.1cm]
				Tobin's Q  & Market value of assets divided by a weighted sum of book value
				of assets ($0.9$) and market value of assets ($0.1$). & (MVA)/(0.9{*}atq +
				0.1{*}MVA)\\[0.1cm]
				Cash to TA  & Cash and short-term investments scaled by total assets &
				cheq/atq\\[0.1cm]
				Cash flow to TA  & Sum of income before extraordinary items and
				depreciation and amortization scaled by total assets & (ibq + dpq)/atq\\[0.1cm]
				PPE to TA  & Property, plant, and equipment scaled by total assets &
				ppentq/atq\\[0.1cm]
				R\&D to TA  & Research and development expense scaled by total assets &
				xrdq/atq\\[0.1cm]
				Size & Natural logarithm of total assets & log(atq)\\[0.1cm]
				Median industry growth & Median change in the log of total assets
				within each industry by quarter & \\[0.1cm]
				Median industry leverage  & Median debt to asset ratios within
				each industry by quarter & \\[0.1cm]\hline
		\end{tabular} }
\end{center}

%

\end{table}%

\paragraph{Large-scale asset purchases (LSAPs).}

In our empirical analysis, our preferred measure of LSAPs is the total gross
amount of U.S. Treasuries and agency mortgage-backed securities (MBS)
purchased by the Fed. To construct this quantitative measure, we obtain data
from the New York Fed's website. U.S. Treasuries' purchases include notes,
bonds, and to a much smaller extent Treasury Inflation-Protected Securities
(TIPS). We also report results using qualitative measures of LSAPs. In this
case, our policy variable is a set of dummy variables equal to one during
policy on periods and zero otherwise. To construct these variables we obtain
information on the operation dates from the New York Fed's website. Further
details are given in Table \ref{Table_LSAP} which provides a short summary of
the Fed's asset purchase programs until 2018, including the dates of implementation.%

\begin{table}[H]
\caption{\textbf{Description of the major large-scale asset purchase programs}}\label{Table_LSAP}
\vspace{-0.2cm}
\footnotesize
The dates and description of the various Fed's interventions are obtained from
the New York Fed's website
(https://www.newyorkfed.org/markets/programs-archive/large-scale-asset-purchases).
See also \cite{kuttner2018} and \cite{swanson2021}. MEP stands for Maturity
Extension Program, also known as Operation Twist. MBS stands for
mortgage-backed securities.
\normalsize
\vspace{-0.5cm}%

\begin{center}%
\begin{tabular}
[c]{cccp{10cm}}\hline
Program & Start Date & End Date & \multicolumn{1}{c}{Description}\\\hline
QE1 & Nov 2008 & Mar 2010 & The Fed purchased \$175 billion (bn) in agency
debt, \$1,250bn in agency MBS, and \$300bn in longer-term Treasury
securities.\\[0.1cm]%
QE2 & Nov 2010 & Jun 2011 & The Fed purchased \$$600$bn of longer-dated
Treasuries.\\[0.1cm]%
MEP & Sep 2011 & Dec 2012 & The Fed purchased \$$667$bn of $6$- to $30$-year
Treasuries offset by sales of \$634bn in Treasuries with remaining maturities
less or equal to 3 years and \$33 billion of Treasuries' redemptions.
Principal payments from agency debt and MBS were also reinvested.\\[0.1cm]%
QE3 & Sep 2012 & Oct 2014 & The Fed purchased \$$40$bn in agency MBS per month
from Sep 2012 until Dec 2013, and \$$45$bn of long-term Treasuries per month
throughout 2013. In Jan 2014 the purchases of MBS and long-term Treasuries
dropped to \$35bn and \$40bn per month, respectively. Both purchases decreased
by \$5bn after each FOMC meeting until October 2014. In total, the Fed
purchased \$790bn in Treasury securities and \$823bn in agency MBS.\\\hline
\end{tabular}

\end{center}

%

\end{table}%

\begin{figure}[H]
	\caption{\textbf{Fed's large-scale asset purchases}}
	\renewcommand\thefigure{2}  \centering{\includegraphics
[scale=0.50]{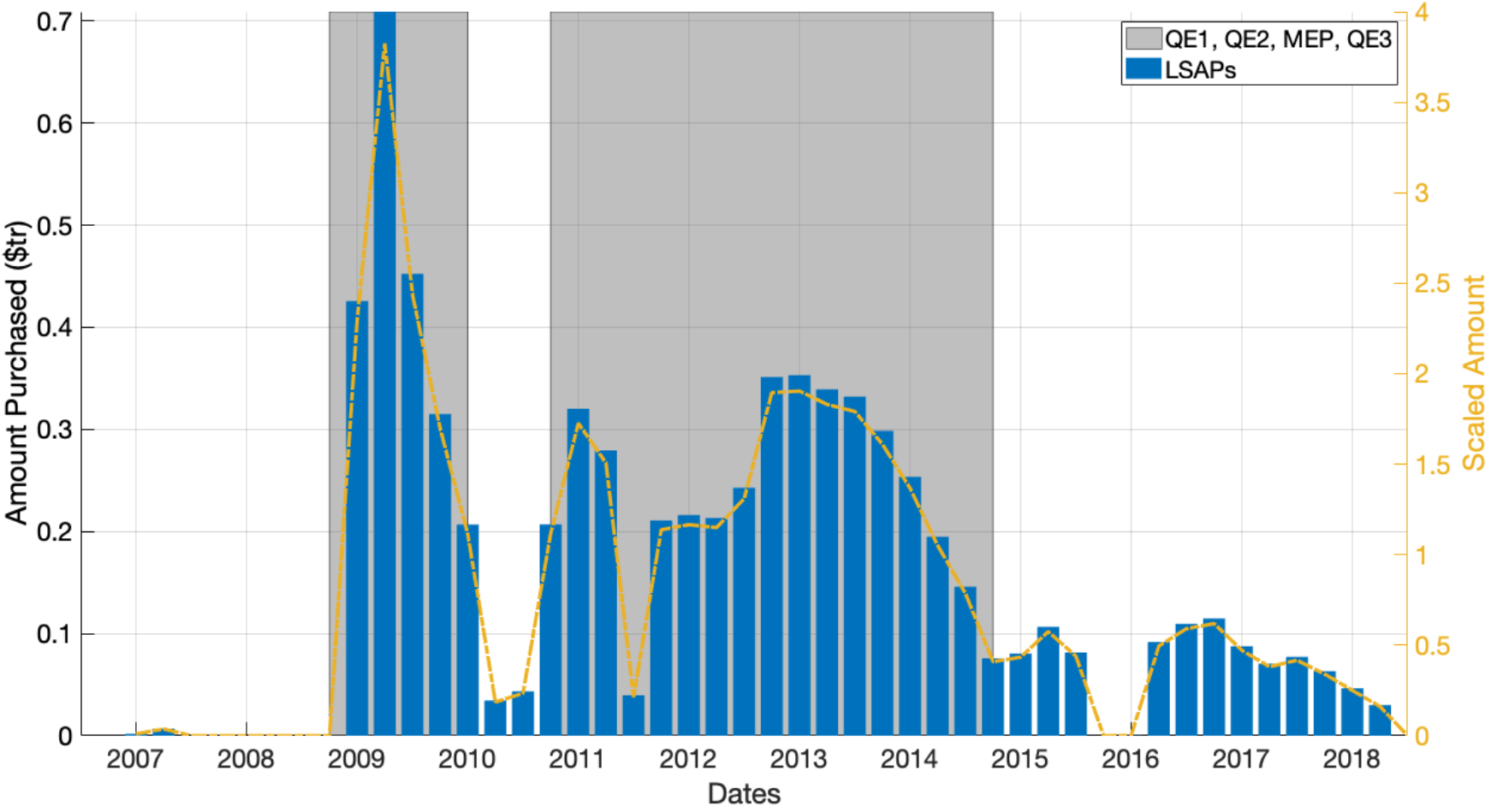}}
	\parbox{0.9\textwidth}{\footnotesize
{The blue bars display quarterly purchases (in trillion dollars) of U.S. Treasuries and agency mortgage-backed securities by the Federal Reserve. The yellow line shows our scaled amount of LSAPs measured on the right-hand side y-axis. The scale used is such that its average value is unity over the period where purchases took place. The shaded grey areas denote the main Fed's interventions over the sample period considered, as described in Table \ref
{Table_LSAP}. Source: New York Fed.}} \label{fig:LSAP_vs_Dummy}
\end{figure}

To make our quantitative measure of LSAPs directly comparable to the
qualitative (dummy) policy variables, we scale the former so that its average
value is unity over the policy sample. This scaling also facilitates the
interpretations of the estimation results by removing the unit of measurement
of the variable. The dynamics of LSAPs are depicted in Figure
\ref{fig:LSAP_vs_Dummy}.%

\paragraph{Macroeconomic indicators.}

In addition, to linear trends we also employ the following macroeconomic indicators:

\begin{itemize}
\item \textit{Real GDP growth} is the percent changes from preceding quarter
in real gross domestic product obtained from the U.S. Bureau of Economic
Analysis. The extracted data are already seasonally adjusted, and the percent
changes are expressed at annual rates.

\item \textit{World real GDP growth} is the annualized log difference of
(seasonally adjusted) world real GDP obtained from the World Bank.

\item \textit{Unemployment rate} is the U.S (seasonally adjusted) unemployment
rate in percent, obtained from the Federal Reserve Bank of St. Louis (FRED).
Quarterly data are obtained by averaging monthly observations within a quarter.

\item \textit{Term spread} is the difference between the 10-year and the
3-month Treasury bond yields. The 10-year yield is the market yield on U.S.
Treasury securities at 10-year constant maturity, quoted on investment basis,
obtained from the Federal Reserve System's website. The data are available on
a daily frequency and are converted into a quarterly frequency by averaging
over a quarter.

\item \textit{Expected inflation} denotes expectations (i.e. median forecasts)
for one-year-ahead annual average CPI inflation. The series is contained in
the Survey of Professional Forecasters conducted by the Federal Reserve Bank
of Philadelphia.
\end{itemize}

\subsection{\bigskip Data filters and sample selection
\label{Sample_Sel_Appdx}}

To align our analysis with previous studies (e.g., \cite{leary2014}), we
disregard observations from financial firms (SIC 6000-6999) and regulated
utilities (SIC 4900-4999) whose financing choices may dictated by regulatory
considerations, as well as from firms belonging to the non-classifiable sector
(SIC codes above or equal to 9900) which in our sample mainly consists of
non-operating firms (i.e. firms that operate no assets on their own).

The sample period includes years from $2007$-Q1 to $2018$-Q3. We drop firms
with gaps in between periods for the following variables: (i) total debt to
total assets (TA), (ii) cash to TA, (iii) market to book, (iv) property plant
and equipment (PPE) to TA, and (v) size.

We select only firms with at least $5$ consecutive time observations based on
the above firm characteristics. This choice is dictated by our econometric
strategy which uses autoregressive distributed lag (PanARDL) models.

We exclude firms with total debt to TA greater than one, and exclude firms
with negative total debt. In total there is only one firm with negative debt
which we remove. Finally, we note that the following variables - debt to
equity (DE), market to book (MB), cash flow to TA (CF2A), and R\&D to TA -
take implausible values for a small number of firms. This is shown in Table
\ref{Table_Drop2}. In the upper panel, it reports various percentiles for the
above mentioned firm characteristics. The lower panel shows the number of
firms associated with those percentiles. To remove the effects of these
outliers we proceed as follows. First, we drop firms with DE and CF2A below
the $0.05\%$ or above the $99.95\%$ percentiles, as well as firms with MB and
R\&D to TA above the $99.95\%$ percentiles. We also drop firms with negative
R\&D. We then winsorize DE and CF2A at the $1$st and $99$th percentiles, and
both MB and R\&D to TA at the $99$th percentile.

Table \ref{Table_Drop} reports the number of firms dropped after removing the outliers.%

\begin{table}[H]
\caption{\textbf{Percentiles (\%) and number of firms by percentiles after applying
			all filters but before removing outliers}}\label{Table_Drop2}
\vspace{-0.2cm}
\footnotesize
The upper panel reports various percentiles for those firm characteristics
which show implausible values. The lower panel displays the number of firms
with values below (above) the lower (upper) percentiles. For example, after
applying all filters but before removing the outliers for market to book,
there are $22$ firms with market to book above $2746.21$, the $99.95\%$ percentile.

\begin{center}%
\begin{tabular}
[c]{c|cccc|cccc}\hline
Variable \textbackslash{} Percentile (\%) & \textbf{min} & \textbf{0.05} &
\textbf{0.1} & \textbf{0.2} & \textbf{99.8} & \textbf{99.9} & \textbf{99.95} &
\textbf{max}\\\hline
\textbf{Debt to equity} & -2995.95 & -198.22 & -101.93 & -51.54 & 67.43 &
148.02 & 268.38 & 38732.00\\
\textbf{Market to book} & 0.03 & 0.10 & 0.16 & 0.23 & 294.83 & 873.28 &
2746.21 & 146344.76\\
\textbf{Cash flow to TA} & -855.55 & -9.39 & -5.00 & -2.55 & 0.37 & 0.55 &
0.82 & 105.00\\
\textbf{R\&D to TA} & -1.09 & 0.00 & 0.00 & 0.00 & 0.49 & 0.70 & 0.95 &
41.00\\\hline
\multicolumn{9}{c}{}\\\hline
& \multicolumn{4}{c|}{\textbf{N. Firms with values} $<p_{\tau}$} &
\multicolumn{4}{c}{\textbf{N. Firms with values} $>p_{\tau}$}\\
&  & \textbf{0.05} & \textbf{0.1} & \textbf{0.2} & \textbf{99.8} &
\textbf{99.9} & \textbf{99.95} & \\\hline
\textbf{Debt to equity} &  & 46 & 76 & 127 & 134 & 83 & 48 & \\
\textbf{Market to book} &  & 9 & 21 & 42 & 52 & 36 & 22 & \\
\textbf{Cash flow to TA} &  & 29 & 50 & 90 & 157 & 83 & 46 & \\
\textbf{R\&D to TA} &  & 42 & 58 & 58 & 91 & 54 & 29 & \\\hline
\end{tabular}

\end{center}

%

\end{table}%
%

\begin{table}[H]
\caption{\textbf{Number of firms dropped while removing outliers}}\label{Table_Drop}
\vspace{-0.2cm}
\footnotesize
We (sequentially) drop firms whose debt to equity ratio is lower (greater)
than the $0.05\%$ ($99.95\%$) percentile, firms whose market to book ratio is
greater than the $99.95\%$ percentile, and firms with cash flow to TA lower
(greater) than the $0.05\%$ ($99.95\%$) percentile. We also exclude firms with
negative R\&D to TA as well as firms with R\&D to TA greater than the
$99.95\%$ percentile. TA stands for total assets.
\normalsize

\begin{center}%
\begin{tabular}
[c]{lcccc}\hline
& \multicolumn{2}{c}{Lower Tail} & \multicolumn{2}{c}{Upper Tail}\\
& Drop if & N. Drops & Drop if & N. Drops\\\hline
\textbf{Debt to equity} & $<0.05\%$ & 46 & $>99.95\%$ & 36\\
\textbf{Market to book} &  &  & $>99.95\%$ & 20\\
\textbf{Cash flow to TA} & $<0.05\%$ & 21 & $>99.95\%$ & 45\\
\textbf{R\&D to TA} & $<0$ & 54 & $>99.95\%$ & 14\\\hline
Tot. &  & 121 &  & 115\\\hline
\end{tabular}

\end{center}

%

\end{table}%

\begin{figure}[H]
	\caption{\textbf{Sample selection}}
	\renewcommand\thefigure{3}  \centering{\includegraphics
[scale=0.55]{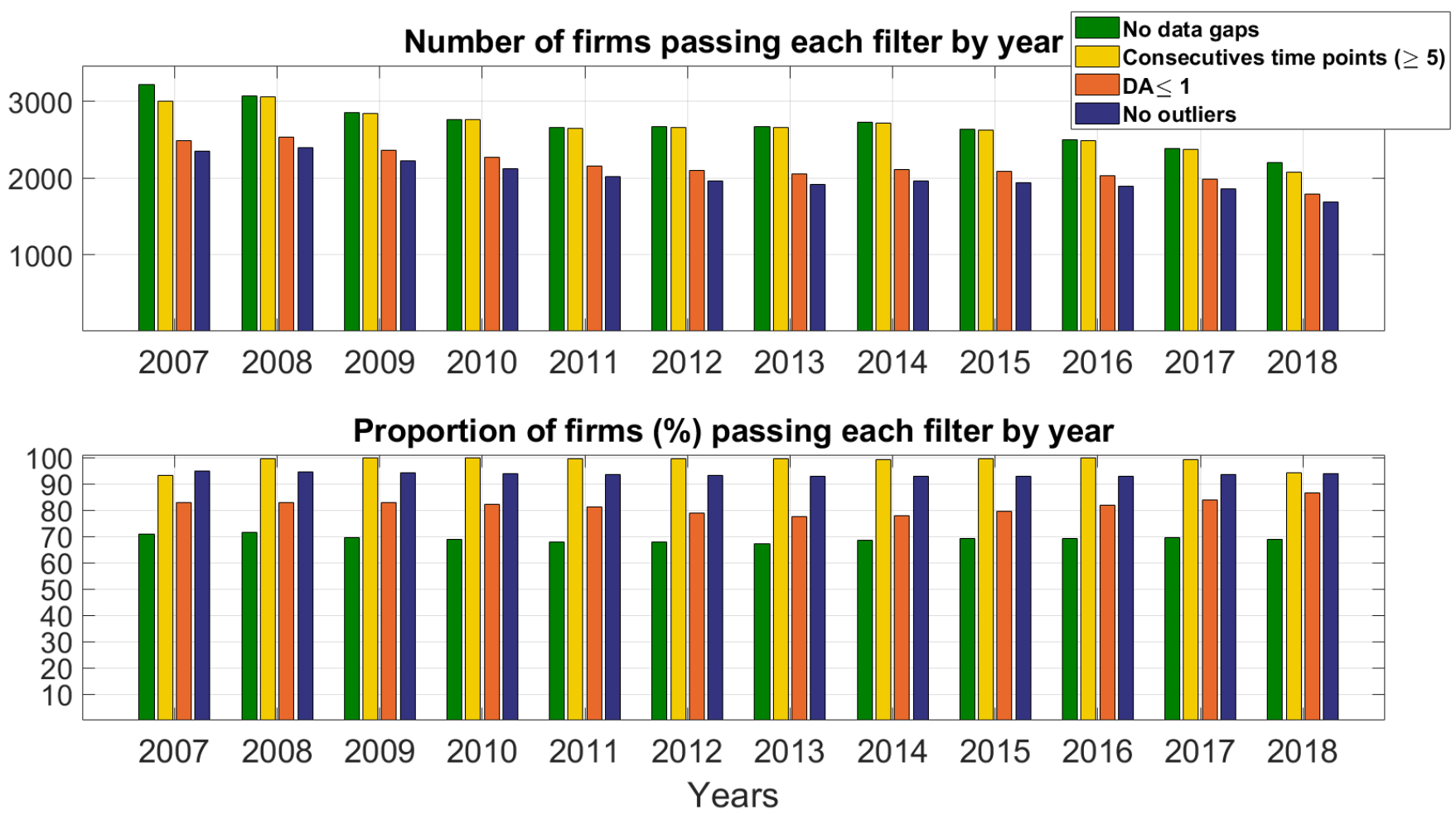}}
	\parbox{0.9\textwidth}{\footnotesize
{Annual statistics are obtained by averaging quarterly statistics
			within each year. The upper panel shows the number of firms available by year after applying each filter.
			The lower panel displays the percentage of firms that pass each filter by year. We consider four filters.
			First, we drop firms with data gaps in between period in total debt to total assets (TA), cash to TA, market to book,
			PPE to TA, and size (green bars). Second, we drop firms with less than $5$ consecutive time observations (yellow bars).
			Third, we exclude firms with a ratio of debt to assets greater than $1$ (orange bars).
			Finally, we remove firms with outliers (blues bars).}} \label
{fig:Sample_Selection}
\end{figure}

To summarise our sample selection screens, Figure \ref{fig:Sample_Selection}
displays the number of firms and percentage of firms selected each year in our
sample after applying each filter. Annual statistics are obtained by averaging
quarterly statistics within each year.%

More details are provided in Table \ref{Table}, where we report the empirical
frequency distribution of firms by year as well as the percentage of firms
that pass each filter by year.%

\begin{table}[H]
\caption{\textbf{Empirical frequency distribution of firms by year}}\label{Table}
\vspace{-0.2cm}
\footnotesize
Columns $2$ to $5$ display the number of firms per year after applying each
filter. Annual statistics are obtained by averaging quarterly statistics
within each year. The columns \textit{\% Pass F1}, \textit{\% Pass F2}, and
\textit{\% Pass F3} report the percentage of firms that pass the first filter
(no data gaps), the percentage of firms remaining after applying the second of
filter ($\geq5$ time points), and the percentage of firms that pass the third
filter (debt to asset (DA) ratios less or equal to $1$), respectively. Column
\textit{\% Pass F4} shows the percentage of selected firms with no outliers.
Column \textit{\% All Filters} denotes the percentage of firms meeting all
four filters, computed as the ratio of the total number of selected firms to
the total number of firms available before applying any filter, in percentage
terms.
\vspace{-0.5cm}%

\begin{center}
\scalebox{0.6}{
			\begin{tabular}
				[c]{cccccccccc}\hline
				\textbf{Year} & \textbf{No data gaps} & \textbf{Consecutive time points
					($\geq5$)} & \textbf{DA $\leq1$} & \textbf{No outliers} & \textbf{\% Pass
					F1} & \textbf{\% Pass F2} & \textbf{\% Pass F3} & \textbf{\% Pass F4} &
				\textbf{\% All filters}\\\hline
				2007 & 3213.5 & 2995.8 & 2485.0 & 2352.5 & 70.8 & 93.3 & 83.0 & 94.7 & 51.9 \\
				2008 & 3069.8 & 3058.0 & 2530.3 & 2389.8 & 71.4 & 99.6 & 82.7 & 94.4 & 55.6 \\
				2009 & 2850.0 & 2845.5 & 2359.8 & 2218.8 & 69.6 & 99.8 & 82.9 & 94.0 & 54.2 \\
				2010 & 2764.5 & 2758.0 & 2268.3 & 2125.3 & 68.8 & 99.8 & 82.2 & 93.7 & 52.9 \\
				2011 & 2660.3 & 2650.0 & 2154.3 & 2014.8 & 67.7 & 99.6 & 81.3 & 93.5 & 51.3 \\
				2012 & 2674.5 & 2658.8 & 2100.5 & 1956.3 & 67.7 & 99.4 & 79.0 & 93.1 & 49.5 \\
				2013 & 2668.5 & 2653.5 & 2057.3 & 1910.8 & 67.2 & 99.4 & 77.5 & 92.9 & 48.1 \\
				2014 & 2727.8 & 2709.3 & 2106.3 & 1956.8 & 68.6 & 99.3 & 77.7 & 92.9 & 49.2 \\
				2015 & 2635.0 & 2626.0 & 2086.3 & 1936.8 & 69.0 & 99.7 & 79.5 & 92.8 & 50.8 \\
				2016 & 2496.0 & 2490.8 & 2034.0 & 1891.5 & 69.3 & 99.8 & 81.7 & 93.0 & 52.5 \\
				2017 & 2388.8 & 2367.5 & 1981.8 & 1855.5 & 69.5 & 99.1 & 83.7 & 93.6 & 54.0 \\
				2018 & 2201.7 & 2075.7 & 1793.3 & 1685.0 & 69.0 & 94.2 & 86.5 & 94.0 & 52.8 \\
				\hline
				\textbf{Tot. num. firms} & 5,666 & 4,946 & 3,883 & 3,647 &  &  &  &  &  \\
				\textbf{Min num. quarters} & 1 & 5 & 5 & 5 &  &  &  &  &  \\
				\textbf{Mean num. quarters} & 22.4 & 25.4 & 26.3 & 26.2 &  &  &  &  &  \\
				\textbf{Median num. quarters} & 18 & 22 & 22 & 22 &  &  &  &  &  \\
				\textbf{Max num. quarters} & 47 & 47 & 47 & 47 &  &  &  &  &  \\
				\textbf{Tot. firm-quarter obs.} & 127,199 & 125,479 & 102,034 & 95,489 &  &  &  &  &  \\
				\hline
		\end{tabular} }
\end{center}

%

\end{table}%

After applying all filters, we end up with a sample of $3,647$ distinct firms.
The total number of firm-quarter observations is $95,489$.\footnote{The actual
number of firm-quarter observations used in our empirical analysis is slightly
lower due to the presence of lagged dependent and explanatory variables.} The
panel data is unbalanced with the number of time series data points available
by firm varying between $5$ and $47$, on average $26.2$ quarters.

\subsection{Summary statistics \label{Stats}}

This subsection provides some summary statistics. Table \ref{Sum_Stats} shows
how data on firm characteristics considered change after applying each data
filter. Summary statistics for selected firm characteristics computed on the
final filtered sample are reported in Table \ref{Sum_Stats_Filtered_Sample}.
Table \ref{Table_Freq_Obs} reports the frequency of firms by number of
consecutive data points (based on the filtered sample).%

\begin{table}[H]
\caption{\textbf{Summary statistics after applying each filter}}\label{Sum_Stats}
\vspace{-0.2cm}
\footnotesize
Summary statistics for selected firm characteristics after applying each
filter. TA stands for total assets.
\vspace{-0.5cm}%

\begin{center}
\scalebox{0.75}{
			\begin{tabular}
				[c]{cccccccccccc}\hline
				& \textbf{Filter} & \textbf{Obs.} & \textbf{mean} & \textbf{std} &
				\textbf{min} & \textbf{max} & \textbf{1\%} & \textbf{25\%} & \textbf{50\%} &
				\textbf{75\%} & \textbf{99\%}\\\hline
				& None & 178897 & 2.03 & 61.60 & -0.05 & 18116.00 & 0.00 & 0.01 & 0.19 & 0.41 & 16.21 \\
				& No Gaps & 127199 & 1.69 & 31.47 & -0.05 & 5319.00 & 0.00 & 0.01 & 0.19 & 0.41 & 16.82 \\
				\textbf{Total debt to TA} & 5 Cont. Obs. & 125479 & 1.60 & 26.83 & -0.05 & 3172.48 & 0.00 & 0.01 & 0.19 & 0.41 & 16.53 \\
				& TD2A$\leq1$ & 102034 & 0.20 & 0.20 & 0.00 & 1.00 & 0.00 & 0.00 & 0.15 & 0.33 & 0.79 \\
				& No Outliers & 95489 & 0.20 & 0.20 & 0.00 & 1.00 & 0.00 & 0.00 & 0.15 & 0.32 & 0.76 \\
				& Winsoriz. & 95489 & 0.20 & 0.20 & 0.00 & 1.00 & 0.00 & 0.00 & 0.15 & 0.32 & 0.76 \\
				\hline
				& None & 182475 & 0.36 & 7.72 & -0.12 & 2071.00 & 0.00 & 0.00 & 0.10 & 0.31 & 2.00 \\
				& No Gaps & 127199 & 0.35 & 5.89 & -0.12 & 836.50 & 0.00 & 0.00 & 0.08 & 0.29 & 2.24 \\
				\textbf{Long-term debt to TA} & 5 Cont. Obs. & 125479 & 0.35 & 5.93 & -0.12 & 836.50 & 0.00 & 0.00 & 0.08 & 0.29 & 2.19 \\
				& TD2A$\leq1$ & 102034 & 0.16 & 0.19 & 0.00 & 1.00 & 0.00 & 0.00 & 0.09 & 0.27 & 0.74 \\
				& No Outliers & 95489 & 0.16 & 0.18 & 0.00 & 1.00 & 0.00 & 0.00 & 0.09 & 0.27 & 0.71 \\
				& Winsoriz. & 95489 & 0.16 & 0.18 & 0.00 & 1.00 & 0.00 & 0.00 & 0.09 & 0.27 & 0.71 \\
				\hline
				& None & 179284 & 1.67 & 60.74 & -0.07 & 18116.00 & 0.00 & 0.00 & 0.01 & 0.05 & 12.74 \\
				& No Gaps & 127199 & 1.34 & 30.24 & -0.07 & 5319.00 & 0.00 & 0.00 & 0.01 & 0.06 & 12.96 \\
				\textbf{Short-term debt to TA} & 5 Cont. Obs. & 125479 & 1.26 & 25.35 & -0.07 & 3172.31 & 0.00 & 0.00 & 0.01 & 0.06 & 12.80 \\
				& TD2A$\leq1$ & 102034 & 0.04 & 0.09 & 0.00 & 1.00 & 0.00 & 0.00 & 0.01 & 0.04 & 0.48 \\
				& No Outliers & 95489 & 0.04 & 0.09 & 0.00 & 1.00 & 0.00 & 0.00 & 0.01 & 0.04 & 0.46 \\
				& Winsoriz. & 95489 & 0.04 & 0.09 & 0.00 & 1.00 & 0.00 & 0.00 & 0.01 & 0.04 & 0.46 \\
				\hline
				& None & 179505 & 1.28 & 290.80 & -16305.61 & 110579.73 & -12.98 & 0.00 & 0.14 & 0.71 & 16.78 \\
				& No Gaps & 127191 & 0.99 & 131.46 & -12846.64 & 38732.00 & -11.96 & 0.00 & 0.13 & 0.68 & 15.12 \\
				\textbf{Debt to equity} & 5 Cont. Obs. & 125473 & 1.00 & 132.35 & -12846.64 & 38732.00 & -11.96 & 0.00 & 0.13 & 0.69 & 15.10 \\
				& TD2A$\leq1$ & 102030 & 1.40 & 139.43 & -2995.95 & 38732.00 & -8.65 & 0.00 & 0.22 & 0.76 & 12.86 \\
				& No Outliers & 95488 & 0.65 & 6.31 & -195.93 & 264.72 & -5.82 & 0.00 & 0.24 & 0.75 & 9.96 \\
				& Winsoriz. & 95488 & 0.61 & 1.63 & -5.82 & 9.96 & -5.82 & 0.00 & 0.24 & 0.75 & 9.96 \\
				\hline
				& None & 161328 & 84.45 & 2867.63 & 0.01 & 597663.23 & 0.49 & 1.15 & 1.70 & 3.23 & 532.49 \\
				& No Gaps & 127199 & 69.41 & 2608.83 & 0.03 & 597663.23 & 0.47 & 1.15 & 1.71 & 3.36 & 383.43 \\
				\textbf{Market to book} & 5 Cont. Obs. & 125479 & 53.01 & 1547.64 & 0.03 & 172747.00 & 0.47 & 1.15 & 1.71 & 3.34 & 350.99 \\
				& TD2A$\leq1$ & 102034 & 12.83 & 824.11 & 0.03 & 146344.76 & 0.45 & 1.08 & 1.52 & 2.46 & 28.53 \\
				& No Outliers & 95489 & 2.89 & 20.45 & 0.03 & 2686.65 & 0.44 & 1.08 & 1.50 & 2.38 & 15.69 \\
				& Winsoriz. & 95489 & 2.20 & 2.23 & 0.03 & 15.69 & 0.44 & 1.08 & 1.50 & 2.38 & 15.69 \\
				\hline
				& None & 176011 & 2.34 & 2.01 & 0.01 & 10.14 & 0.54 & 1.15 & 1.60 & 2.61 & 9.91 \\
				& No Gaps & 127199 & 2.39 & 2.05 & 0.04 & 10.14 & 0.51 & 1.14 & 1.61 & 2.72 & 9.77 \\
				\textbf{Tobin's Q} & 5 Cont. Obs. & 125479 & 2.38 & 2.04 & 0.04 & 10.14 & 0.51 & 1.14 & 1.60 & 2.71 & 9.75 \\
				& TD2A$\leq1$ & 102034 & 1.85 & 1.29 & 0.04 & 10.00 & 0.49 & 1.08 & 1.45 & 2.15 & 7.60 \\
				& No Outliers & 95489 & 1.78 & 1.15 & 0.04 & 9.97 & 0.48 & 1.08 & 1.43 & 2.10 & 6.36 \\
				& Winsoriz. & 95489 & 1.78 & 1.15 & 0.04 & 9.97 & 0.48 & 1.08 & 1.43 & 2.10 & 6.36 \\
				\hline
				& None & 184051 & 0.24 & 0.27 & -1.18 & 1.00 & 0.00 & 0.04 & 0.13 & 0.35 & 0.98 \\
				& No Gaps & 127199 & 0.24 & 0.28 & -1.18 & 1.00 & 0.00 & 0.03 & 0.12 & 0.36 & 0.98 \\
				\textbf{Cash to TA} & 5 Cont. Obs. & 125479 & 0.24 & 0.27 & -1.18 & 1.00 & 0.00 & 0.03 & 0.12 & 0.36 & 0.98 \\
				& TD2A$\leq1$ & 102034 & 0.23 & 0.26 & -0.08 & 1.00 & 0.00 & 0.04 & 0.12 & 0.33 & 0.97 \\
				& No Outliers & 95489 & 0.22 & 0.26 & -0.08 & 1.00 & 0.00 & 0.04 & 0.12 & 0.32 & 0.97 \\
				& Winsoriz. & 95489 & 0.22 & 0.26 & -0.08 & 1.00 & 0.00 & 0.04 & 0.12 & 0.32 & 0.97 \\
				\hline
				& None & 179224 & -2.18 & 328.77 & -127324.00 & 2203.00 & -8.02 & -0.04 & 0.01 & 0.03 & 0.19 \\
				& No Gaps & 123824 & -0.69 & 34.25 & -9045.50 & 2203.00 & -7.55 & -0.07 & 0.01 & 0.03 & 0.22 \\
				\textbf{Cash flow to TA} & 5 Cont. Obs. & 122191 & -0.67 & 34.01 & -9045.50 & 2203.00 & -7.33 & -0.07 & 0.01 & 0.03 & 0.22 \\
				& TD2A$\leq1$ & 99780 & -0.05 & 3.21 & -855.55 & 105.00 & -0.74 & -0.02 & 0.02 & 0.03 & 0.13 \\
				& No Outliers & 93397 & -0.02 & 0.17 & -7.47 & 0.81 & -0.51 & -0.01 & 0.02 & 0.03 & 0.12 \\
				& Winsoriz. & 93397 & -0.01 & 0.09 & -0.51 & 0.12 & -0.51 & -0.01 & 0.02 & 0.03 & 0.12 \\
				\hline
				& None & 183854 & 0.23 & 0.25 & 0.00 & 2.45 & 0.00 & 0.05 & 0.13 & 0.34 & 0.93 \\
				& No Gaps & 127199 & 0.24 & 0.26 & 0.00 & 2.45 & 0.00 & 0.04 & 0.14 & 0.36 & 0.94 \\
				\textbf{PPE to TA} & 5 Cont. Obs. & 125479 & 0.24 & 0.26 & 0.00 & 2.45 & 0.00 & 0.04 & 0.14 & 0.37 & 0.94 \\
				& TD2A$\leq1$ & 102034 & 0.25 & 0.25 & 0.00 & 1.00 & 0.00 & 0.05 & 0.15 & 0.37 & 0.93 \\
				& No Outliers & 95489 & 0.25 & 0.25 & 0.00 & 1.00 & 0.00 & 0.06 & 0.16 & 0.37 & 0.93 \\
				& Winsoriz. & 95489 & 0.25 & 0.25 & 0.00 & 1.00 & 0.00 & 0.06 & 0.16 & 0.37 & 0.93 \\
				\hline
				& None & 184102 & 0.13 & 20.85 & -6.92 & 8825.00 & 0.00 & 0.00 & 0.00 & 0.02 & 0.47 \\
				& No Gaps & 127199 & 0.12 & 24.79 & -3.41 & 8825.00 & 0.00 & 0.00 & 0.00 & 0.02 & 0.54 \\
				\textbf{R\&D to TA} & 5 Cont. Obs. & 125479 & 0.12 & 24.96 & -3.41 & 8825.00 & 0.00 & 0.00 & 0.00 & 0.02 & 0.54 \\
				& TD2A$\leq1$ & 102034 & 0.02 & 0.16 & -1.09 & 41.00 & 0.00 & 0.00 & 0.00 & 0.02 & 0.24 \\
				& No Outliers & 95489 & 0.02 & 0.04 & 0.00 & 0.94 & 0.00 & 0.00 & 0.00 & 0.02 & 0.20 \\
				& Winsoriz. & 95489 & 0.02 & 0.04 & 0.00 & 0.20 & 0.00 & 0.00 & 0.00 & 0.02 & 0.20 \\
				\hline
				& None & 184102 & 5.12 & 3.03 & -6.91 & 13.19 & -3.91 & 3.38 & 5.49 & 7.25 & 10.88 \\
				& No Gaps & 127199 & 4.73 & 2.96 & -6.91 & 13.19 & -3.24 & 2.92 & 4.94 & 6.86 & 10.76 \\
				\textbf{Size} (log of TA) & 5 Cont. Obs. & 125479 & 4.75 & 2.96 & -6.91 & 13.19 & -3.17 & 2.92 & 4.95 & 6.87 & 10.78 \\
				& TD2A$\leq1$ & 102034 & 5.46 & 2.42 & -6.91 & 13.19 & -0.23 & 3.75 & 5.48 & 7.19 & 10.91 \\
				& No Outliers & 95489 & 5.58 & 2.33 & -5.30 & 13.19 & 0.34 & 3.89 & 5.58 & 7.25 & 10.95 \\
				& Winsoriz. & 95489 & 5.58 & 2.33 & -5.30 & 13.19 & 0.34 & 3.89 & 5.58 & 7.25 & 10.95 \\ \hline
		\end{tabular} }
\end{center}

%

\end{table}%

%

\begin{table}[H]
\caption{\textbf{Summary statistics based on the filtered sample}}\label{Sum_Stats_Filtered_Sample}
\vspace{-0.2cm}
\footnotesize
This table presents the number of observations, mean, standard deviation and
different percentiles for selected firm characteristics computed after
applying all filters. LT and ST stand for long-term and short-term,
respectively. TA stands for total assets.

\begin{center}%
\begin{tabular}
[c]{lcccccccccc}\hline
& \textbf{N. obs.} & \textbf{mean} & \textbf{std} & \textbf{min} &
\textbf{max} & \textbf{5\%} & \textbf{25\%} & \textbf{50\%} & \textbf{75\%} &
\textbf{95\%}\\\hline
\textbf{Tot. debt to TA} & 95,489 & 0.20 & 0.20 & 0.00 & 1.00 & 0.00 & 0.00 &
0.15 & 0.32 & 0.58\\
\textbf{LT debt to TA} & 95,489 & 0.16 & 0.18 & 0.00 & 1.00 & 0.00 & 0.00 &
0.09 & 0.27 & 0.52\\
\textbf{ST debt to TA} & 95,489 & 0.04 & 0.09 & 0.00 & 1.00 & 0.00 & 0.00 &
0.01 & 0.04 & 0.21\\
\textbf{Market to book} & 95,489 & 2.20 & 2.23 & 0.03 & 15.69 & 0.70 & 1.08 &
1.50 & 2.38 & 5.97\\
\textbf{Tobin's Q} & 95,489 & 1.78 & 1.15 & 0.04 & 9.97 & 0.74 & 1.08 & 1.43 &
2.10 & 3.99\\
\textbf{Cash to TA} & 95,489 & 0.22 & 0.26 & -0.08 & 1.00 & 0.00 & 0.04 &
0.12 & 0.32 & 0.84\\
\textbf{Cash flow to TA} & 93,397 & -0.01 & 0.09 & -0.51 & 0.12 & -0.19 &
-0.01 & 0.02 & 0.03 & 0.06\\
\textbf{PPE to TA} & 95,489 & 0.25 & 0.25 & 0.00 & 1.00 & 0.01 & 0.06 & 0.16 &
0.37 & 0.80\\
\textbf{R\&D to TA} & 95,489 & 0.02 & 0.04 & 0.00 & 0.20 & 0.00 & 0.00 &
0.00 & 0.02 & 0.09\\
\textbf{Size} (log of TA) & 95,489 & 5.58 & 2.33 & -5.30 & 13.19 & 1.86 &
3.89 & 5.58 & 7.25 & 9.35\\\hline
\end{tabular}

\end{center}

%

\end{table}%
%

\begin{table}[H]
\caption{\textbf{Empirical frequency distribution of firms by number of consecutive
			time observations (based on the filtered sample)}}\label{Table_Freq_Obs}
\vspace{-0.2cm}
\footnotesize
The first column, \textit{N. Obs.}, indicates the number of time period
observations. The columns \textit{N. firms} and \textit{\% of firms} report
the frequency and percentage of firms by number of consecutive observations
available, respectively. The column (firms) \textit{with $\geq x$ obs.} shows
the frequency of firms that have at least $5,6,7,...$ number of consecutive
data points.

\begin{center}%
\begin{tabular}
[c]{cccc|cccc}\hline
\textbf{N. Obs.} & \textbf{N. firms} & \textbf{\% of firms} & \textbf{$\geq x$
obs.} & \textbf{N. obs.} & \textbf{N. firms} & \textbf{\% of firms} &
\textbf{$\geq x$ obs.}\\\hline
5 & 128 & 3.5 & 3647 & 27 & 60 & 1.6 & 1594\\
6 & 130 & 3.6 & 3519 & 28 & 34 & 0.9 & 1534\\
7 & 153 & 4.2 & 3389 & 29 & 34 & 0.9 & 1500\\
8 & 98 & 2.7 & 3236 & 30 & 40 & 1.1 & 1466\\
9 & 127 & 3.5 & 3138 & 31 & 53 & 1.5 & 1426\\
10 & 108 & 3.0 & 3011 & 32 & 34 & 0.9 & 1373\\
11 & 109 & 3.0 & 2903 & 33 & 29 & 0.8 & 1339\\
12 & 77 & 2.1 & 2794 & 34 & 36 & 1.0 & 1310\\
13 & 105 & 2.9 & 2717 & 35 & 39 & 1.1 & 1274\\
14 & 89 & 2.4 & 2612 & 36 & 34 & 0.9 & 1235\\
15 & 121 & 3.3 & 2523 & 37 & 29 & 0.8 & 1201\\
16 & 88 & 2.4 & 2402 & 38 & 24 & 0.7 & 1172\\
17 & 92 & 2.5 & 2314 & 39 & 27 & 0.7 & 1148\\
18 & 96 & 2.6 & 2222 & 40 & 30 & 0.8 & 1121\\
19 & 98 & 2.7 & 2126 & 41 & 23 & 0.6 & 1091\\
20 & 74 & 2.0 & 2028 & 42 & 27 & 0.7 & 1068\\
21 & 72 & 2.0 & 1954 & 43 & 24 & 0.7 & 1041\\
22 & 82 & 2.2 & 1882 & 44 & 54 & 1.5 & 1017\\
23 & 68 & 1.9 & 1800 & 45 & 40 & 1.1 & 963\\
24 & 51 & 1.4 & 1732 & 46 & 82 & 2.2 & 923\\
25 & 47 & 1.3 & 1681 & 47 & 841 & 23.1 & 841\\\cline{5-8}\cline{6-8}%
\cline{7-8}\cline{8-8}%
26 & 40 & 1.1 & 1634 & Tot. & 3647 & 100 & \\\hline
\end{tabular}

\end{center}

\bigskip%

\end{table}%

\subsection{Industrial classification \label{Industry_Stats}}

In this subsection, we describe the grouping of firms into various industries
based on the three-digit Standard Industrial Classification (SIC). Because
some industries in our sample only include a handful of firms, we require each
industry to contain at least 20 distinct firms. Three-digit SIC industries
with less than 20 firms are grouped together within each two-digit SIC industry.

As shown in Table \ref{SIC}, some industries in our sample contain less than
20 firms also at the two-digit SIC level. As a result, these industries are
grouped together within each division, and no further sub-grouping (at the
three-digit) is undertaken. To illustrate, the division \textit{Mining}
(two-digit SIC $10 - 14$) contains four two-digit SIC industries. The first
group, metal mining (SIC $10$), contains $52$ distinct firms. The second
group, coal mining (SIC $12$), includes $19$ firms. The third, oil and gas
extraction (SIC $13$), comprises $221$ firms, while the fourth, non-metallic
minerals except fuels (SIC $14$), only contains $13$ firms. Based on the
criterion mentioned above, we group firms in SIC $12$ and $14$ together. We
denote this new group of all the remaining two-digit SIC industries within the
mining division as ``mining (others)''. For this group, we do not undertake
further three-digit SIC sub-grouping.%

\begin{table}[H]
\caption{\textbf{Number of firms and two-digit SIC industries within each division}}\label{SIC}
\vspace{-0.2cm}
\footnotesize
The first row (\textit{$\#$ of firms}) reports the number of firms within each
major division. The second row (\textit{$\#$ of 2-dig SIC industries}) shows
the number of non-empty 2-digit SIC industries within each division. The third
(fourth) row displays the number of 2-digit industries with less (more) than
20 firms. The last row reports the number of 2-digit SIC industries within
each division after regrouping industries with less than $20$ firms. Note that
we do not sub-group firms into 2-dig SIC industries for the division
agriculture (SIC $01-09$) and construction (SIC $15-17$) because the number of
firms within these divisions is not large enough.
\vspace{-0.5cm}%

\begin{center}
\scalebox{0.85}{
			\begin{tabular}
				[c]{lcccccccc}\hline
				& \multicolumn{8}{c}{\textbf{Industry divisions (by SIC)}}\\
				\textbf{Division} & A & B & C & D & E & F & G & I \\
				\textbf{Division name} & Agr. & Mining & Construct. & Manuf. &
				Transp. & Wholesale & Retail & Services
				\\
				\textbf{2-dig SIC range} & 01 - 09 & 10 - 14 & 15 - 17 & 20 - 39
				& 40 - 49 & 50 - 51 & 52 - 59 & 70 - 88 \\
				\hline
				Number (\#) of firms & 23 & 305 & 43 & 1872 & 216 & 142 & 253 & 793 \\
				\# of 2-dig SIC industries &  & 4 &  & 20 & 8 & 2 & 8 & 11 \\
				\# of 2-dig SIC with $0<$ \# firms $<20$ &  & 2 &  & 4 & 4 & 0 & 3 & 6 \\
				\# of 2-dig SIC with \# firms $\geq20$ &  & 2 &  & 16 & 4 & 2 & 5 & 5 \\
				\# of 2-dig SIC after regrouping & 1 & 3 & 1 & 17 & 5 & 2 & 6 & 6 \\ \hline
		\end{tabular} }
\end{center}

%

\end{table}%

In total, firms in our sample can be divided into $67$ three-digit SIC
industries. These are listed in Table \ref{table_3dig_sic}, where we report
information on the SIC codes, number of firms within each three-digit SIC
industry, as well as information on the corresponding two-digit SIC
industries. To illustrate, the two-digit SIC industry \textit{Machinery \&
Equipment} (SIC 35), containing in total $161$ firms, can be divided into 4
three-digit SIC industries of which \textit{Machinery \& Equipment (others)}
consists of several three-digit SIC industries each composed of less than $20$ firms.%

\begin{table}[H]
\caption{\textbf{Three-digit SIC industry classification}}\label{table_3dig_sic}
\vspace{-0.2cm}
\footnotesize
The first column enumerates the three-digit SIC industries in our sample.
Column \textit{3-dig SIC} and \textit{3-dig SIC description} report the
three-digit Standard Industrial Classification (SIC) codes and the
corresponding industry group names, respectively, while column \textit{$\#$
(3-dig)} displays the number of firms within each group. Columns \textit{2-dig
SIC} and \textit{2-dig SIC description} provide the two-digit SIC codes and
the major group names to which the three-digit SIC industries belong,
respectively. Column \textit{$\#$ (2-dig)} reports the total number of firms
within each two-digit SIC industry.
\vspace{-0.1cm}%

\begin{center}
\scalebox{0.80}{
			\begin{tabular}{cp{3.5cm}p{5cm}cccc}
				\hline
				\textbf{n.} & \textbf{3-dig SIC } & \textbf{3-dig SIC description} & \textbf{\# (3-dig)} & \textbf{2-dig SIC} & \textbf{2-dig SIC description} & \textbf{\# (2-dig)} \\
				\hline
				1 & 010; 020; 070 & Agriculture & 23 & 01; 02; 07 & Agriculture & 23 \\
				\hline
				2 & 104 & Gold \& Silver Ores & 30 & 10 & Metal Mining & 52 \\
				3 & 100; 109 & Metal Mining (others) & 22 &  &  &  \\
				\hline
				4 & 131 & Crude Petrol. \& Natural Gas & 180 & 13 & Oil \& Gas Extraction & 221 \\
				5 & 138 & Oil \& Gas Field Services & 41 &  &  &  \\
				\hline
				6 & 122; 140 & Mining (others) & 32 & 12; 14 & Mining (others) & 32 \\
				\hline
				7 & 152; 153; 154; 160; & Construction & 43 & 15; 16; 17 & Construction & 43 \\
				& 162; 170; 173 &  &  &  &  &  \\
				\hline
				8 & 208 & Beverages & 27 & 20 & Food and Kindred & 99 \\
				9 & 200; 201; 202; 203; & Food \& Kindred (others) & 72 &  &  &  \\
				& 204; 205; 206; 207; &  &  &  &  &  \\
				& 209 &  &  &  &  &  \\
				\hline
				10 & 230; 232; & Apparel \& Textile Products & 33 & 23 & Apparel \& Textile Products & 33 \\
				& 233; 234; &  &  &  &  &  \\
				& 239 &  &  &  &  &  \\
				\hline
				11 & 240; 242; 243; 245 & Lumber \& Wood Prod. & 24 & 24 & Lumber \& Wood Prod. & 24 \\
				\hline
				12 & 261; 262; 263; 265; & Paper Prod. & 31 & 26 & Paper Prod. & 31 \\
				& 267 &  &  &  &  &  \\
				\hline
				13 & 271; 272; 273; 274; & Printing \& Publishing & 26 & 27 & Printing \& Publishing & 26 \\
				& 275; 276; 278; 279 &  &  &  &  &  \\
				\hline
				14 & 283 & Drugs & 516 & 28 & Chemicals & 640 \\
				15 & 284 & Soaps, Clean. \& Toilet Goods & 24 &  &  &  \\
				16 & 286 & Industrial Organic Chemicals & 25 &  &  &  \\
				17 & 280; 281; 282; 285; & Chemicals (others) & 75 &  &  &  \\
				& 287; 289 &  &  &  &  &  \\
				\hline
				18 & 291; 299 & Petroleum \& Coal Prod. & 28 & 29 & Petroleum \& Coal Prod. & 28 \\
				\hline
				19 & 301; 302; 306; 308 & Rubber \& Plastics Prod. & 30 & 30 & Rubber \& Plastics Prod. & 30 \\
				\hline
				20 & 321; 322; 324; 325; & Stone, Clay \& Glass Prod. & 20 & 32 & Stone, Clay \& Glass Prod. & 20 \\
				& 326; 327; 329 &  &  &  &  &  \\
				\hline
				21 & 331 & Furnace \& Basic Steel Prod. & 20 & 33 & Primary Metal & 43 \\
				22 & 333; 334; 335; 336; & Primary Metal (others) & 23 &  &  &  \\
				& 339 &  &  &  &  &  \\
				\hline
				23 & 342; 344; 345; 346; & Fabricated Metal Prod. & 44 & 34 & Fabricated Metal Prod. & 44 \\
				& 347; 348; 349 &  &  &  &  &  \\
				\hline
				24 & 353 & Construct. \& Relat. Machinery & 28 & 35 & Machinery \& Equipment & 161 \\
				25 & 356 & General Industrial Machinery & 22 &  &  &  \\
				26 & 357 & Computer \& Office Equipment & 60 &  &  &  \\
				27 & 351; 352; 354; 355; & Machinery \& Equip. (others) & 51 &  &  &  \\
				& 358; 359 &  &  &  &  &  \\
				
				\hline
		\end{tabular} }
\end{center}

%

\end{table}%
%

\setcounter{table}{9}%
%

\begin{table}[H]
\caption{(cont.)}
\vspace{-0.2cm}
\footnotesize
%

\vspace{-0.25cm}%

\begin{center}
\scalebox{0.80}{
			\begin{tabular}{cp{3.5cm}p{5cm}cccc}
				\hline
				\textbf{n.} & \textbf{3-dig SIC } & \textbf{3-dig SIC description} & \textbf{\# (3-dig)} & \textbf{2-dig SIC} & \textbf{2-dig SIC description} & \textbf{\# (2-dig)} \\
				\hline
				28 & 362 & Electrical Industrial Apparatus & 24 & 36 & Electronic & 294 \\
				29 & 366 & Communications Equipment & 80 &  &  &  \\
				30 & 367 & Electronic Comp. \& Accessory & 132 &  &  &  \\
				31 & 369 & Misc. Electr. Equip. \& Supplies & 24 &  &  &  \\
				32 & 360; 361; 363; 364; & Electronic (others) & 34 &  &  &  \\
				& 365 &  &  &  &  &  \\
				\hline
				33 & 371 & Motor Vehicles \& Equipment & 42 & 37 & Transp. Equip. & 81 \\
				34 & 372; 373; 374; 375; & Transp. Equip. (others) & 39 &  &  &  \\
				& 376; 379 &  &  &  &  &  \\
				\hline
				35 &  381; 382; 385; 386; & Instruments (others) & 85 & 38 & Instruments & 246 \\
				& 387 &  &  &  &  &  \\
				36 & 384 & Medic. Instruments \& Supplies & 161 &  &  &  \\
				\hline
				37 & 391; 393; 394; 395; & Misc. Manufacturing & 29 & 39 & Misc. Manufacturing & 29 \\
				& 399 &  &  &  &  &  \\
				\hline
				38 & 210; 211; 220; 221; & Manufacturing (others) & 43 & 21; 22; 25; 31 & Manufacturing (others) & 43 \\
				& 222; 227; 251; 252; &  &  &  &  &  \\
				& 253; 254; 259; 310; &  &  &  &  &  \\
				& 314 &  &  &  &  &  \\
				\hline
				39 & 421 & Trucking \& Warehousing & 26 & 42 & Trucking \& Warehousing & 26 \\
				\hline
				40 & 451; 452; 458 & Air Transportation & 32 & 45 & Air Transportation & 32 \\
				\hline
				41 & 470; 473 & Transp. Service & 20 & 47 & Transp. Service & 20 \\
				\hline
				42 & 481 & Telephone Communication & 31 & 48 & Communications & 102 \\
				43 & 489 & Communications Services & 41 &  &  &  \\
				44 & 483; 484; 488 & Communications (others) & 30 &  &  &  \\
				\hline
				45 & 401; 410; 440; 441; & Transportation (others) & 36 & 40; 41; 44; 46 & Transportation (others) & 36 \\
				& 461 &  &  &  &  &  \\
				\hline
				46 & 500; 501; 503; 504; & Wholesale Durable Goods & 78 & 50 & Wholesale Durable Goods & 78 \\
				& 505; 506; 507; 508; &  &  &  &  &  \\
				& 509 &  &  &  &  &  \\
				\hline
				47 & 517 & Petrol. \& Petroleum Products & 22 & 51 & Wholesale Non-Dur. Goods & 64 \\
				48 & 511; 512; 513; 514; & Wholesale Non-Dur. Goods & 42 &  &  &  \\
				& 515; 516; 518; 519 &  &  &  &  &  \\
				\hline
				49 & 540; 541 & Food Stores & 25 & 54 & Food Stores & 25 \\
				\hline
				50 & 550; 553 & Automotive Dealers & 24 & 55 & Automotive Dealers & 24 \\
				\hline
				51 & 560; 562; 565; 566 & Apparel Stores & 39 & 56 & Apparel Stores & 39 \\
				\hline
				52 & 581 & Eating/Drinking Places & 56 & 58 & Eating/Drinking Places & 56 \\
				\hline
				53 & 596 & Nonstore Retailers & 29 & 59 & Miscellaneous Retail & 74 \\
				54 & 590; 591; 594; 599 & Miscellaneous Retail (others) & 45 &  &  &  \\
				\hline
				55 & 520; 521; 531; 533; & Retail (others) & 35 & 52; 53; 57 & Retail (others) & 35 \\
				& 539; 570; 571; 573 &  &  &  &  &  \\
				\hline
				56 & 736 & Personnel Supply Services & 20 & 73 & Business Services & 522 \\
				57 & 737 & Comput. \& Data Proc. Services & 431 &  &  &  \\
				58 & 738 & Misc. Business Services & 34 &  &  &  \\
				59 & 731; 732; 733; 734; & Business Services (others) & 37 &  &  &  \\
				& 735 &  &  &  &  &  \\
				\hline
				60 & 790; 794; 799 & Recreation Services & 41 & 79 & Recreation Services & 41 \\
				\hline
				61 & 809 & Misc. Health \& Allied Services & 26 & 80 & Health Services & 88 \\
				62 & 800; 801; 805; 806; & Health Services (others) & 62 &  &  &  \\
				& 807; 808 &  &  &  &  &  \\
				\hline
				63 & 820 & Educational Services & 23 & 82 & Educational Services & 23 \\
				\hline
				64 & 873 & Research \& Testing Services & 21 & 87 & Engineering Services &  \\
				65 & 874 & Manag. \& Public Relations & 24 &  &  &  \\
				66 & 870; 871; 872 & Engineering Services (others) & 28 &  &  &  \\
				\hline
				67 & 701; 720; 750; 751; & Services (others) & 46 & 70; 72; 75; 78 & Services (others) & 46 \\
				& 781; 782; 783; 811; &  &  & 81; 83 &  &  \\
				& 830; 835 &  &  &  &  &  \\
				\hline
		\end{tabular} }
\end{center}

%

\end{table}%


Table \ref{freq_byindustry} reports some statistics on the empirical frequency
distribution of firms by year across the three-digit SIC industries.

%

\begin{table}[H]
\caption{\textbf{Frequency of firms across three-digit SIC industries and over time}}\label{freq_byindustry}
\vspace{-0.2cm}
\footnotesize
Annual statistics obtained by averaging quarterly statistics within each year.
Columns \textit{min} and \textit{max} report the minimum and maximum number of
firms in an industry over time, respectively. Columns \textit{med} and
\textit{mean} display the median and average number of firms in an industry in
a particular year; \textit{std} measures the standard deviation across all
industries at each point in time.
\normalsize

\begin{center}%
\begin{tabular}
[c]{cccccc}\hline
\textbf{Year} & \textbf{min} & \textbf{max} & \textbf{med} & \textbf{mean} &
\textbf{std}\\\hline
2007 & 9.3 & 256.8 & 23.8 & 35.1 & 41.9\\
2008 & 10.0 & 250.5 & 24.3 & 35.7 & 41.2\\
2009 & 10.0 & 221.0 & 22.3 & 33.1 & 36.4\\
2010 & 9.8 & 203.8 & 21.0 & 31.7 & 34.1\\
2011 & 9.3 & 193.0 & 19.8 & 30.1 & 32.8\\
2012 & 8.5 & 187.5 & 19.0 & 29.2 & 31.9\\
2013 & 9.5 & 192.5 & 19.8 & 28.5 & 32.6\\
2014 & 9.3 & 242.8 & 19.3 & 29.2 & 37.7\\
2015 & 9.0 & 267.8 & 18.8 & 28.9 & 39.5\\
2016 & 8.5 & 287.5 & 18.0 & 28.2 & 40.6\\
2017 & 7.0 & 300.0 & 18.0 & 27.7 & 41.3\\
2018 & 6.0 & 272.0 & 16.3 & 25.1 & 37.3\\\hline
\end{tabular}

\end{center}

%

\end{table}%


\subsection{Three-digit SIC industry characteristics \label{[Stat_Industry]}}

This subsection provides some summary statistics for selected variables at the industry-level.

In Panel A of Table \ref{tab: Sort_by_Quintiles}, we report on differences in
industry characteristics (such as industry median leverage, size,
profitability, etc.), according to different degree of financial leverage. In
particular, industry-quarter observations are sorted into quintiles based on
debt to asset ratios. For each of these quantiles, we report the average of
the selected industry-specific characteristics. As can be seen firms in higher
leverage industries tend to be larger and have more tangible assets, whilst
firms in lower leverage industries tend to be characterised by both higher
cash holdings and also higher market to book ratios as well as larger Tobin's
Q. The relation between leverage and age or industry growth is more nonlinear.%

\begin{table}[H]
\caption{\textbf{Industry characteristics sorted into quintiles}}\label{tab: Sort_by_Quintiles}
\vspace{-0.2cm}
\footnotesize
The statistics in this table are obtained as follows. First, at each point in
time, we compute the median of selected firm characteristics within each
three-digit SIC industry. These industry-quarter observations are then sorted
into quintiles based on debt to assets (panel A), cash to assets (panel B), or
size (panel C). For each quintile we then report the average of the selected
characteristics (listed in the first column). TA denotes total assets. A
description of the variables considered can be found in Table
\ref{table:Definition_Variable}.
\normalsize

\begin{center}
\scalebox{1}{
			\begin{tabular}{lccccc}
				& \multicolumn{5}{c}{Panel A: Sorting by debt to assets} \\
				\hline
				& \multicolumn{5}{c}{Debt to assets quintile} \\
				& 1 & 2 & 3 & 4 & 5 \\
				\hline
				Tot. debt to TA & 0.05 & 0.14 & 0.20 & 0.25 & 0.36 \\
				Size (log of TA) & 4.85 & 5.39 & 5.99 & 6.51 & 6.80 \\
				Age (years) & 14.29 & 17.27 & 18.66 & 18.47 & 16.02 \\
				Cash flow to TA & 0.00 & 0.02 & 0.02 & 0.02 & 0.02 \\
				Cash to TA & 0.23 & 0.11 & 0.07 & 0.06 & 0.06 \\
				PPE to TA & 0.18 & 0.20 & 0.28 & 0.33 & 0.37 \\
				Market to book & 1.80 & 1.48 & 1.43 & 1.38 & 1.35 \\
				Tobin's Q & 1.64 & 1.40 & 1.37 & 1.33 & 1.31 \\
				Industry growth (\%) & 0.21 & 0.62 & 0.61 & 0.48 & 0.54 \\
				\hline
				& \multicolumn{5}{c}{} \\
				& \multicolumn{5}{c}{Panel B: Sorting by cash to assets} \\
				\hline
				& \multicolumn{5}{c}{Cash to assets quintile} \\
				& 1 & 2 & 3 & 4 & 5 \\
				\hline
				Tot. debt to TA & 0.29 & 0.24 & 0.20 & 0.18 & 0.08 \\
				Size (log of TA) & 6.69 & 6.30 & 5.89 & 5.62 & 5.04 \\
				Age (years)  & 17.30 & 17.93 & 17.95 & 17.14 & 14.43 \\
				Cash flow to TA & 0.02 & 0.02 & 0.02 & 0.02 & 0.01 \\
				Cash to TA & 0.03 & 0.06 & 0.08 & 0.11 & 0.26 \\
				PPE to TA & 0.38 & 0.31 & 0.26 & 0.23 & 0.16 \\
				Market to book & 1.32 & 1.41 & 1.45 & 1.47 & 1.77 \\
				Tobin's Q & 1.28 & 1.35 & 1.39 & 1.40 & 1.62 \\
				Industry growth (\%) & 0.69 & 0.58 & 0.36 & 0.53 & 0.32 \\
				\hline
				& \multicolumn{5}{c}{} \\
				& \multicolumn{5}{c}{Panel C: Sorting by size} \\
				\hline
				& \multicolumn{5}{c}{Size quintile} \\
				& 1 & 2 & 3 & 4 & 5 \\
				\hline
				Tot. debt to TA & 0.11 & 0.14 & 0.20 & 0.24 & 0.30 \\
				Size (log of TA) & 4.31 & 5.25 & 5.92 & 6.56 & 7.49 \\
				Age (years)  & 14.72 & 15.37 & 17.37 & 18.58 & 18.70 \\
				Cash flow to TA & 0.00 & 0.02 & 0.02 & 0.02 & 0.02 \\
				Cash to TA & 0.18 & 0.14 & 0.09 & 0.07 & 0.06 \\
				PPE to TA & 0.20 & 0.20 & 0.27 & 0.28 & 0.41 \\
				Market to book & 1.77 & 1.49 & 1.39 & 1.47 & 1.30 \\
				Tobin's Q & 1.62 & 1.41 & 1.34 & 1.41 & 1.27 \\
				Industry growth (\%) & 0.01 & 0.60 & 0.54 & 0.73 & 0.58 \\
				\hline
		\end{tabular} }
\end{center}

%

\end{table}%

\begin{figure}[H]
	\caption{\textbf{Industry characteristics across debt to assets deciles}}
	\renewcommand\thefigure{4} \centering{\includegraphics
[scale=0.55]{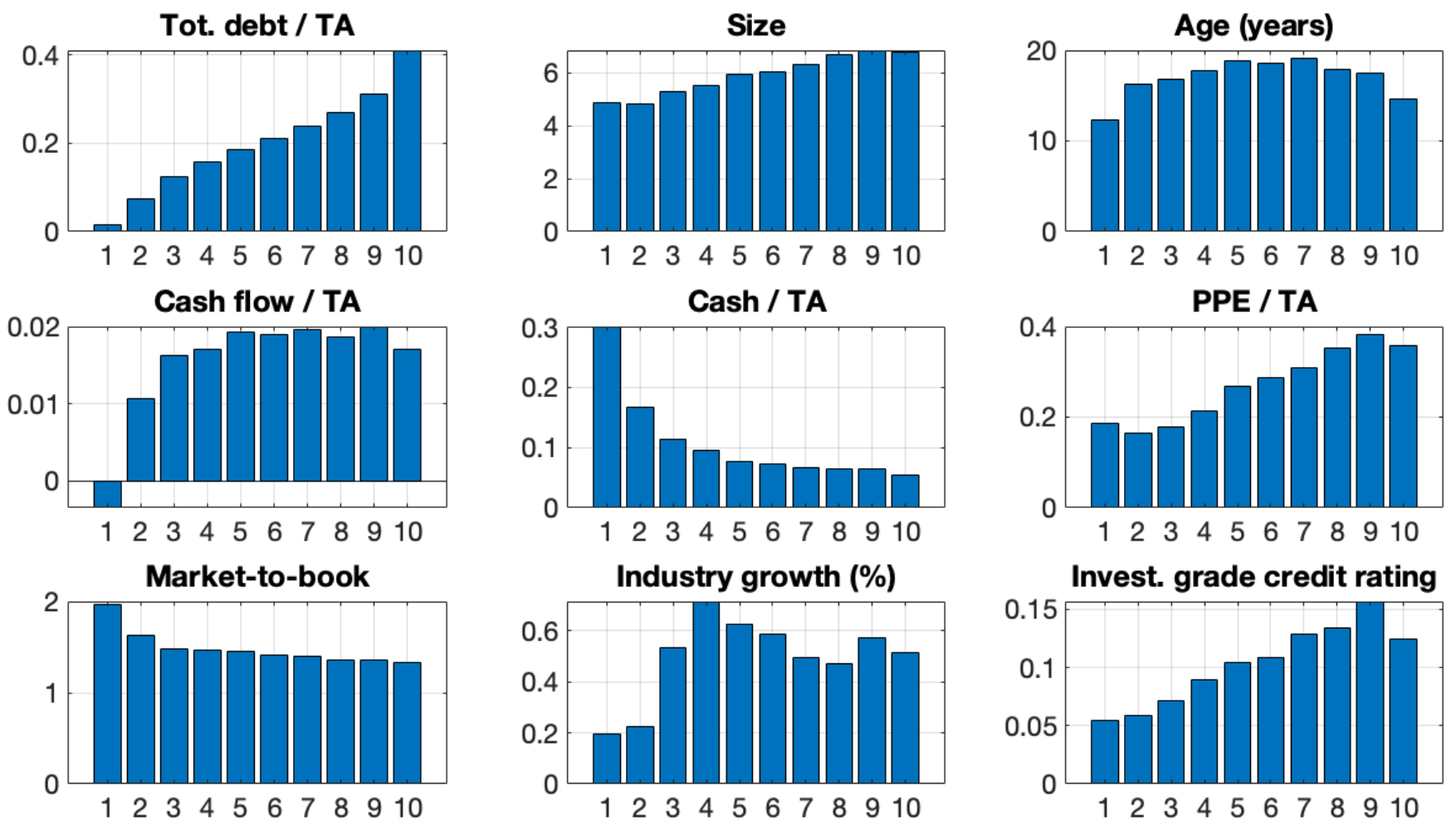}}
	\parbox{0.9\textwidth}{\footnotesize
{Industry characteristics across total debt to total assets deciles. TA denotes total assets. A description of the variables can be found in Table \ref
{table:Definition_Variable}
			. Invest. grade credit rating is the proportion of firms in an industry with investment grade credit rating.}
	} \label{fig:sort_by_leverage}
\end{figure}

It is interesting to note that some of the above documented patterns at the
industry-level also hold at the firm-level, as documented by \cite{graham2011}%
. Similar conclusions hold when sorting industry-quarter observations by cash
to assets (Panel B) or size quintiles (Panel C). These relations are also
illustrated in Figure \ref{fig:sort_by_leverage} which shows the average of
several industry-quarter observations, sorted into deciles from lowest to
highest industry median leverage.%

\begin{figure}[H]
	\caption{\textbf{Leverage across three-digit SIC industries}}
	\renewcommand\thefigure{5} \centering{\includegraphics
[scale=0.55]{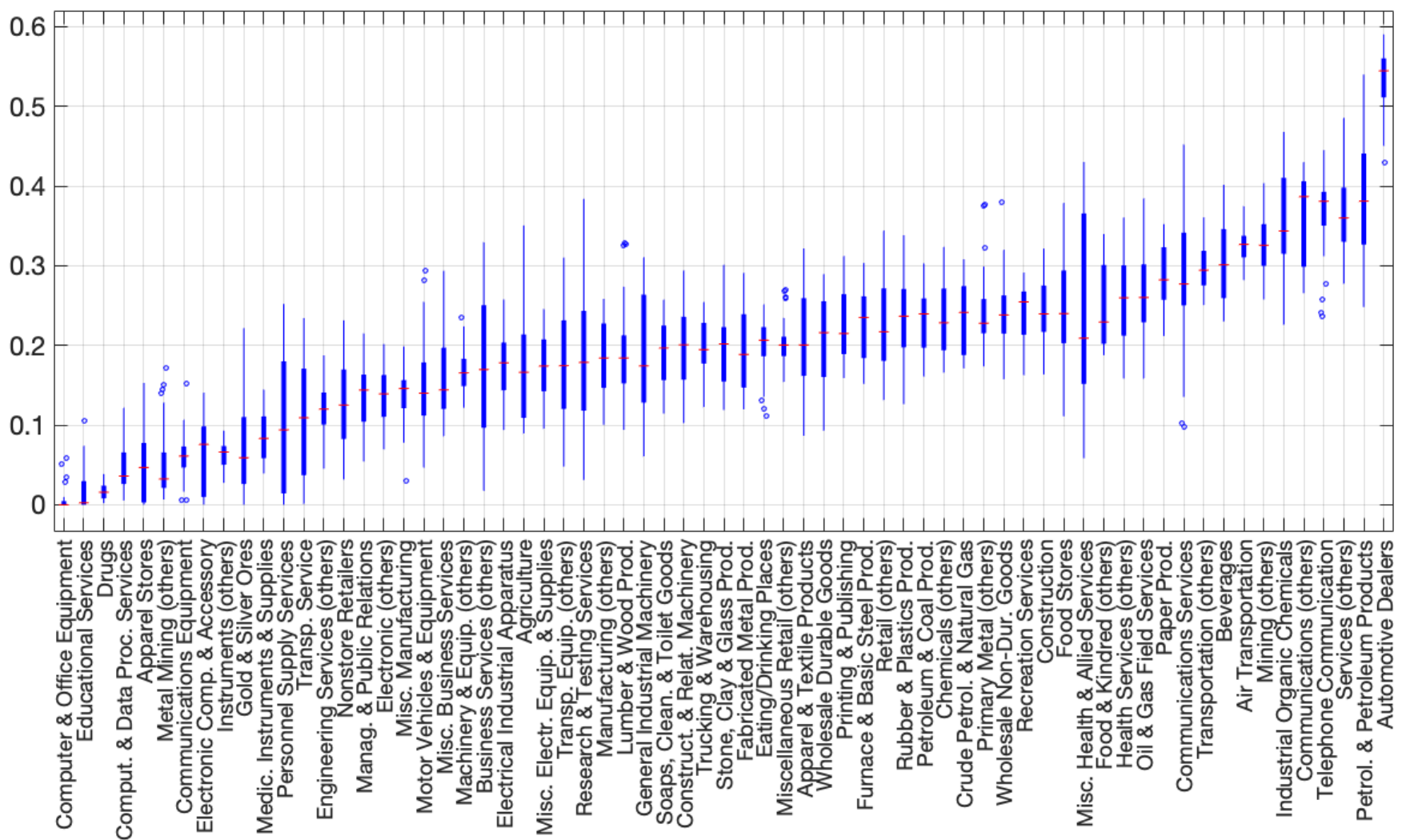}}
	\parbox{0.9\textwidth}{\footnotesize
{Box plots for industry median leverage (where leverage is defined as total debt to total assets). On each box, the central mark indicates
			the median, and the bottom and top edges of the (dark blue) box display the $25$th and $75$th
			percentiles,respectively. The x-axis reports the three-digit SIC industries sorted from
			smallest to largest industry median leverage (averaged over time).}}
	\label{fig:Ind_Median_DA}
\end{figure}

Figure \ref{fig:Ind_Median_DA} displays the box plots for industry median
leverage for each three-digit SIC industry, sorted from smallest to largest
industry median leverage (averaged over time). It shows a significant degree
of heterogeneity in the use of leverage across industries. It is also readily
apparent that industry median leverages tend to vary over time.%

\section{Identification strategy and estimation
\label{Appendix_Threshold_Optim}}

We now provide some additional information on our identification and the
estimation strategies discussed in Sections \ref{sec: Ident_Strategy} and
\ref{ARDL} of the paper, respectively.

\subsection{Cross-industry variation to identify the policy effects
\label{cross-variation}}

\begin{figure}[H]
	\caption{\textbf
{Proportion of firms with debt to asset ratios below the upper quartile by industry}
	}
	\renewcommand\thefigure{6} \centering{\includegraphics
[scale=0.55]{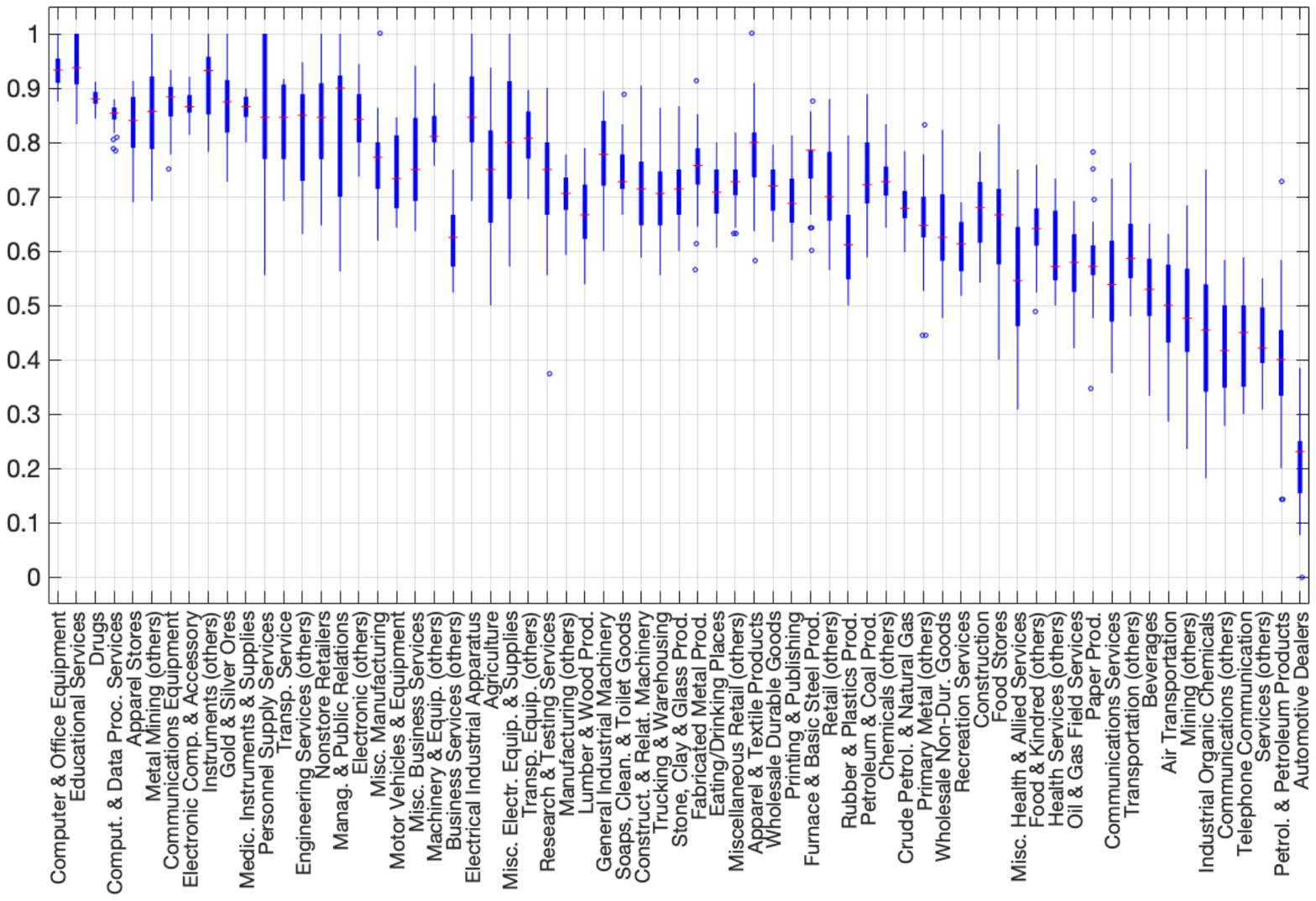}}
	\parbox{0.9\textwidth}{\footnotesize
{Box plots for the proportion of firms in each industry with
			debt to asset ratios (DA) below the upper quartile ($\pi_{st}(75)$).
			On each box, the central mark indicates the median, and the bottom and top edges of the (dark blue)
			box display the $25$th and $75$th percentiles, respectively. The x-axis reports the three-digit SIC
			industries sorted from smallest to largest industry median leverage (averaged over time).}
	}\label{fig:Sec_Hetero}
\end{figure}

As discussed in the paper, identification of the policy effectiveness
coefficient, $\beta_{1}$, in equation (\ref{panel 1}), requires a sufficient
degree of variations in $q_{t}$ over time and $\pi_{st}(\gamma)$, defined in
equation (\ref{Frac_debt}), across industries. We graphically demonstrate that
there is a high degree of variation in $\pi_{st}(\gamma)$ across industries.
To this end, in Figure \ref{fig:Sec_Hetero}, we report the box plots for
$\pi_{st}(75)$, the proportions of firms with debt to asset ratios (DA) below
the upper quartile (sorted from smallest to largest industry median leverage),
across the three-digit SIC industries to illustrate that they show significant
variation across industries and also over time.%

\subsection{Quantile threshold parameter estimates\label{appx_thresh_est}}

As discussed in the paper, the grid search procedure used to estimate $\gamma$
consists in selecting many values of $\gamma$ along a grid, compute the sum of
squared residuals (SSR) for each of these values, to then choose as estimates
the value that provides the smallest SSR. We calculate the SSR for all values
of $0.25\leq\gamma\leq0.9$ in increments of $0.01$.

\begin{figure}[H]
	\caption{\textbf{Histogram plot of $\pi_{st}(\gamma)$ for selected
			values of $\gamma$}}
	\renewcommand\thefigure{8}  \centering{\includegraphics
[scale=0.55]{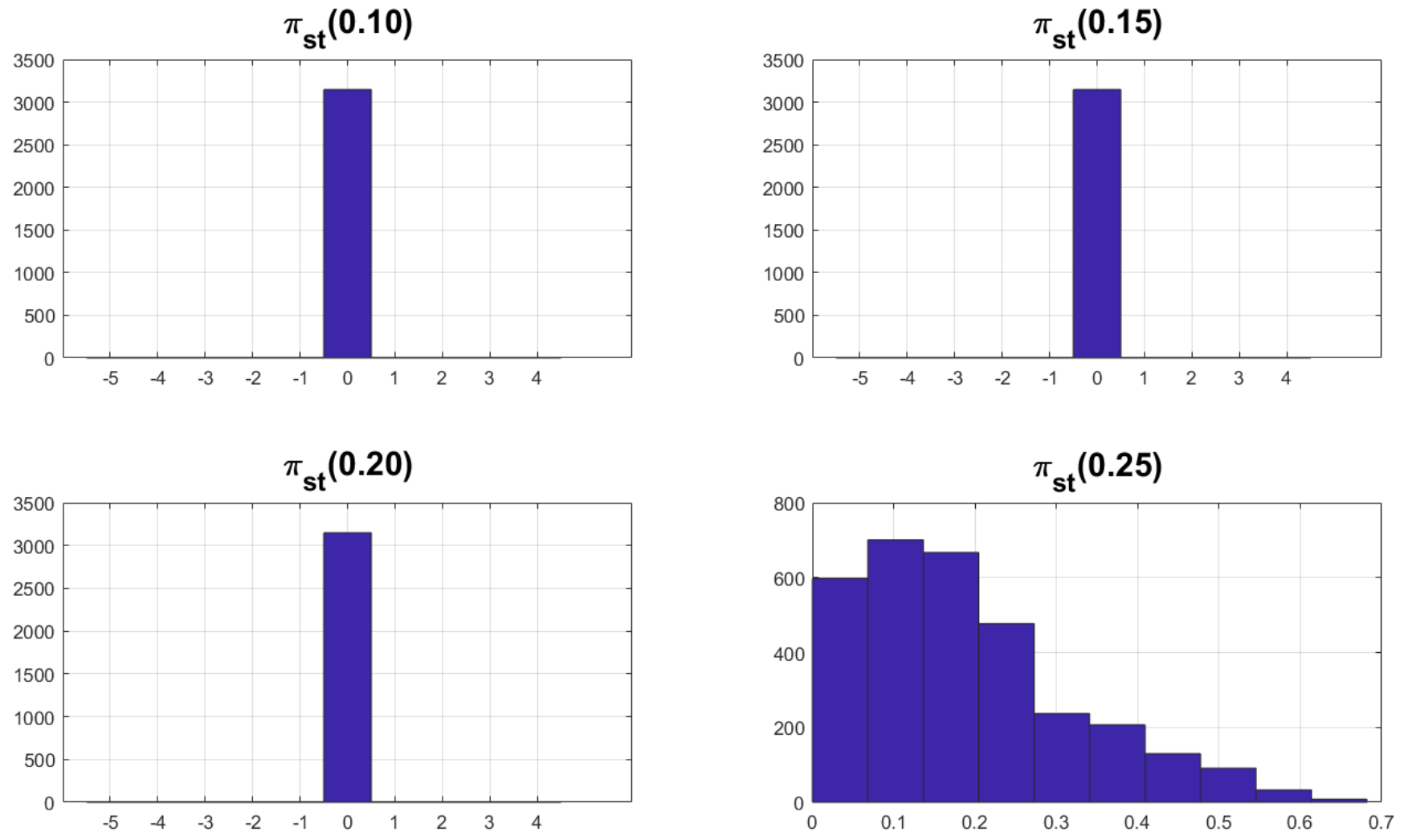}}
	\parbox{0.9\textwidth}{\footnotesize
{Each panel displays the sample distribution of $\pi_{st}(\gamma)$
			(across sectors and over time) for different values of
			$\gamma\in\left\{ 0.10,0.15,0.20,0.25\right\}$.
			In both upper panels as well as in the left-hand side bottom panel, $\pi
_{st}(\gamma)=0$ for all $s$ and $t$.}}	\label{fig:check_hist}
\end{figure}

Here we show why we choose to start the grid search at $0.25$ instead of
$0.1$. To do so, in Figure \ref{fig:check_hist} we display the sample
distribution of $\pi_{st}(\gamma)$ for $\gamma=0.10,0.15,0.20,0.25$. It is
clear that we cannot start the grid search from $0.1$ because by construction,
$\pi_{st}(\gamma)=0$ whenever $g_{t}(\gamma)=0$, and given that the $q$-th
quantile of DA is equal to zero for all values of $q$ below $0.21$.%

\section{Computing half-life and mean lag length \label{Appx_HalfL_MeanL}}

In this section, we discuss how to compute the mean lag of a PanARDL(p), and
other related measures. For clarity of exposition, and without loss of
generality, we rewrite the panel regression model given by equation
(\ref{ARDL_model_lsap}) in the main body of the paper, abstracting from all
regressors but the interaction of LSAPs with our industry-specific measure of
debt capacity, $\chi_{st}=\chi_{st}(\gamma_{post})=q_{t}\times\pi
_{s,t-1}(\gamma_{post})$. Hence, focusing solely on the policy coefficients of
interest, the PanARDL(p) model is given by
\begin{equation}
\lambda(L)y_{is,t}=B(L)\chi_{st}+u_{is,t}, \label{ARDL_model_simple}%
\end{equation}
where $\lambda(L)=1-\lambda_{1}L-...-\lambda_{p}L^{p}$ and $B(L)=\beta
_{0}+\beta_{1}L-...+\beta_{p}L^{p}$.

Multiplying both sides of (\ref{ARDL_model_simple}) by $\lambda(L)^{-1}$, we
get
\begin{equation}
y_{is,t}=C(L)\chi_{st}+\lambda(L)^{-1}u_{is,t},
\end{equation}
where $C(L)=\lambda(L)^{-1}B(L)=\sum_{h=0}^{\infty}\varphi_{h}L^{h}$.

The $\varphi_{h}$ ($h=0,1,2,...$) can be obtained by noting that
\begin{equation}
\left(  1-\lambda_{1}L-...-\lambda_{p}L^{p}\right)  \left(  \varphi
_{0}+\varphi_{1}L+\varphi_{2}L^{2}+...\right)  =\beta_{0}+\beta_{1}%
L-...+\beta_{p}L^{p}, \label{equality_ARMA}%
\end{equation}
where the left-hand side can be rewritten as
\[
\varphi_{0}+\left(  \varphi_{1}-\varphi_{0}\lambda_{1}\right)  L+\left(
\varphi_{2}-\varphi_{1}\lambda_{1}-\varphi_{0}\lambda_{2}\right)
L^{2}+...+\left(  \varphi_{h}-\sum_{j=1}^{h}\varphi_{h-j}\lambda_{j}\right)
L^{h}+...
\]

Thus, from (\ref{equality_ARMA}) we have
\begin{equation}%
\begin{array}
[c]{ccl}%
\varphi_{h}=\beta_{h}+\sum_{j=1}^{h}\lambda_{j}\varphi_{h-j}, &  & h=1,2,...
\end{array}
, \label{varphi}%
\end{equation}
where $\varphi_{0}=\beta_{0}$, $\lambda_{j}=0$ for $j>p$, and $\beta_{h}=0$
for $h>p$.

\paragraph{Scalar lag distribution of $\chi$.}

The scalar lag distribution of $\chi$ is given by
\[
\frac{C(L)}{C(1)}=\frac{\sum_{i=0}^{\infty}\varphi_{h}L^{h}}{\sum
_{h=0}^{\infty}\varphi_{h}}=\sum_{h=0}^{\infty}\rho_{h}L^{h},
\]
where
\begin{equation}
\rho_{h}=\frac{\varphi_{h}}{\sum_{h=0}^{\infty}\varphi_{h}}. \label{rho}%
\end{equation}

By construction $\sum_{h=0}^{\infty}\rho_{h}=1$.

\paragraph{Mean lag and half-life.}

The mean lag between $\chi$ and $y$ is given by
\begin{equation}
W^{\prime}(1)=\frac{C^{\prime}(1)}{C(1)}=\frac{\sum_{h=0}^{\infty}h\varphi
_{h}}{\sum_{h=0}^{\infty}\varphi_{h}}=\sum_{h=0}^{\infty}h\rho_{h}.
\label{mean_lag}%
\end{equation}

In the case of the PanARDL(2):
\begin{equation}
W^{\prime}(1)=\frac{\beta_{1}+2\beta_{2}}{\beta_{0}+\beta_{1}+\beta_{2}}%
+\frac{\lambda_{1}+2\lambda_{2}}{1-\lambda_{1}-\lambda_{2}}.
\label{mean_lag_ardl2}%
\end{equation}

The half-life is defined as the number of periods required for the peak
response of $y$ to $\chi$ to dissipate by one half. In other words, it is the
value of $h$ such that
\begin{equation}
sup\left(  \varphi_{h}\right)  \geq\frac{\varphi_{max}}{2}, \label{halflife}%
\end{equation}
where $\varphi_{max}$ denotes the peak response.

\section{Estimation results for the main specification where $q_{t}$ denotes
the total size of LSAPs\label{OS_Main_Results}}

In this section, we report estimation results for the threshold PanARDL(2)
specification. We also provide additional details on the industry-specific
weights used to compute the policy effects at the national level. We also show
that our empirical results are robust to a number alternative choices.

\subsection{Full estimation results\label{Appdx_BenchFullResults}}

This subsection reports additional estimation results for the panel regression
model given by equation (\ref{ARDL_model_lsap}) in the paper.%

\begin{table}[H]
\caption{\textbf{FE--TE estimates of the effects of LSAPs on non-financial firm's debt to asset ratios based on the PanARDL(2) model}}\label{tab:ARDL2_bench}
\vspace{-0.2cm}
\footnotesize
Estimates of the coefficients of the PanARDL(2) model described in equation
(\ref{ARDL_model_lsap}). The dependent variable is debt to asset ratio (DA).
$q_{t}$ is the (scaled) amount of U.S. Treasuries and agency MBS purchased by
the Fed; $\pi_{s,t}(\hat{\gamma})$ denotes the proportion of firms in an
industry with DA below the $\hat{\gamma}$-th quantile. The first three columns
report results for the single-threshold panel regression model, where
$\gamma_{pre}=\gamma_{post}$. The last three columns report results for the
two-threshold panel regression, where $\gamma_{pre}\neq\gamma_{post}$. The
estimated quantile threshold parameters are shown in Table
\ref{tab: thresholds_LSAPs}. All regressions include both firm-specific
effects and time effects. Columns (1) and (4) include industry-specific linear
time trends, columns (2) and (5) include the interaction of industry dummies
and real GDP growth, while columns (3) and (6) include both. The sample
consists of an unbalanced panel of $3,647$ U.S. publicly traded non-financial
firms observed at a quarterly frequency over the period $2007$:Q1 - $2018$:Q3.
Robust standard errors in parentheses (*** $p<0.01$, ** $p<0.05$, * $p<0.1$).

\begin{center}%
\begin{tabular}
[c]{l|ccc|ccc}%
\multicolumn{1}{c}{} & \multicolumn{6}{c}{Dependent variable: debt to assets
($DA_{t}$)}\\\cline{2-7}\cline{3-7}\cline{4-7}\cline{5-7}\cline{6-7}%
\multicolumn{1}{c}{} & \multicolumn{3}{c|}{$\gamma_{pre}=\gamma_{post}=\gamma
$} & \multicolumn{3}{c}{$\gamma_{pre}\neq\gamma_{post}$}\\
\multicolumn{1}{c}{} & (1) & (2) & (3) & (4) & (5) & (6)\\\hline
$\pi_{s,t-1}(\hat{\gamma}_{pre})$ & 0.0125{*}{*} & 0.0124{*}{*} & 0.0158{*}
{*}{*} & 0.0176{*}{*}{*} & 0.0144{*}{*}{*} & 0.0181{*}{*}{*}\\
& (0.0052) & (0.0052) & (0.0053) & (0.0050) & (0.0049) & (0.0051)\\
$q_{t}\times\pi_{s,t-1}(\hat{\gamma}_{post})$ & 0.0037{*} & 0.0028{*} &
0.0031{*} & 0.0058{*}{*}{*} & 0.0054{*}{*}{*} & 0.0062{*}{*}{*}\\
& (0.0021) & (0.0016) & (0.0016) & (0.0020) & (0.0020) & (0.0020)\\
$\pi_{s,t-2}(\hat{\gamma}_{pre})$ & -0.0021 & -0.0114{*}{*} & -0.0095{*} &
-0.0086 & -0.0106{*} & -0.0088\\
& (0.0061) & (0.0058) & (0.0058) & (0.0055) & (0.0055) & (0.0055)\\
$q_{t-1}\times\pi_{s,t-2}(\hat{\gamma}_{post})$ & 0.0012 & 0.0009 & 0.0007 &
0.0001 & -0.0002 & -0.0002\\
& (0.0026) & (0.0020) & (0.0020) & (0.0025) & (0.0025) & (0.0025)\\
$\pi_{s,t-3}(\hat{\gamma}_{pre})$ & 0.0052 & 0.0091{*}{*} & 0.0104{*}{*} &
0.0096{*}{*} & 0.0086{*}{*} & 0.0100{*}{*}\\
& (0.0050) & (0.0045) & (0.0046) & (0.0043) & (0.0043) & (0.0044)\\
$q_{t-2}\times\pi_{s,t-3}(\hat{\gamma}_{post})$ & 0.0019 & -0.0001 & 0.0002 &
0.003 & 0.0008 & 0.0017\\
& (0.0021) & (0.0017) & (0.0017) & (0.0019) & (0.0020) & (0.0020)\\
$DA_{t-1}$ & 0.8123{*}{*}{*} & 0.8136{*}{*}{*} & 0.8121{*}{*}{*} &
0.8125{*}{*}{*} & 0.8138{*}{*}{*} & 0.8122{*}{*}{*}\\
& (0.0091) & (0.0091) & (0.0091) & (0.0091) & (0.0091) & (0.0091)\\
$DA_{t-2}$ & 0.0263{*}{*}{*} & 0.0271{*}{*}{*} & 0.0264{*}{*}{*} &
0.0261{*}{*}{*} & 0.0271{*}{*}{*} & 0.0265{*}{*}{*}\\
& (0.0077) & (0.0077) & (0.0077) & (0.0077) & (0.0077) & (0.0077)\\
$\left(  Cash/A\right)  _{t}$ & -0.0930{*}{*}{*} & -0.0932{*}{*}{*} &
-0.0929{*}{*}{*} & -0.0929{*}{*}{*} & -0.0933{*}{*}{*} & -0.0929{*}{*}{*}\\
& (0.0078) & (0.0078) & (0.0078) & (0.0078) & (0.0078) & (0.0078)\\
$\left(  Cash/A\right)  _{t-1}$ & 0.0532{*}{*}{*} & 0.0530{*}{*}{*} &
0.0529{*}{*}{*} & 0.0531{*}{*}{*} & 0.0530{*}{*}{*} & 0.0529{*}{*}{*}\\
& (0.0080) & (0.0081) & (0.0080) & (0.0080) & (0.0081) & (0.0080)\\
$\left(  Cash/A\right)  _{t-2}$ & 0.0034 & 0.0034 & 0.0035 & 0.0034 & 0.0035 &
0.0035\\
& (0.0044) & (0.0044) & (0.0044) & (0.0044) & (0.0044) & (0.0044)\\
$\left(  PPE/A\right)  _{t}$ & 0.0640{*}{*}{*} & 0.0645{*}{*}{*} & 0.0648{*}
{*}{*} & 0.0640{*}{*}{*} & 0.0644{*}{*}{*} & 0.0647{*}{*}{*}\\
& (0.0168) & (0.0169) & (0.0168) & (0.0168) & (0.0169) & (0.0168)\\
$\left(  PPE/A\right)  _{t-1}$ & -0.0336{*} & -0.0344{*} & -0.0342{*} &
-0.0336{*} & -0.0345{*} & -0.0343{*}\\
& (0.0180) & (0.0180) & (0.0180) & (0.0180) & (0.0180) & (0.0180)\\
$\left(  PPE/A\right)  _{t-2}$ & -0.0085 & -0.0092 & -0.0087 & -0.0084 &
-0.0092 & -0.0086\\
& (0.0088) & (0.0088) & (0.0088) & (0.0088) & (0.0088) & (0.0088)\\\hline
\end{tabular}

\end{center}

%

\footnotesize
Continued on next page.
\end{table}%

%

\setcounter{table}{12}%
%

\begin{table}[H]
\caption{(cont.)}
\vspace{-0.75cm}
\footnotesize

\begin{center}%
\begin{tabular}
[c]{l|ccc|ccc}%
\multicolumn{1}{c}{} & \multicolumn{6}{c}{Dependent variable: debt to assets
($DA_{t}$)}\\\cline{2-7}\cline{3-7}\cline{4-7}\cline{5-7}\cline{6-7}%
\multicolumn{1}{c}{} & \multicolumn{3}{c|}{$\gamma_{pre}=\gamma_{post}=\gamma
$} & \multicolumn{3}{c}{$\gamma_{pre}\neq\gamma_{post}$}\\
\multicolumn{1}{c}{} & (1) & (2) & (3) & (4) & (5) & (6)\\\hline
$Size_{t}$ & 0.0287{*}{*}{*} & 0.0290{*}{*}{*} & 0.0287{*}{*}{*} &
0.0287{*}{*}{*} & 0.0290{*}{*}{*} & 0.0287{*}{*}{*}\\
& (0.0038) & (0.0038) & (0.0038) & (0.0038) & (0.0038) & (0.0038)\\
$Size_{t-1}$ & -0.0298{*}{*}{*} & -0.0298{*}{*}{*} & -0.0298{*}{*}{*} &
-0.0298{*}{*}{*} & -0.0298{*}{*}{*} & -0.0298{*}{*}{*}\\
& (0.0040) & (0.0041) & (0.0040) & (0.0040) & (0.0041) & (0.0040)\\
$Size_{t-2}$ & 0.0045{*}{*}{*} & 0.0044{*}{*}{*} & 0.0045{*}{*}{*} &
0.0045{*}{*}{*} & 0.0044{*}{*}{*} & 0.0045{*}{*}{*}\\
& (0.0017) & (0.0017) & (0.0017) & (0.0017) & (0.0017) & (0.0017)\\
$Industry\;leverage_{t}$ & 0.2154{*}{*}{*} & 0.2095{*}{*}{*} & 0.2113{*}{*}{*}
& 0.2152{*}{*}{*} & 0.2100{*}{*}{*} & 0.2119{*}{*}{*}\\
& (0.0098) & (0.0098) & (0.0100) & (0.0099) & (0.0098) & (0.0100)\\
$Industry\;leverage_{t-1}$ & -0.1488{*}{*}{*} & -0.1434{*}{*}{*} &
-0.1379{*}{*}{*} & -0.1377{*}{*}{*} & -0.1411{*}{*}{*} & -0.1349{*}{*}{*}\\
& (0.0114) & (0.0119) & (0.0119) & (0.0119) & (0.0118) & (0.0119)\\
$Industry\;leverage_{t-2}$ & -0.0039 & -0.0142 & -0.0089 & -0.0064 & -0.0140 &
-0.0082\\
& (0.0090) & (0.0097) & (0.0099) & (0.0098) & (0.0097) & (0.0099)\\
$Industry\;growth_{t}$ & -0.0689{*}{*}{*} & -0.0767{*}{*}{*} & -0.0672{*}
{*}{*} & -0.0707{*}{*}{*} & -0.0771{*}{*}{*} & -0.0681{*}{*}{*}\\
& (0.0137) & (0.0136) & (0.0139) & (0.0137) & (0.0136) & (0.0139)\\
$Industry\;growth_{t-1}$ & -0.0276{*}{*} & -0.0371{*}{*}{*} & -0.0280{*}{*} &
-0.0294{*}{*} & -0.0385{*}{*}{*} & -0.0296{*}{*}\\
& (0.0121) & (0.0123) & (0.0125) & (0.0121) & (0.0123) & (0.0125)\\
$Industry\;growth_{t-2}$ & -0.0039 & -0.0196{*} & -0.0101 & -0.0063 &
-0.0210{*} & -0.0116\\
& (0.0115) & (0.0115) & (0.0117) & (0.0115) & (0.0115) & (0.0117)\\\hline
Fixed effects & Yes & Yes & Yes & Yes & Yes & Yes\\
Time effects & Yes & Yes & Yes & Yes & Yes & Yes\\
Industry linear trends & Yes & No & Yes & Yes & No & Yes\\
Ind. dummy$\times$RGDP & No & Yes & Yes & No & Yes & Yes\\
Observations & 84548 & 84548 & 84548 & 84548 & 84548 & 84548\\
$N$ & 3647 & 3647 & 3647 & 3647 & 3647 & 3647\\
$max(T_{i})$ & 44 & 44 & 44 & 44 & 44 & 44\\
$avg(T_{i})$ & 23.2 & 23.2 & 23.2 & 23.2 & 23.2 & 23.2\\
$med(T_{i})$ & 19 & 19 & 19 & 19 & 19 & 19\\
$min(T_{i})$ & 2 & 2 & 2 & 2 & 2 & 2\\\hline
\end{tabular}

\end{center}

%

\end{table}%


\newpage

\subsection{Long-run effects of LSAPs\label{Apx_LR_bench}}

Table \ref{tab:LR_bench_apx} reports the estimated long-run effects of LSAPs
and other determinants on firms' debt to asset ratios.%

\begin{table}[H]
\caption{\textbf{FE--TE estimates of the long-run effects of LSAPs on debt to asset ratios of non-financial firms}}\label{tab:LR_bench_apx}
\vspace{-0.2cm}
\footnotesize
Estimates of long-run effects of LSAPs, defined in equation
(\ref{LR_definition}), on firms' debt to asset ratios (DA) as well as the
long-run effects of both firm- and industry-specific variables on DA, for the
PanARDL(2) model described in equation (\ref{ARDL_model_lsap}). The first
three columns report results for the single-threshold panel regression model,
where $\gamma_{pre}=\gamma_{post}$. The last three columns report results for
the two-threshold panel regression, where $\gamma_{pre}\neq\gamma_{post}$. The
estimated quantile threshold parameters are shown in Table
\ref{tab: thresholds_LSAPs}. All regressions include both firm-specific
effects and time effects. Columns (1) and (4) include industry-specific linear
time trends, columns (2) and (5) include the interaction of industry dummies
and real GDP growth, while columns (3) and (6) include both. $LSAP$ is the
(scaled) amount of U.S. Treasuries and agency MBS purchased by the Fed;
$\pi_{-1}(\gamma)$ denotes the one-quarter lagged proportion of firms in an
industry with DA below the $\gamma$-th quantile. The sample consists of an
unbalanced panel of $3,647$ U.S. publicly traded non-financial firms observed
at a quarterly frequency over the period $2007$:Q1 - $2018$:Q3. Robust
standard errors (in parentheses) are computed using the delta method (***
$p<0.01$, ** $p<0.05$, * $p<0.1$).
\normalsize

\begin{center}
\scalebox{0.85}{
			\begin{tabular}{l|ccc|ccc}
				\multicolumn{1}{c}{} & \multicolumn{6}{c}{Dependent variable: debt to assets (DA)} \\
				\cline{2-7} \cline{3-7} \cline{4-7} \cline{5-7} \cline{6-7} \cline{7-7}
				\multicolumn{1}{c}{} & \multicolumn{3}{c|}{$\gamma_{pre}=\gamma_{post}=\gamma$} & \multicolumn{3}{c}{$\gamma_{pre}\neq\gamma_{post}$} \\
				\multicolumn{1}{c}{} & (1) & (2) & (3) & (4) & (5) & (6) \\
				\hline
				$\pi_{-1}(\hat{\gamma}_{pre})$ & 0.0966{*}{*}{*} & 0.0637{*}{*} & 0.1033{*}{*}{*} & 0.1152{*}{*}{*} & 0.0772{*}{*}{*} & 0.1203{*}{*}{*} \\
				& (0.0306) & (0.0300) & (0.0335) & (0.0316) & (0.0291) & (0.0321) \\
				$LSAP\times\pi_{-1}(\hat{\gamma}_{post})$ & 0.0424{*}{*}{*} & 0.0220{*}{*} & 0.0254{*}{*}{*} & 0.0546{*}{*}{*} & 0.0379{*}{*}{*} & 0.0475{*}{*}{*} \\
				& (0.0116) & (0.0087) & (0.0092) & (0.0108) & (0.0106) & (0.0112) \\
				Cash to assets & -0.2260{*}{*}{*} & -0.2311{*}{*}{*} & -0.2261{*}{*}{*} & -0.2256{*}{*}{*} & -0.2313{*}{*}{*} & -0.2259{*}{*}{*} \\
				& (0.0179) & (0.0179) & (0.0179) & (0.0179) & (0.0179) & (0.0179) \\
				PPE to assets & 0.1354{*}{*}{*} & 0.1306{*}{*}{*} & 0.1353{*}{*}{*} & 0.1361{*}{*}{*} & 0.1304{*}{*}{*} & 0.1359{*}{*}{*} \\
				& (0.0290) & (0.0287) & (0.0290) & (0.0290) & (0.0288) & (0.0290) \\
				Size & 0.0213{*}{*}{*} & 0.0231{*}{*}{*} & 0.0213{*}{*}{*} & 0.0213{*}{*}{*} & 0.0231{*}{*}{*} & 0.0214{*}{*}{*} \\
				& (0.0046) & (0.0046) & (0.0046) & (0.0046) & (0.0046) & (0.0046) \\
				Industry Leverage & 0.3882{*}{*}{*} & 0.3258{*}{*}{*} & 0.3997{*}{*}{*} & 0.4402{*}{*}{*} & 0.3443{*}{*}{*} & 0.4264{*}{*}{*} \\
				& (0.0460) & (0.0497) & (0.0571) & (0.0563) & (0.0501) & (0.0575) \\
				Industry Growth & -0.6223{*}{*}{*} & -0.8377{*}{*}{*} & -0.6524{*}{*}{*} & -0.6596{*}{*}{*} & -0.8588{*}{*}{*} & -0.6775{*}{*}{*} \\
				& (0.1309) & (0.1321) & (0.1368) & (0.1317) & (0.1326) & (0.1373) \\
				\hline
				Fixed effects & Yes & Yes & Yes & Yes & Yes & Yes \\
				Time effects & Yes & Yes & Yes & Yes & Yes & Yes \\
				Industry linear trends & Yes & No & Yes & Yes & No & Yes \\
				Ind. dummy$\times$RGDP & No & Yes & Yes & No & Yes & Yes \\
				Observations & 84548 & 84548 & 84548 & 84548 & 84548 & 84548 \\
				$N$ & 3647 & 3647 & 3647 & 3647 & 3647 & 3647 \\
				$max(T_{i})$ & 44 & 44 & 44 & 44 & 44 & 44 \\
				$avg(T_{i})$ & 23.2 & 23.2 & 23.2 & 23.2 & 23.2 & 23.2 \\
				$med(T_{i})$ & 19 & 19 & 19 & 19 & 19 & 19 \\
				$min(T_{i})$ & 2 & 2 & 2 & 2 & 2 & 2 \\
				\hline
		\end{tabular}	}
\end{center}

%

\end{table}%

\subsection{Description of the industry weights used to compute the effects of
LSAPs at national level\label{PE_appx}}

We now provide some additional details related to the computation of the
average policy effects at the national level described in equation (\ref{PE})
of the paper.

\begin{figure}[H]
	\caption{\textbf{Industry-specific weights based on firm size and employment}%
}
	\renewcommand\thefigure{10} \centering{\includegraphics
[scale=0.55]{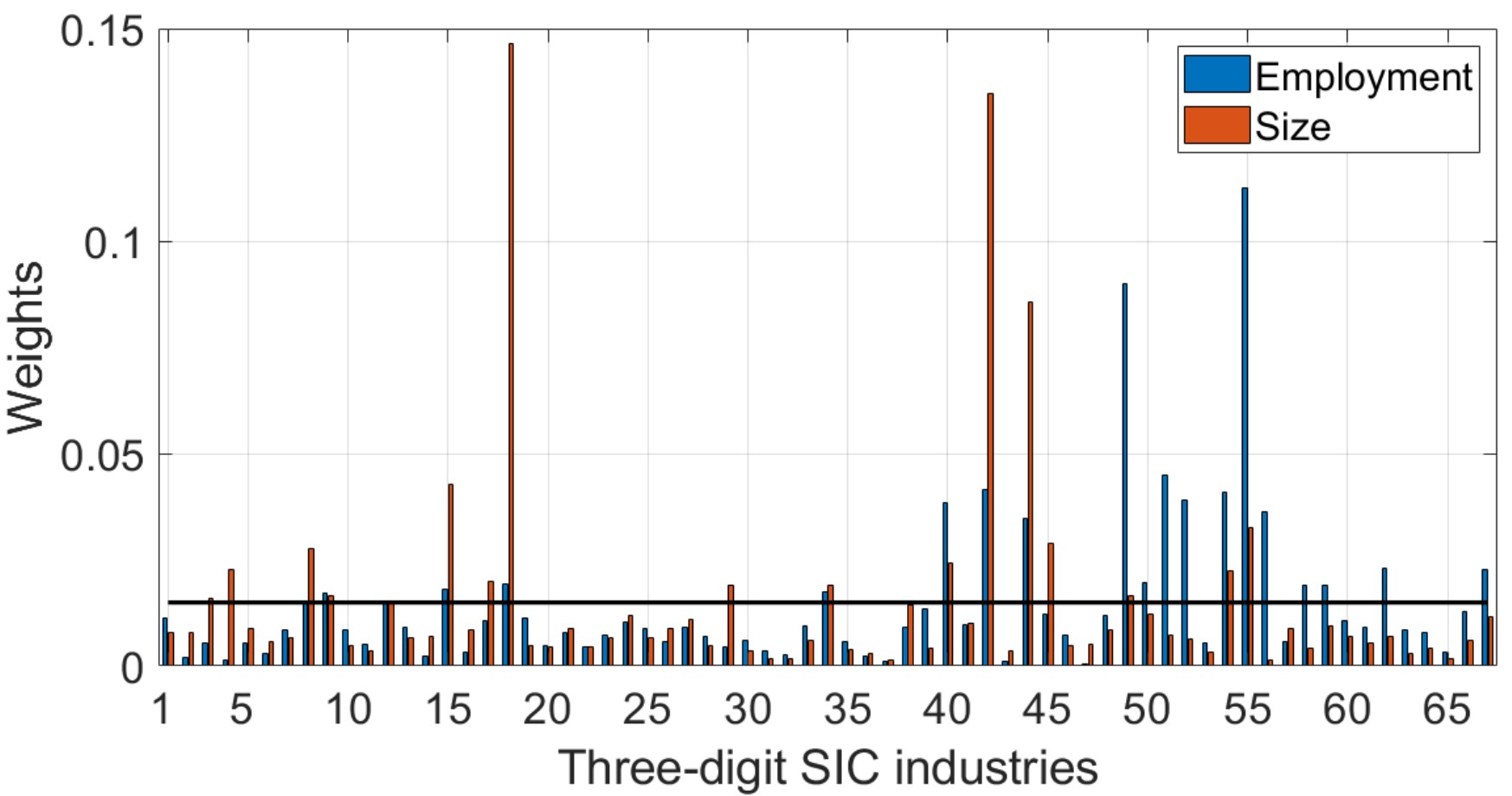}}
	\parbox{0.9\textwidth}{\footnotesize
{This figure displays the industry-specific weights used to compute the average per quarter policy effect at the national level. The blue bars indicate industry shares based on the average number of employees per firm within an industry. The orange bars report the weights based on average firm size within an industry. The black horizontal line shows the weights
based on a simple average (i.e. giving the same weight to each industry).}}
\label{fig:constant_weights}
\end{figure}

To calculate the average per quarter policy effect at the national level, we
need to compute the share of industry $s$ in the economy. To this extent, we
use two measures: (i) employment (measured as the average number of employees
per firm within an industry), and (ii) size (measured as firm's total asset,
in millions of dollars, averaged across firms and over time, within an
industry). The industry-specific weights obtained from both measures are shown
in Figure \ref{fig:constant_weights}.%

The estimates of the average policy effects (APE) at the industry and national
level described in equation (\ref{PE_s}) and (\ref{PE}) for the preferred
two-threshold PanARDL(2) model are reported in the paper.


\subsection{Additional control variables \label{AddVar}}

In this subsection, we demonstrate that our empirical results are robust to
the inclusion of an even larger set of both firm- and industry-level
regressors. Table \ref{tab:SR_bench_Addvar} reports the estimated net
short-run effects of LSAPs for the two-threshold PanARDL(2) model augmented
with additional explanatory variables. Similar conclusions hold for the
single-threshold model.%

\begin{table}[H]
\caption{\textbf{FE--TE estimates of the net short-run effects of LSAPs on debt to asset ratios of non-financial firms}}\label{tab:SR_bench_Addvar}
\vspace{-0.2cm}
\footnotesize
Estimates of net short-run effects of LSAPs on firms' debt to asset ratios
(DA) for the two-threshold PanARDL(2) model described in equation
(\ref{ARDL_model_lsap}). Net short-run effects are defined as the sum of the
estimated coefficients of current and lagged values of the regressor under
consideration. The estimated quantile threshold parameters are shown in Table
\ref{tab: thresholds_LSAPs}. All regressions include both firm-specific
effects and time effects. Columns (1) and (2) include industry-specific linear
time trends, columns (3) and (4) include the interaction of industry dummies
and real GDP growth, while columns (5) and (6) include both. $LSAP$ is the
(scaled) amount of U.S. Treasuries and agency MBS purchased by the Fed;
$\pi_{-1}(\gamma)$ denotes the one-quarter lagged proportion of firms in an
industry with DA below the $\gamma$-th quantile. The sample consists of an
unbalanced panel of $3,647$ U.S. publicly traded non-financial firms observed
at a quarterly frequency over the period $2007$:Q1 - $2018$:Q3. Robust
standard errors (in parentheses) are computed using the delta method (***
$p<0.01$, ** $p<0.05$, * $p<0.1$).
\normalsize

\begin{center}
\scalebox{0.85}{
			\begin{tabular}{l|cccccc}
				\multicolumn{1}{c}{} & \multicolumn{6}{c}{Dependent variable: debt to assets (DA)} \\
				\hline
				\multicolumn{1}{c}{} & (1) & (2) & (3) & (4) & (5) & (6) \\
				\hline
				$\pi_{-1}(\hat{\gamma}_{pre})$ & 0.0200{*}{*}{*} & 0.0201{*}{*}{*} & 0.0126{*}{*}{*} & 0.0146{*}{*}{*} & 0.0205{*}{*}{*} & 0.0206{*}{*}{*} \\
				& (0.0051) & (0.0051) & (0.0046) & (0.0047) & (0.0052) & (0.0052) \\
				$LSAP\times\pi_{-1}(\hat{\gamma}_{post})$ & 0.0087{*}{*}{*} & 0.0088{*}{*}{*} & 0.0064{*}{*}{*} & 0.0068{*}{*}{*} & 0.0073{*}{*}{*} & 0.0073{*}{*}{*} \\
				& (0.0017) & (0.0017) & (0.0017) & (0.0017) & (0.0018) & (0.0018) \\
				\hline
				\multicolumn{1}{c}{} & \multicolumn{6}{c}{\textit{Firm-specific variables}} \\
				\hline
				Lagged DA & Yes & Yes & Yes & Yes & Yes & Yes \\
				Cash to assets & Yes & Yes & Yes & Yes & Yes & Yes \\
				MB &  & Yes &  & Yes &  & Yes \\
				PPE to assets & Yes & Yes & Yes & Yes & Yes & Yes \\
				R\&D to assets &  & Yes &  & Yes &  & Yes \\
				Size & Yes & Yes & Yes & Yes & Yes & Yes \\
				\hline
				\multicolumn{1}{c}{} & \multicolumn{6}{c}{\textit{Industry-specific variables}} \\
				\hline
				Industry Leverage & Yes & Yes & Yes & Yes & Yes & Yes \\
				Industry Growth & Yes & Yes & Yes & Yes & Yes & Yes \\
				Industry Q & Yes &  & Yes &  & Yes &  \\
				Industry Cash/TA & Yes & Yes & Yes & Yes & Yes & Yes \\
				Industry MB &  & Yes &  & Yes &  & Yes \\
				Industry PPE/TA & Yes & Yes & Yes & Yes & Yes & Yes \\
				Industry R\&D/TA &  & Yes &  & Yes &  & Yes \\
				Industry Size & Yes & Yes & Yes & Yes & Yes & Yes \\
				\hline
				Fixed effects & Yes & Yes & Yes & Yes & Yes & Yes \\
				Time effects & Yes & Yes & Yes & Yes & Yes & Yes \\
				Industry linear trends & Yes & Yes & No & No & Yes & Yes \\
				Ind. dummy$\times$RGDP & No & No & Yes & Yes & Yes & Yes \\
				\hline
		\end{tabular}	}
\end{center}

%

\end{table}%

\newpage

\subsection{Observed macroeconomic indicators as proxies for $f_{t}$
\label{MacroVar}}

As discussed in the main paper, we use real GDP growth and/or linear trends as
proxies for $f_{t}$. This subsection reports estimation results using
alternative observed macroeconomic indicators to those used in the paper. We
consider four main alternative macroeconomic indicators: (i) growth in real
world output, (ii) the U.S. unemployment rate, (iii) the term spread (computed
as the difference between 10-year and 3-month Treasury bond yields), and (iv)
the one-year-ahead expected inflation. We re-estimate the threshold
parameters, and report the net short-run effects of LSAPs on firms' capital
structure below.

\subsubsection{Quantile threshold parameter estimates}

We re-estimate the threshold parameters associated with $\pi(\gamma)$, the
proportion of firms in an industry with DA below the $\gamma$-th quantile, for
different choices of $f_{t}$. The estimated thresholds are shown in Table
\ref{tab: thresholds_LSAPs_apx_ft}.%

\begin{table}[H]
\caption{\textbf{Estimated quantile threshold parameters}}\label{tab: thresholds_LSAPs_apx_ft}
\vspace{-0.2cm}
\footnotesize
Estimates of the quantile threshold parameters from a grid search procedure
for the PanARDL(2) model described in equation (\ref{ARDL_model_lsap}). Panel
A shows the estimated threshold parameters for the single-threshold panel
regression model, where $\gamma_{pre}=\gamma_{post}$. Panel B displays results
for the two-threshold model, where $\gamma_{pre}\neq\gamma_{post}$. In column
(1) we use the real world GDP growth as a proxy for $f_{t}$. In column (2),
$f_{t}$ denotes the unemployment rate. Column (3) includes three
macro-indicators interacted with industry dummies: U.S. real GDP growth, the
term spread, and expected inflation. The estimation sample consists of an
unbalanced panel of $3,647$ U.S. publicly traded non-financial firms observed
at a quarterly frequency over the period $2007$:Q1 - $2018$:Q3.
\normalsize

\begin{center}%
\begin{tabular}
[c]{cccc}\cline{2-4}\cline{3-4}\cline{4-4}
& (1) & (2) & (3)\\\hline
\textit{Panel A:} & \multicolumn{3}{c}{$\gamma_{pre}=\gamma_{post}=\gamma$}\\
$\hat{\gamma}$ & 0.56 & 0.56 & 0.56\\[0.2cm]%
\textit{Panel B:} & \multicolumn{3}{c}{$\gamma_{pre}\neq\gamma_{post}$}\\
$\hat{\gamma}_{pre}$ & 0.56 & 0.56 & 0.56\\
$\hat{\gamma}_{post}$ & 0.77 & 0.77 & 0.78\\\hline
$f_{t}$ & WGDP & Unemp & Multi\\\hline
\end{tabular}

\end{center}

%

\end{table}%

The first and second columns use real world output and U.S. unemployment rate
as a proxy for $f_{t}$, respectively. In the third column, we consider we
consider a model with multiple observed factors by using three macroeconomic
indicators, namely (i) growth in real GDP, (ii) the term spread, and (iii) the
one-year-ahead expected inflation. The choice of term spread and inflation
expectation is motivated by Frank and Goyal (2009).

\subsubsection{Short-run effects of LSAPs using alternative proxies for
$f_{t}$}

Table \ref{tab:SR_macroVar} reports the estimates of the net short-run effects
of LSAPs and other firm- and industry-specific characteristics on firms' leverage.%

\begin{table}[H]
\caption{\textbf{FE--TE estimates of the net short-run effects of LSAPs on debt to asset ratios of non-financial firms }}\label{tab:SR_macroVar}
\vspace{-0.2cm}
\footnotesize
Estimates of net short-run effects of LSAPs on firms' debt to asset ratios
(DA) for the PanARDL(2) model described in equation (\ref{ARDL_model_lsap}).
Net short-run effects are defined as the sum of the estimated coefficients of
current and lagged values of the regressor under consideration. The first
three columns report results for the single-threshold panel regression model,
where $\gamma_{pre}=\gamma_{post}$. The last three columns report results for
the two-threshold panel regression, where $\gamma_{pre}\neq\gamma_{post}$. The
estimated quantile threshold parameters are shown in Table
\ref{tab: thresholds_LSAPs_apx_ft}. All regressions include both firm-specific
effects and time effects. Columns (1) and (4) include the interaction of
industry dummies and world real GDP growth, columns (2) and (5) include the
interaction of industry dummies and unemployment rate, while columns (3) and
(6) include three macro-indicators interacted with industry dummies: U.S. real
GDP growth, the term spread, and expected inflation. $LSAP$ is the (scaled)
amount of U.S. Treasuries and agency MBS purchased by the Fed; $\pi
_{-1}(\gamma)$ denotes the one-quarter lagged proportion of firms in an
industry with DA below the $\gamma$-th quantile. The sample consists of an
unbalanced panel of $3,647$ U.S. publicly traded non-financial firms observed
at a quarterly frequency over the period $2007$:Q1 - $2018$:Q3. Robust
standard errors (in parentheses) are computed using the delta method (***
$p<0.01$, ** $p<0.05$, * $p<0.1$).
\normalsize

\begin{center}
\scalebox{0.85}{
			\begin{tabular}{l|ccc|ccc}
				\multicolumn{1}{c}{} & \multicolumn{6}{c}{Dependent variable: debt to assets (DA)} \\
				\cline{2-7} \cline{3-7} \cline{4-7} \cline{5-7} \cline{6-7} \cline{7-7}
				\multicolumn{1}{c}{} & \multicolumn{3}{c|}{$\gamma_{pre}=\gamma_{post}=\gamma$} & \multicolumn{3}{c}{$\gamma_{pre}\neq\gamma_{post}$} \\
				\multicolumn{1}{c}{} & (1) & (2) & (3) & (4) & (5) & (6) \\
				\hline
				$\pi_{-1}(\hat{\gamma}_{pre})$ & 0.0096{*}{*} & 0.0127{*}{*} & 0.0132{*}{*} & 0.0122{*}{*}{*} & 0.0160{*}{*}{*} & 0.0148{*}{*}{*} \\
				& (0.0047) & (0.0051) & (0.0053) & (0.0045) & (0.0048) & (0.0049) \\
				$LSAP\times\pi_{-1}(\hat{\gamma}_{post})$ & 0.0040{*}{*}{*} & 0.0055{*}{*}{*} & 0.0028 & 0.0062{*}{*}{*} & 0.0092{*}{*}{*} & 0.0061{*}{*}{*} \\
				& (0.0014) & (0.0018) & (0.0018) & (0.0017) & (0.0022) & (0.0021) \\
				\hline
				Fixed effects & Yes & Yes & Yes & Yes & Yes & Yes \\
				Time effects & Yes & Yes & Yes & Yes & Yes & Yes \\
				$f_{t}$ & WGDP & Unemp. & Multi & WGDP & Unemp. & Multi \\
				Observations & 84548 & 84548 & 84548 & 84548 & 84548 & 84548 \\
				$N$ & 3647 & 3647 & 3647 & 3647 & 3647 & 3647 \\
				$max(T_{i})$ & 44 & 44 & 44 & 44 & 44 & 44 \\
				$avg(T_{i})$ & 23.2 & 23.2 & 23.2 & 23.2 & 23.2 & 23.2 \\
				$med(T_{i})$ & 19 & 19 & 19 & 19 & 19 & 19 \\
				$min(T_{i})$ & 2 & 2 & 2 & 2 & 2 & 2 \\
				\hline
		\end{tabular}	}
\end{center}

%

\end{table}%


\subsection{Small-T bias and half-panel jackknife FE-TE estimation
\label{small-T_bias_OS}}

In this subsection, we report estimation results for the PanARDL(2) model
described in equation (\ref{ARDL_model_lsap}) after correcting for potential
small-sample bias arising from the fact that we employ a dynamic panel model
with fixed effects where the number of time series observations for some of
the firms in our sample is small.

Subsection \ref{appx_mint8} and \ref{appx_mint10} report the estimated
short-run effects after dropping firms with few time series observations,
namely firms with less than $8$ and $10$ time observations, respectively.

The number of firms available after selecting only firms with at least $8$
time observations is $3,236$ ($88.7\%$ of the initial sample). In this case,
after removing the pre-sample, the minimum, average, and maximum $T$ are equal
to $5$, $25.7$, and $44$, respectively. Instead, when selecting firms with at
least $10$ observations, the number of firms included in the sample is equal
to $3,011$ ($82.6\%$ of the initial sample), and the minimum, average, and
maximum $T$ after excluding the pre-sample are equal to $7$, $27.2$, and $44$, respectively.

Subsection \ref{appx_jackk} reports estimation results after correcting for
the small-$T$ bias by applying the half-panel jackknife method.

In the context of linear dynamic panel data models with possibly weakly
exogenous regressors, with $N$ (the number of cross-sections) large relative
to $T$ (the number of time observations), Chudik et al. (2018) show that the
bias of the half-panel jackknife FE--TE estimator is of order $T^{-2}$ and it
only requires that $N/T^{3}\rightarrow0$, as $N,T\rightarrow\infty$ for valid
inference. Instead the FE--E estimator requires $N/T\rightarrow0$, as
$N,T\rightarrow\infty$ jointly, and thus a larger $T$ to avoid potentially
biased estimation and size distortions.

\subsubsection{Short-run effects of LSAPs for firms with at least 8 time
observations\label{appx_mint8}}%

\begin{table}[H]
\caption{\textbf{FE--TE estimates of the net short-run effects of LSAPs on debt to asset ratios of non-financial firms with at least $8$ observations}}\label{tab:SR_bench_mint8}
\vspace{-0.2cm}
\footnotesize
Estimates of net short-run effects of LSAPs on firms' debt to asset ratios
(DA) as well as the effects of both firm- and industry-specific variables on
DA, for the PanARDL(2) model described in equation (\ref{ARDL_model_lsap}).
Net short-run effects are defined as the sum of the estimated coefficients of
current and lagged values of the regressor under consideration. The first
three columns report results for the single-threshold panel regression model,
where $\gamma_{pre}=\gamma_{post}$. The last three columns report results for
the two-threshold panel regression, where $\gamma_{pre}\neq\gamma_{post}$. The
estimated quantile threshold parameters are shown in Table
\ref{tab: thresholds_LSAPs}. All regressions include both firm-specific
effects and time effects. Columns (1) and (4) include industry-specific linear
time trends, columns (2) and (5) include the interaction of industry dummies
and real GDP growth, while columns (3) and (6) include both. $LSAP$ is the
(scaled) amount of U.S. Treasuries and agency MBS purchased by the Fed;
$\pi_{-1}(\gamma)$ denotes the one-quarter lagged proportion of firms in an
industry with DA below the $\gamma$-th quantile. The sample only includes
firms with at least $8$ time observations, resulting in an unbalanced panel of
$3,236$ U.S. publicly traded non-financial firms observed at a quarterly
frequency over the period $2007$:Q1 - $2018$:Q3. Robust standard errors (in
parentheses) are computed using the delta method (*** $p<0.01$, ** $p<0.05$, *
$p<0.1$).
\normalsize

\begin{center}
\scalebox{0.85}{
			\begin{tabular}{l|ccc|ccc}
				\multicolumn{1}{c}{} & \multicolumn{6}{c}{Dependent variable: debt to assets (DA)} \\
				\cline{2-7} \cline{3-7} \cline{4-7} \cline{5-7} \cline{6-7} \cline{7-7}
				\multicolumn{1}{c}{} & \multicolumn{3}{c|}{$\gamma_{pre}=\gamma_{post}=\gamma$} & \multicolumn{3}{c}{$\gamma_{pre}\neq\gamma_{post}$} \\
				\multicolumn{1}{c}{} & (1) & (2) & (3) & (4) & (5) & (6) \\
				\hline
				$\pi_{-1}(\hat{\gamma}_{pre})$ & 0.0155{*}{*}{*} & 0.0100{*}{*} & 0.0164{*}{*}{*} & 0.0181{*}{*}{*} & 0.0121{*}{*}{*} & 0.0190{*}{*}{*} \\
				& (0.0049) & (0.0047) & (0.0053) & (0.0050) & (0.0046) & (0.0051) \\
				$LSAP\times\pi_{-1}(\hat{\gamma}_{post})$ & 0.0069{*}{*}{*} & 0.0035{*}{*} & 0.0040{*}{*}{*} & 0.0089{*}{*}{*} & 0.0060{*}{*}{*} & 0.0076{*}{*}{*} \\
				& (0.0019) & (0.0014) & (0.0015) & (0.0017) & (0.0017) & (0.0018) \\
				Lagged DA & 0.8415{*}{*}{*} & 0.8436{*}{*}{*} & 0.8414{*}{*}{*} & 0.8415{*}{*}{*} & 0.8438{*}{*}{*} & 0.8416{*}{*}{*} \\
				& (0.0050) & (0.0050) & (0.0050) & (0.0050) & (0.0050) & (0.0050) \\
				Cash to assets & -0.0360{*}{*}{*} & -0.0363{*}{*}{*} & -0.0361{*}{*}{*} & -0.0359{*}{*}{*} & -0.0363{*}{*}{*} & -0.0360{*}{*}{*} \\
				& (0.0030) & (0.0029) & (0.0030) & (0.0030) & (0.0029) & (0.0030) \\
				PPE to assets & 0.0209{*}{*}{*} & 0.0198{*}{*}{*} & 0.0208{*}{*}{*} & 0.0210{*}{*}{*} & 0.0198{*}{*}{*} & 0.0209{*}{*}{*} \\
				& (0.0047) & (0.0046) & (0.0047) & (0.0047) & (0.0046) & (0.0047) \\
				Size & 0.0034{*}{*}{*} & 0.0036{*}{*}{*} & 0.0034{*}{*}{*} & 0.0034{*}{*}{*} & 0.0036{*}{*}{*} & 0.0034{*}{*}{*} \\
				& (0.0008) & (0.0007) & (0.0008) & (0.0008) & (0.0007) & (0.0008) \\
				Industry Leverage & 0.0623{*}{*}{*} & 0.0515{*}{*}{*} & 0.0642{*}{*}{*} & 0.0704{*}{*}{*} & 0.0544{*}{*}{*} & 0.0684{*}{*}{*} \\
				& (0.0075) & (0.0079) & (0.0092) & (0.0090) & (0.0080) & (0.0092) \\
				Industry Growth & -0.0974{*}{*}{*} & -0.1304{*}{*}{*} & -0.1032{*}{*}{*} & -0.1033{*}{*}{*} & -0.1336{*}{*}{*} & -0.1072{*}{*}{*} \\
				& (0.0209) & (0.0206) & (0.0219) & (0.0210) & (0.0207) & (0.0219) \\
				\hline
				Fixed effects & Yes & Yes & Yes & Yes & Yes & Yes \\
				Time effects & Yes & Yes & Yes & Yes & Yes & Yes \\
				Industry linear trends & Yes & No & Yes & Yes & No & Yes \\
				Ind. dummy$\times$RGDP & No & Yes & Yes & No & Yes & Yes \\
				Observations & 83290 & 83290 & 83290 & 83290 & 83290 & 83290 \\
				$N$ & 3236 & 3236 & 3236 & 3236 & 3236 & 3236 \\
				$max(T_{i})$ & 44 & 44 & 44 & 44 & 44 & 44 \\
				$avg(T_{i})$ & 25.7 & 25.7 & 25.7 & 25.7 & 25.7 & 25.7 \\
				$med(T_{i})$ & 23 & 23 & 23 & 23 & 23 & 23 \\
				$min(T_{i})$ & 5 & 5 & 5 & 5 & 5 & 5 \\
				\hline
		\end{tabular}	}
\end{center}

%

\end{table}%

\subsubsection{Short-run effects of LSAPs for firms with at least 10 time
observations \label{appx_mint10}}%

\begin{table}[H]
\caption{\textbf{FE--TE estimates of the net short-run effects of LSAPs on debt to asset ratios of non-financial firms with at least $10$ observations}}\label{tab:SR_bench_mint10}
\vspace{-0.2cm}
\footnotesize
Estimates of net short-run effects of LSAPs on firms' debt to asset ratios
(DA) as well as the effects of both firm- and industry-specific variables on
DA, for the PanARDL(2) model described in equation (\ref{ARDL_model_lsap}).
Net short-run effects are defined as the sum of the estimated coefficients of
current and lagged values of the regressor under consideration. The first
three columns report results for the single-threshold panel regression model,
where $\gamma_{pre}=\gamma_{post}$. The last three columns report results for
the two-threshold panel regression, where $\gamma_{pre}\neq\gamma_{post}$. The
estimated quantile threshold parameters are shown in Table
\ref{tab: thresholds_LSAPs}. All regressions include both firm-specific
effects and time effects. Columns (1) and (4) include industry-specific linear
time trends, columns (2) and (5) include the interaction of industry dummies
and real GDP growth, while columns (3) and (6) include both. $LSAP$ is the
(scaled) amount of U.S. Treasuries and agency MBS purchased by the Fed;
$\pi_{-1}(\gamma)$ denotes the one-quarter lagged proportion of firms in an
industry with DA below the $\gamma$-th quantile. The sample only includes
firms with at least $10$ time observations, resulting in an unbalanced panel
of $3,011$ U.S. publicly traded non-financial firms observed at a quarterly
frequency over the period $2007$:Q1 - $2018$:Q3. Robust standard errors (in
parentheses) are computed using the delta method (*** $p<0.01$, ** $p<0.05$, *
$p<0.1$).
\normalsize

\begin{center}
\scalebox{0.85}{
			\begin{tabular}{l|ccc|ccc}
				\multicolumn{1}{c}{} & \multicolumn{6}{c}{Dependent variable: debt to assets (DA)} \\
				\cline{2-7} \cline{3-7} \cline{4-7} \cline{5-7} \cline{6-7} \cline{7-7}
				\multicolumn{1}{c}{} & \multicolumn{3}{c|}{$\gamma_{pre}=\gamma_{post}=\gamma$} & \multicolumn{3}{c}{$\gamma_{pre}\neq\gamma_{post}$} \\
				\multicolumn{1}{c}{} & (1) & (2) & (3) & (4) & (5) & (6) \\
				\hline
				$\pi_{-1}(\hat{\gamma}_{pre})$ & 0.0153{*}{*}{*} & 0.0098{*}{*} & 0.0160{*}{*}{*} & 0.0178{*}{*}{*} & 0.0120{*}{*}{*} & 0.0188{*}{*}{*} \\
				& (0.0049) & (0.0047) & (0.0053) & (0.0050) & (0.0046) & (0.0051) \\
				$LSAP\times\pi_{-1}(\hat{\gamma}_{post})$ & 0.0070{*}{*}{*} & 0.0036{*}{*}{*} & 0.0041{*}{*}{*} & 0.0090{*}{*}{*} & 0.0062{*}{*}{*} & 0.0076{*}{*}{*} \\
				& (0.0018) & (0.0014) & (0.0015) & (0.0017) & (0.0017) & (0.0018) \\
				Lagged DA & 0.8447{*}{*}{*} & 0.8467{*}{*}{*} & 0.8446{*}{*}{*} & 0.8447{*}{*}{*} & 0.8469{*}{*}{*} & 0.8447{*}{*}{*} \\
				& (0.0050) & (0.0050) & (0.0050) & (0.0050) & (0.0050) & (0.0050) \\
				Cash to assets & -0.0355{*}{*}{*} & -0.0358{*}{*}{*} & -0.0356{*}{*}{*} & -0.0355{*}{*}{*} & -0.0358{*}{*}{*} & -0.0355{*}{*}{*} \\
				& (0.0030) & (0.0029) & (0.0030) & (0.0030) & (0.0029) & (0.0030) \\
				PPE to assets & 0.0197{*}{*}{*} & 0.0188{*}{*}{*} & 0.0196{*}{*}{*} & 0.0198{*}{*}{*} & 0.0187{*}{*}{*} & 0.0196{*}{*}{*} \\
				& (0.0046) & (0.0045) & (0.0046) & (0.0046) & (0.0045) & (0.0046) \\
				Size & 0.0035{*}{*}{*} & 0.0038{*}{*}{*} & 0.0035{*}{*}{*} & 0.0035{*}{*}{*} & 0.0038{*}{*}{*} & 0.0035{*}{*}{*} \\
				& (0.0008) & (0.0007) & (0.0008) & (0.0008) & (0.0007) & (0.0008) \\
				Industry Leverage & 0.0605{*}{*}{*} & 0.0494{*}{*}{*} & 0.0618{*}{*}{*} & 0.0685{*}{*}{*} & 0.0524{*}{*}{*} & 0.0661{*}{*}{*} \\
				& (0.0075) & (0.0079) & (0.0092) & (0.0090) & (0.0079) & (0.0092) \\
				Industry Growth & -0.0960{*}{*}{*} & -0.1297{*}{*}{*} & -0.1027{*}{*}{*} & -0.1017{*}{*}{*} & -0.1331{*}{*}{*} & -0.1066{*}{*}{*} \\
				& (0.0211) & (0.0207) & (0.0220) & (0.0212) & (0.0208) & (0.0220) \\
				\hline
				Fixed effects & Yes & Yes & Yes & Yes & Yes & Yes \\
				Time effects & Yes & Yes & Yes & Yes & Yes & Yes \\
				Industry linear trends & Yes & No & Yes & Yes & No & Yes \\
				Ind. dummy$\times$RGDP & No & Yes & Yes & No & Yes & Yes \\
				Observations & 82038 & 82038 & 82038 & 82038 & 82038 & 82038 \\
				$N$ & 3011 & 3011 & 3011 & 3011 & 3011 & 3011 \\
				$max(T_{i})$ & 44 & 44 & 44 & 44 & 44 & 44 \\
				$avg(T_{i})$ & 27.2 & 27.2 & 27.2 & 27.2 & 27.2 & 27.2 \\
				$med(T_{i})$ & 25 & 25 & 25 & 25 & 25 & 25 \\
				$min(T_{i})$ & 7 & 7 & 7 & 7 & 7 & 7 \\
				\hline
		\end{tabular}	}
\end{center}

%

\end{table}%


\subsubsection{Half-panel jackknife FE-TE estimates \label{appx_jackk}}

%

\begin{table}[H]
\caption{\textbf{Half-panel jackknife FE--TE estimates of the effects of LSAPs on non-financial firm's debt to asset ratios based on the PanARDL(2) model}}\label{tab:ARDL2_benchJackk}
\vspace{-0.2cm}
\footnotesize
Estimates of the coefficients of the PanARDL(2) model described in equation
(\ref{ARDL_model_lsap}). The dependent variable is debt to asset ratio (DA).
$q_{t}$ is the (scaled) amount of U.S. Treasuries and agency MBS purchased by
the Fed; $\pi_{s,t}(\hat{\gamma})$ denotes the proportion of firms in an
industry with DA below the $\hat{\gamma}$-th quantile. The first three columns
report results for the single-threshold panel regression model, where
$\gamma_{pre}=\gamma_{post}$. The last three columns report results for the
two-threshold panel regression, where $\gamma_{pre}\neq\gamma_{post}$. The
estimated quantile threshold parameters are shown in Table
\ref{tab: thresholds_LSAPs}. All regressions include both firm-specific
effects and time effects. Columns (1) and (4) include industry-specific linear
time trends, columns (2) and (5) include the interaction of industry dummies
and real GDP growth, while columns (3) and (6) include both. The sample only
includes firms with at least $8$ time observations, resulting in an unbalanced
panel of $3,236$ U.S. publicly traded non-financial firms observed at a
quarterly frequency over the period $2007$:Q1 - $2018$:Q3. Robust standard
errors in parentheses (*** $p<0.01$, ** $p<0.05$, * $p<0.1$).

\begin{center}%
\begin{tabular}
[c]{l|ccc|ccc}%
\multicolumn{1}{c}{} & \multicolumn{6}{c}{Dependent variable: debt to assets
($DA_{t}$)}\\\cline{2-7}\cline{3-7}\cline{4-7}\cline{5-7}\cline{6-7}%
\multicolumn{1}{c}{} & \multicolumn{3}{c|}{$\gamma_{pre}=\gamma_{post}=\gamma
$} & \multicolumn{3}{c}{$\gamma_{pre}\neq\gamma_{post}$}\\
\multicolumn{1}{c}{} & (1) & (2) & (3) & (4) & (5) & (6)\\\hline
$\pi_{s,t-1}(\hat{\gamma}_{pre})$ & 0.0118{*}{*} & 0.0122{*}{*} & 0.0137{*}{*}
& 0.0125{*}{*} & 0.0147{*}{*}{*} & 0.0159{*}{*}{*}\\
& (0.0056) & (0.0057) & (0.0059) & (0.0055) & (0.0055) & (0.0056)\\
$q_{t}\times\pi_{s,t-1}(\hat{\gamma}_{post})$ & 0.0018 & 0.0024 & 0.0021 &
0.0039{*} & 0.0043{*} & 0.0043{*}\\
& (0.0023) & (0.0018) & (0.0018) & (0.0022) & (0.0022) & (0.0022)\\
$\pi_{s,t-2}(\hat{\gamma}_{pre})$ & -0.0070 & -0.0157{*}{*}{*} & -0.0155{*}{*}
& -0.0131{*}{*} & -0.0140{*}{*} & -0.0139{*}{*}\\
& (0.0063) & (0.0060) & (0.0060) & (0.0058) & (0.0058) & (0.0058)\\
$q_{t-1}\times\pi_{s,t-2}(\hat{\gamma}_{post})$ & 0.0038 & 0.0022 & 0.0021 &
0.0019 & 0.0015 & 0.0016\\
& (0.0026) & (0.0021) & (0.0021) & (0.0025) & (0.0025) & (0.0025)\\
$\pi_{s,t-3}(\hat{\gamma}_{pre})$ & 0.0038 & 0.0085{*} & 0.0077 & 0.0068 &
0.0084{*} & 0.0075\\
& (0.0054) & (0.0048) & (0.0050) & (0.0046) & (0.0046) & (0.0047)\\
$q_{t-2}\times\pi_{s,t-3}(\hat{\gamma}_{post})$ & 0.0013 & 0.0005 & 0.0005 &
0.0022 & 0.0004 & 0.0008\\
& (0.0022) & (0.0018) & (0.0018) & (0.0021) & (0.0022) & (0.0022)\\
$DA_{t-1}$ & 0.9120{*}{*}{*} & 0.9120{*}{*}{*} & 0.9118{*}{*}{*} &
0.9121{*}{*}{*} & 0.9121{*}{*}{*} & 0.9118{*}{*}{*}\\
& (0.0102) & (0.0102) & (0.0102) & (0.0102) & (0.0102) & (0.0102)\\
$DA_{t-2}$ & 0.0376{*}{*}{*} & 0.0381{*}{*}{*} & 0.0377{*}{*}{*} &
0.0375{*}{*}{*} & 0.0382{*}{*}{*} & 0.0377{*}{*}{*}\\
& (0.0089) & (0.0089) & (0.0089) & (0.0089) & (0.0089) & (0.0089)\\
$\left(  Cash/A\right)  _{t}$ & -0.0945{*}{*}{*} & -0.0946{*}{*}{*} &
-0.0943{*}{*}{*} & -0.0945{*}{*}{*} & -0.0946{*}{*}{*} & -0.0944{*}{*}{*}\\
& (0.0090) & (0.0090) & (0.0090) & (0.0090) & (0.0090) & (0.0090)\\
$\left(  Cash/A\right)  _{t-1}$ & 0.0615{*}{*}{*} & 0.0614{*}{*}{*} &
0.0611{*}{*}{*} & 0.0615{*}{*}{*} & 0.0615{*}{*}{*} & 0.0612{*}{*}{*}\\
& (0.0089) & (0.0089) & (0.0089) & (0.0089) & (0.0089) & (0.0089)\\
$\left(  Cash/A\right)  _{t-2}$ & 0.0091{*} & 0.0090{*} & 0.0091{*} &
0.0091{*} & 0.0090{*} & 0.0091{*}\\
& (0.0052) & (0.0052) & (0.0052) & (0.0052) & (0.0052) & (0.0052)\\\hline
\end{tabular}

\end{center}

%

\footnotesize
Continued on next page.
\end{table}%

%

\setcounter{table}{19}%
%

\begin{table}[H]
\caption{(cont.)}
\vspace{-0.75cm}
\footnotesize

\begin{center}%
\begin{tabular}
[c]{l|ccc|ccc}%
\multicolumn{1}{c}{} & \multicolumn{6}{c}{Dependent variable: debt to assets
($DA_{t}$)}\\\cline{2-7}\cline{3-7}\cline{4-7}\cline{5-7}\cline{6-7}%
\multicolumn{1}{c}{} & \multicolumn{3}{c|}{$\gamma_{pre}=\gamma_{post}=\gamma
$} & \multicolumn{3}{c}{$\gamma_{pre}\neq\gamma_{post}$}\\
\multicolumn{1}{c}{} & (1) & (2) & (3) & (4) & (5) & (6)\\\hline
$\left(  PPE/A\right)  _{t}$ & 0.0614{*}{*}{*} & 0.0620{*}{*}{*} &
0.0624{*}{*}{*} & 0.0613{*}{*}{*} & 0.0620{*}{*}{*} & 0.0624{*}{*}{*}\\
& (0.0197) & (0.0198) & (0.0198) & (0.0197) & (0.0198) & (0.0198)\\
$\left(  PPE/A\right)  _{t-1}$ & -0.0438{*}{*} & -0.0449{*}{*} & -0.0451{*}{*}
& -0.0438{*}{*} & -0.0448{*}{*} & -0.0451{*}{*}\\
& (0.0202) & (0.0202) & (0.0202) & (0.0202) & (0.0202) & (0.0202)\\
$\left(  PPE/A\right)  _{t-2}$ & -0.0193{*} & -0.0195{*} & -0.0191{*} &
-0.0190{*} & -0.0195{*} & -0.0191{*}\\
& (0.0106) & (0.0106) & (0.0106) & (0.0106) & (0.0106) & (0.0106)\\
$Size_{t}$ & 0.0276{*}{*}{*} & 0.0282{*}{*}{*} & 0.0277{*}{*}{*} &
0.0276{*}{*}{*} & 0.0282{*}{*}{*} & 0.0277{*}{*}{*}\\
& (0.0045) & (0.0045) & (0.0045) & (0.0045) & (0.0045) & (0.0045)\\
$Size_{t-1}$ & -0.0332{*}{*}{*} & -0.0332{*}{*}{*} & -0.0333{*}{*}{*} &
-0.0333{*}{*}{*} & -0.0332{*}{*}{*} & -0.0333{*}{*}{*}\\
& (0.0046) & (0.0046) & (0.0046) & (0.0046) & (0.0046) & (0.0046)\\
$Size_{t-2}$ & 0.0061{*}{*}{*} & 0.0061{*}{*}{*} & 0.0061{*}{*}{*} &
0.0061{*}{*}{*} & 0.0061{*}{*}{*} & 0.0061{*}{*}{*}\\
& (0.0020) & (0.0020) & (0.0020) & (0.0020) & (0.0020) & (0.0020)\\
$Industry\;leverage_{t}$ & 0.2166{*}{*}{*} & 0.2118{*}{*}{*} & 0.2146{*}{*}{*}
& 0.2167{*}{*}{*} & 0.2126{*}{*}{*} & 0.2151{*}{*}{*}\\
& (0.0109) & (0.0110) & (0.0111) & (0.0110) & (0.0110) & (0.0111)\\
$Industry\;leverage_{t-1}$ & -0.1613{*}{*}{*} & -0.1570{*}{*}{*} &
-0.1542{*}{*}{*} & -0.1546{*}{*}{*} & -0.1542{*}{*}{*} & -0.1513{*}{*}{*}\\
& (0.0119) & (0.0125) & (0.0126) & (0.0126) & (0.0125) & (0.0126)\\
$Industry\;leverage_{t-2}$ & -0.0119 & -0.0242{*}{*} & -0.0196{*} & -0.0167 &
-0.0228{*}{*} & -0.0184{*}\\
& (0.0098) & (0.0107) & (0.0107) & (0.0107) & (0.0107) & (0.0107)\\
$Industry\;growth_{t}$ & -0.0706{*}{*}{*} & -0.0821{*}{*}{*} & -0.0712{*}%
{*}{*} & -0.0719{*}{*}{*} & -0.0827{*}{*}{*} & -0.0723{*}{*}{*}\\
& (0.0156) & (0.0159) & (0.0156) & (0.0156) & (0.0159) & (0.0156)\\
$Industry\;growth_{t-1}$ & -0.0315{*}{*} & -0.0440{*}{*}{*} & -0.0326{*}{*} &
-0.0317{*}{*} & -0.0453{*}{*}{*} & -0.0339{*}{*}\\
& (0.0137) & (0.0141) & (0.0140) & (0.0137) & (0.0141) & (0.0139)\\
$Industry\;growth_{t-2}$ & -0.0064 & -0.0195 & -0.0103 & -0.0064 & -0.0208 &
-0.0118\\
& (0.0131) & (0.0133) & (0.0132) & (0.0131) & (0.0133) & (0.0132)\\\hline
Fixed effects & Yes & Yes & Yes & Yes & Yes & Yes\\
Time effects & Yes & Yes & Yes & Yes & Yes & Yes\\
Industry linear trends & Yes & No & Yes & Yes & No & Yes\\
Ind. dummy$\times$RGDP & No & Yes & Yes & No & Yes & Yes\\
Observations & 82092 & 82092 & 82092 & 82092 & 82092 & 82092\\
$N$ & 3236 & 3236 & 3236 & 3236 & 3236 & 3236\\
$max(T_{i})$ & 44 & 44 & 44 & 44 & 44 & 44\\
$avg(T_{i})$ & 25.4 & 25.4 & 25.4 & 25.4 & 25.4 & 25.4\\
$med(T_{i})$ & 22 & 22 & 22 & 22 & 22 & 22\\
$min(T_{i})$ & 4 & 4 & 4 & 4 & 4 & 4\\\hline
\end{tabular}

\end{center}

%

\end{table}%

%

\begin{table}[H]
\caption{\textbf{Half-panel jackknife FE--TE estimates of the net short-run effects of LSAPs on debt to asset ratios of non-financial firms}}\label{tab:SR_bench_Jackk}
\vspace{-0.2cm}
\footnotesize
Estimates of net short-run effects of LSAPs on firms' debt to asset ratios
(DA) as well as the effects of both firm- and industry-specific variables on
DA, for the PanARDL(2) model described in equation (\ref{ARDL_model_lsap}).
Net short-run effects are defined as the sum of the estimated coefficients of
current and lagged values of the regressor under consideration. The first
three columns report results for the single-threshold panel regression model,
where $\gamma_{pre}=\gamma_{post}$. The last three columns report results for
the two-threshold panel regression, where $\gamma_{pre}\neq\gamma_{post}$. The
estimated quantile threshold parameters are shown in Table
\ref{tab: thresholds_LSAPs}. All regressions include both firm-specific
effects and time effects. Columns (1) and (4) include industry-specific linear
time trends, columns (2) and (5) include the interaction of industry dummies
and real GDP growth, while columns (3) and (6) include both. $LSAP$ is the
(scaled) amount of U.S. Treasuries and agency MBS purchased by the Fed;
$\pi_{-1}(\gamma)$ denotes the one-quarter lagged proportion of firms in an
industry with DA below the $\gamma$-th quantile. The sample only includes
firms with at least $8$ time observations, resulting in an unbalanced panel of
$3,236$ U.S. publicly traded non-financial firms observed at a quarterly
frequency over the period $2007$:Q1 - $2018$:Q3. Robust standard errors (in
parentheses) are computed using the delta method (*** $p<0.01$, ** $p<0.05$, *
$p<0.1$).
\normalsize

\begin{center}
\scalebox{0.85}{
			\begin{tabular}{l|ccc|ccc}
				\multicolumn{1}{c}{} & \multicolumn{6}{c}{Dependent variable: debt to assets (DA)} \\
				\cline{2-7} \cline{3-7} \cline{4-7} \cline{5-7} \cline{6-7} \cline{7-7}
				\multicolumn{1}{c}{} & \multicolumn{3}{c|}{$\gamma_{pre}=\gamma_{post}=\gamma$} & \multicolumn{3}{c}{$\gamma_{pre}\neq\gamma_{post}$} \\
				\multicolumn{1}{c}{} & (1) & (2) & (3) & (4) & (5) & (6) \\
				\hline
				$\pi_{-1}(\hat{\gamma}_{pre})$ & 0.0085 & 0.0049 & 0.0059 & 0.0063 & 0.0091 & 0.0095 \\
				& (0.0062) & (0.0063) & (0.0067) & (0.0064) & (0.0062) & (0.0065) \\
				$LSAP\times\pi_{-1}(\hat{\gamma}_{post})$ & 0.0068{*}{*}{*} & 0.0051{*}{*}{*} & 0.0047{*}{*} & 0.0080{*}{*}{*} & 0.0061{*}{*}{*} & 0.0066{*}{*}{*} \\
				& (0.0022) & (0.0018) & (0.0018) & (0.0021) & (0.0021) & (0.0022) \\
				Lagged DA & 0.9496{*}{*}{*} & 0.9500{*}{*}{*} & 0.9494{*}{*}{*} & 0.9496{*}{*}{*} & 0.9502{*}{*}{*} & 0.9495{*}{*}{*} \\
				& (0.0067) & (0.0067) & (0.0067) & (0.0067) & (0.0067) & (0.0067) \\
				Cash to assets & -0.0239{*}{*}{*} & -0.0241{*}{*}{*} & -0.0241{*}{*}{*} & -0.0239{*}{*}{*} & -0.0241{*}{*}{*} & -0.0241{*}{*}{*} \\
				& (0.0043) & (0.0042) & (0.0043) & (0.0043) & (0.0042) & (0.0043) \\
				PPE to assets & -0.0017 & -0.0024 & -0.0018 & -0.0014 & -0.0023 & -0.0017 \\
				& (0.0072) & (0.0071) & (0.0072) & (0.0072) & (0.0071) & (0.0072) \\
				Size & 0.0005 & 0.0010 & 0.0005 & 0.0005 & 0.0010 & 0.0005 \\
				& (0.0012) & (0.0012) & (0.0012) & (0.0012) & (0.0012) & (0.0012) \\
				Industry Leverage & 0.0434{*}{*}{*} & 0.0307{*}{*}{*} & 0.0408{*}{*}{*} & 0.0454{*}{*}{*} & 0.0356{*}{*}{*} & 0.0454{*}{*}{*} \\
				& (0.0097) & (0.0113) & (0.0119) & (0.0117) & (0.0114) & (0.0120) \\
				Industry Growth & -0.1085{*}{*}{*} & -0.1456{*}{*}{*} & -0.1141{*}{*}{*} & -0.1100{*}{*}{*} & -0.1489{*}{*}{*} & -0.1180{*}{*}{*} \\
				& (0.0267) & (0.0281) & (0.0273) & (0.0268) & (0.0282) & (0.0274) \\
				\hline
				Fixed effects & Yes & Yes & Yes & Yes & Yes & Yes \\
				Time effects & Yes & Yes & Yes & Yes & Yes & Yes \\
				Industry linear trends & Yes & No & Yes & Yes & No & Yes \\
				Ind. dummy$\times$RGDP & No & Yes & Yes & No & Yes & Yes \\
				Observations & 82092 & 82092 & 82092 & 82092 & 82092 & 82092 \\
				$N$ & 3236 & 3236 & 3236 & 3236 & 3236 & 3236 \\
				$max(T_{i})$ & 44 & 44 & 44 & 44 & 44 & 44 \\
				$avg(T_{i})$ & 25.4 & 25.4 & 25.4 & 25.4 & 25.4 & 25.4 \\
				$med(T_{i})$ & 22 & 22 & 22 & 22 & 22 & 22 \\
				$min(T_{i})$ & 4 & 4 & 4 & 4 & 4 & 4 \\
				\hline
		\end{tabular}	}
\end{center}

%

\end{table}%
%

\begin{table}[H]
\caption{\textbf{Half-panel jackknife FE--TE estimates of the long-run effects of LSAPs on debt to asset ratios of non-financial firms}}\label{tab:LR_bench_Jackk}
\vspace{-0.2cm}
\footnotesize
Estimates of long-run effects of LSAPs, defined in equation
(\ref{LR_definition}), on firms' debt to asset ratios (DA) as well as the
long-run effects of both firm- and industry-specific variables on DA, for the
PanARDL(2) model described in equation (\ref{ARDL_model_lsap}). The first
three columns report results for the single-threshold panel regression model,
where $\gamma_{pre}=\gamma_{post}$. The last three columns report results for
the two-threshold panel regression, where $\gamma_{pre}\neq\gamma_{post}$. The
estimated quantile threshold parameters are shown in Table
\ref{tab: thresholds_LSAPs}. All regressions include both firm-specific
effects and time effects. Columns (1) and (4) include industry-specific linear
time trends, columns (2) and (5) include the interaction of industry dummies
and real GDP growth, while columns (3) and (6) include both. $LSAP$ is the
(scaled) amount of U.S. Treasuries and agency MBS purchased by the Fed;
$\pi_{-1}(\gamma)$ denotes the one-quarter lagged proportion of firms in an
industry with DA below the $\gamma$-th quantile. The sample only includes
firms with at least $8$ time observations, resulting in an unbalanced panel of
$3,236$ U.S. publicly traded non-financial firms observed at a quarterly
frequency over the period $2007$:Q1 - $2018$:Q3. Robust standard errors (in
parentheses) are computed using the delta method (*** $p<0.01$, ** $p<0.05$, *
$p<0.1$).
\normalsize

\begin{center}
\scalebox{0.85}{
			\begin{tabular}{l|ccc|ccc}
				\multicolumn{1}{c}{} & \multicolumn{6}{c}{Dependent variable: debt to assets (DA)} \\
				\cline{2-7} \cline{3-7} \cline{4-7} \cline{5-7} \cline{6-7} \cline{7-7}
				\multicolumn{1}{c}{} & \multicolumn{3}{c|}{$\gamma_{pre}=\gamma_{post}=\gamma$} & \multicolumn{3}{c}{$\gamma_{pre}\neq\gamma_{post}$} \\
				\multicolumn{1}{c}{} & (1) & (2) & (3) & (4) & (5) & (6) \\
				\hline
				$\pi_{-1}(\hat{\gamma}_{pre})$ & 0.1683 & 0.0985 & 0.1174 & 0.1253 & 0.1832 & 0.1884 \\
				& (0.1248) & (0.1285) & (0.1353) & (0.1293) & (0.1285) & (0.1328) \\
				$LSAP\times\pi_{-1}(\hat{\gamma}_{post})$ & 0.1353{*}{*}{*} & 0.1022{*}{*}{*} & 0.0931{*}{*} & 0.1584{*}{*}{*} & 0.1224{*}{*}{*} & 0.1312{*}{*}{*} \\
				& (0.0483) & (0.0382) & (0.0385) & (0.0466) & (0.0461) & (0.0466) \\
				Cash to assets & -0.4745{*}{*}{*} & -0.4832{*}{*}{*} & -0.4772{*}{*}{*} & -0.4742{*}{*}{*} & -0.4839{*}{*}{*} & -0.4778{*}{*}{*} \\
				& (0.0929) & (0.0930) & (0.0929) & (0.0930) & (0.0934) & (0.0931) \\
				PPE to assets & -0.0331 & -0.0481 & -0.036 & -0.0288 & -0.0464 & -0.034 \\
				& (0.1431) & (0.1416) & (0.1428) & (0.1432) & (0.1422) & (0.1431) \\
				Size & 0.0091 & 0.0209 & 0.0102 & 0.0092 & 0.0210 & 0.0104 \\
				& (0.0235) & (0.0230) & (0.0234) & (0.0235) & (0.0231) & (0.0235) \\
				Industry Leverage & 0.8597{*}{*}{*} & 0.6142{*}{*}{*} & 0.8067{*}{*}{*} & 0.9005{*}{*}{*} & 0.7155{*}{*}{*} & 0.8995{*}{*}{*} \\
				& (0.2109) & (0.2326) & (0.2526) & (0.2554) & (0.2395) & (0.2596) \\
				Industry Growth & -2.1510{*}{*}{*} & -2.9141{*}{*}{*} & -2.2566{*}{*}{*} & -2.1827{*}{*}{*} & -2.9914{*}{*}{*} & -2.3394{*}{*}{*} \\
				& (0.6018) & (0.6975) & (0.6184) & (0.6076) & (0.7072) & (0.6263) \\
				\hline
				Fixed effects & Yes & Yes & Yes & Yes & Yes & Yes \\
				Time effects & Yes & Yes & Yes & Yes & Yes & Yes \\
				Industry linear trends & Yes & No & Yes & Yes & No & Yes \\
				Ind. dummy$\times$RGDP & No & Yes & Yes & No & Yes & Yes \\
				Observations & 82092 & 82092 & 82092 & 82092 & 82092 & 82092 \\
				$N$ & 3236 & 3236 & 3236 & 3236 & 3236 & 3236 \\
				$max(T_{i})$ & 44 & 44 & 44 & 44 & 44 & 44 \\
				$avg(T_{i})$ & 25.4 & 25.4 & 25.4 & 25.4 & 25.4 & 25.4 \\
				$med(T_{i})$ & 22 & 22 & 22 & 22 & 22 & 22 \\
				$min(T_{i})$ & 4 & 4 & 4 & 4 & 4 & 4 \\
				\hline
		\end{tabular}	}
\end{center}

%

\end{table}%

\newpage


\subsection{Robustness to the choice of dynamic specification
\label{Apx_Dyn_Specif}}

In the main the paper, we focus on the more general PanARDL(2) specification,
described in equation (\ref{ARDL_model_lsap}). For completeness, we also
provide estimation results for the partial adjustment model, a commonly used
specification in the empirical capital structure research (Graham and Leary
(2011)), and the PanARDL(1) model.

Table \ref{tab: thresholds_LSAPs_dyn} summarises the estimates of the
threshold parameters associated with $\pi(\gamma)$, the proportion of firms in
an industry with DA below the $\gamma$-th quantile, across the three choice of
dynamic specification, including the PanARDL(2) model for ease of reference.
Each panel focuses on a different choice of $f_{t}$.

In Subsection \ref{Appx:PAdj} we report the FE-TE estimates of the
coefficients of the partial adjustment model. We also report the long-run
effects of LSAPs and other regressors on firms' capital structure. Subsection
\ref{Appx:ARDL1} shows the net short-run and long-run effects of LSAPs on
firms' capital structure based on the PanARDL(1) specification.

\subsubsection{Quantile threshold parameter estimates}%

\begin{table}[H]
\caption{\textbf{Estimated quantile threshold parameters}}\label{tab: thresholds_LSAPs_dyn}
\vspace{-0.2cm}
\footnotesize
Estimates of the quantile threshold parameters from a grid search procedure
across both the partial adjustment model and the PanARDL specifications
described in equation (\ref{ARDL_model_lsap}). In Panel A and B, $f_{t}$
denotes linear time trends and real GDP growth, respectively. In Panel C,
$\mathbf{f}_{t}$ includes both linear time trends and real GDP growth. The
estimation sample consists of an unbalanced panel of $3,647$ U.S. publicly
traded non-financial firms observed at a quarterly frequency over the period
$2007$:Q1 - $2018$:Q3.
\normalsize

\begin{center}%
\begin{tabular}
[c]{cccc}
& \multicolumn{3}{c}{\textit{A: scaled linear trends}}\\\cline{2-4}%
\cline{3-4}\cline{4-4}
& Par. Adj. & PanARDL(1) & PanARDL(2)\\\hline
& \multicolumn{3}{c}{$\gamma_{pre}=\gamma_{post}=\gamma$}\\
$\hat{\gamma}$ & 0.56 & 0.76 & 0.76\\[0.1cm]
& \multicolumn{3}{c}{$\gamma_{pre}\neq\gamma_{post}$}\\
$\hat{\gamma}_{pre}$ & 0.56 & 0.56 & 0.56\\
$\hat{\gamma}_{post}$ & 0.77 & 0.77 & 0.77\\\hline
&  &  & \\
& \multicolumn{3}{c}{\textit{B: real GDP growth}}\\\cline{2-4}\cline{3-4}%
\cline{4-4}
& Part. Adj. & PanARDL(1) & PanARDL(2)\\\hline
& \multicolumn{3}{c}{$\gamma_{pre}=\gamma_{post}=\gamma$}\\
$\hat{\gamma}$ & 0.52 & 0.52 & 0.56\\[0.1cm]
& \multicolumn{3}{c}{$\gamma_{pre}\neq\gamma_{post}$}\\
$\hat{\gamma}_{pre}$ & 0.52 & 0.52 & 0.56\\
$\hat{\gamma}_{post}$ & 0.77 & 0.77 & 0.77\\\hline
&  &  & \\
& \multicolumn{3}{c}{\textit{C: lin. trends \& RGDP growth}}\\\cline{2-4}%
\cline{3-4}\cline{4-4}
& Part. Adj. & PanARDL(1) & PanARDL(2)\\\hline
& \multicolumn{3}{c}{$\gamma_{pre}=\gamma_{post}=\gamma$}\\
$\hat{\gamma}$ & 0.56 & 0.69 & 0.56\\[0.1cm]
& \multicolumn{3}{c}{$\gamma_{pre}\neq\gamma_{post}$}\\
$\hat{\gamma}_{pre}$ & 0.56 & 0.56 & 0.56\\
$\hat{\gamma}_{post}$ & 0.77 & 0.77 & 0.77\\\hline
\end{tabular}

\end{center}

%

\end{table}%


\subsubsection{FE-TE estimates based on the partial adjustment model
\label{Appx:PAdj}}%

\begin{table}[H]
\caption{\textbf{FE--TE estimates of the effects of LSAPs on non-financial firm's debt to asset ratios based on the partial adjustment model}}\label{tab:PAdj_bench}
\vspace{-0.2cm}
\footnotesize
Estimates of the coefficients of the partial adjustment model based on
equation (\ref{ARDL_model_lsap}). The dependent variable is debt to asset
ratio (DA). $q_{t}$ is the (scaled) amount of U.S. Treasuries and agency MBS
purchased by the Fed; $\pi_{s,t}(\hat{\gamma})$ denotes the proportion of
firms in an industry with DA below the $\hat{\gamma}$-th quantile. The first
three columns report results for the single-threshold panel regression model,
where $\gamma_{pre}=\gamma_{post}$. The last three columns report results for
the two-threshold panel regression, where $\gamma_{pre}\neq\gamma_{post}$. The
estimated quantile threshold parameters are shown in Table
\ref{tab: thresholds_LSAPs_dyn}. All regressions include both firm-specific
effects and time effects. Columns (1) and (4) include industry-specific linear
time trends, columns (2) and (5) include the interaction of industry dummies
and real GDP growth, while columns (3) and (6) include both. The sample
consists of an unbalanced panel of $3,647$ U.S. publicly traded non-financial
firms observed at a quarterly frequency over the period $2007$:Q1 - $2018$:Q3.
Robust standard errors in parentheses (*** $p<0.01$, ** $p<0.05$, * $p<0.1$).

\begin{center}%
\begin{tabular}
[c]{l|ccc|ccc}%
\multicolumn{1}{c}{} & \multicolumn{6}{c}{Dependent variable: debt to assets
($DA_{t}$)}\\\cline{2-7}\cline{3-7}\cline{4-7}\cline{5-7}\cline{6-7}%
\multicolumn{1}{c}{} & \multicolumn{3}{c|}{$\gamma_{pre}=\gamma_{post}=\gamma
$} & \multicolumn{3}{c}{$\gamma_{pre}\neq\gamma_{post}$}\\
\multicolumn{1}{c}{} & (1) & (2) & (3) & (4) & (5) & (6)\\\hline
$\pi_{s,t-1}(\hat{\gamma}_{pre})$ & 0.0447{*}{*}{*} & 0.0406{*}{*}{*} &
0.0450{*}{*}{*} & 0.0460{*}{*}{*} & 0.0418{*}{*}{*} & 0.0463{*}{*}{*}\\
& (0.0041) & (0.0039) & (0.0042) & (0.0040) & (0.0037) & (0.0040)\\
$q_{t}\times\pi_{s,t-1}(\hat{\gamma}_{post})$ & 0.0033{*}{*}{*} & 0.0023{*}{*}
& 0.0032{*}{*}{*} & 0.0077{*}{*}{*} & 0.0062{*}{*}{*} & 0.0075{*}{*}{*}\\
& (0.0012) & (0.0011) & (0.0012) & (0.0014) & (0.0014) & (0.0015)\\
$DA_{t-1}$ & 0.8264{*}{*}{*} & 0.8287{*}{*}{*} & 0.8265{*}{*}{*} &
0.8266{*}{*}{*} & 0.8289{*}{*}{*} & 0.8267{*}{*}{*}\\
& (0.0053) & (0.0053) & (0.0053) & (0.0053) & (0.0053) & (0.0053)\\
$\left(  Cash/A\right)  _{t}$ & -0.0496{*}{*}{*} & -0.0502{*}{*}{*} &
-0.0496{*}{*}{*} & -0.0496{*}{*}{*} & -0.0503{*}{*}{*} & -0.0496{*}{*}{*}\\
& (0.0034) & (0.0034) & (0.0034) & (0.0034) & (0.0034) & (0.0034)\\
$\left(  PPE/A\right)  _{t}$ & 0.0249{*}{*}{*} & 0.0240{*}{*}{*} &
0.0250{*}{*}{*} & 0.0250{*}{*}{*} & 0.0239{*}{*}{*} & 0.0250{*}{*}{*}\\
& (0.0053) & (0.0052) & (0.0053) & (0.0053) & (0.0052) & (0.0053)\\
$Size_{t}$ & 0.0051{*}{*}{*} & 0.0054{*}{*}{*} & 0.0051{*}{*}{*} &
0.0051{*}{*}{*} & 0.0054{*}{*}{*} & 0.0051{*}{*}{*}\\
& (0.0008) & (0.0008) & (0.0008) & (0.0008) & (0.0008) & (0.0008)\\
$Industry\;leverage_{t}$ & 0.1391{*}{*}{*} & 0.1175{*}{*}{*} & 0.1370{*}{*}{*}
& 0.1414{*}{*}{*} & 0.1196{*}{*}{*} & 0.1394{*}{*}{*}\\
& (0.0075) & (0.0067) & (0.0076) & (0.0075) & (0.0068) & (0.0077)\\
$Industry\;growth_{t}$ & -0.0483{*}{*}{*} & -0.0620{*}{*}{*} & -0.0427{*}%
{*}{*} & -0.0499{*}{*}{*} & -0.0637{*}{*}{*} & -0.0442{*}{*}{*}\\
& (0.0131) & (0.0131) & (0.0134) & (0.0131) & (0.0131) & (0.0134)\\\hline
Fixed effects & Yes & Yes & Yes & Yes & Yes & Yes\\
Time effects & Yes & Yes & Yes & Yes & Yes & Yes\\
Industry linear trends & Yes & No & Yes & Yes & No & Yes\\
Ind. dummy$\times$RGDP & No & Yes & Yes & No & Yes & Yes\\
Observations & 84548 & 84548 & 84548 & 84548 & 84548 & 84548\\
$N$ & 3647 & 3647 & 3647 & 3647 & 3647 & 3647\\
$max(T_{i})$ & 44 & 44 & 44 & 44 & 44 & 44\\
$avg(T_{i})$ & 23.2 & 23.2 & 23.2 & 23.2 & 23.2 & 23.2\\
$med(T_{i})$ & 19 & 19 & 19 & 19 & 19 & 19\\
$min(T_{i})$ & 2 & 2 & 2 & 2 & 2 & 2\\\hline
\end{tabular}

\end{center}

%

\end{table}%
%

\begin{table}[H]
\caption{\textbf{FE--TE estimates of the long-run effects of LSAPs on debt to asset ratios of non-financial firms based on the partial adjustment model}}\label{tab:LR_PAdj}
\vspace{-0.2cm}
\footnotesize
Estimates of long-run effects of LSAPs, defined in equation
(\ref{LR_definition}), on firms' debt to asset ratios (DA) as well as the
long-run effects of both firm- and industry-specific variables on DA, for the
partial adjustment model from equation (\ref{ARDL_model_lsap}). The first
three columns report results for the single-threshold panel regression model,
where $\gamma_{pre}=\gamma_{post}$. The last three columns report results for
the two-threshold panel regression, where $\gamma_{pre}\neq\gamma_{post}$. The
estimated quantile threshold parameters are shown in Table
\ref{tab: thresholds_LSAPs_dyn}. All regressions include both firm-specific
effects and time effects. Columns (1) and (4) include industry-specific linear
time trends, columns (2) and (5) include the interaction of industry dummies
and real GDP growth, while columns (3) and (6) include both. $LSAP$ is the
(scaled) amount of U.S. Treasuries and agency MBS purchased by the Fed;
$\pi_{-1}(\gamma)$ denotes the one-quarter lagged proportion of firms in an
industry with DA below the $\gamma$-th quantile. The sample consists of an
unbalanced panel of $3,647$ U.S. publicly traded non-financial firms observed
at a quarterly frequency over the period $2007$:Q1 - $2018$:Q3. Robust
standard errors (in parentheses) are computed using the delta method (***
$p<0.01$, ** $p<0.05$, * $p<0.1$).
\normalsize

\begin{center}
\scalebox{0.85}{
			\begin{tabular}{l|ccc|ccc}
				\multicolumn{1}{c}{} & \multicolumn{6}{c}{Dependent variable: debt to assets (DA)} \\
				\cline{2-7} \cline{3-7} \cline{4-7} \cline{5-7} \cline{6-7} \cline{7-7}
				\multicolumn{1}{c}{} & \multicolumn{3}{c|}{$\gamma_{pre}=\gamma_{post}=\gamma$} & \multicolumn{3}{c}{$\gamma_{pre}\neq\gamma_{post}$} \\
				\multicolumn{1}{c}{} & (1) & (2) & (3) & (4) & (5) & (6) \\
				\hline
				$\pi_{-1}(\hat{\gamma}_{pre})$ & 0.2576{*}{*}{*} & 0.2373{*}{*}{*} & 0.2594{*}{*}{*} & 0.2654{*}{*}{*} & 0.2443{*}{*}{*} & 0.2671{*}{*}{*} \\
				& (0.0250) & (0.0235) & (0.0253) & (0.0242) & (0.0227) & (0.0245) \\
				$LSAP\times\pi_{-1}(\hat{\gamma}_{post})$ & 0.0192{*}{*}{*} & 0.0136{*}{*} & 0.0186{*}{*}{*} & 0.0441{*}{*}{*} & 0.0363{*}{*}{*} & 0.0433{*}{*}{*} \\
				& (0.0067) & (0.0064) & (0.0067) & (0.0085) & (0.0082) & (0.0085) \\
				Cash to assets & -0.2858{*}{*}{*} & -0.2934{*}{*}{*} & -0.2860{*}{*}{*} & -0.2858{*}{*}{*} & -0.2938{*}{*}{*} & -0.2860{*}{*}{*} \\
				& (0.0189) & (0.0189) & (0.0189) & (0.0189) & (0.0189) & (0.0189) \\
				PPE to assets & 0.1433{*}{*}{*} & 0.1400{*}{*}{*} & 0.1439{*}{*}{*} & 0.1439{*}{*}{*} & 0.1400{*}{*}{*} & 0.1445{*}{*}{*} \\
				& (0.0306) & (0.0302) & (0.0306) & (0.0306) & (0.0302) & (0.0306) \\
				Size & 0.0295{*}{*}{*} & 0.0313{*}{*}{*} & 0.0295{*}{*}{*} & 0.0296{*}{*}{*} & 0.0314{*}{*}{*} & 0.0296{*}{*}{*} \\
				& (0.0046) & (0.0046) & (0.0047) & (0.0046) & (0.0046) & (0.0047) \\
				Industry Leverage & 0.8013{*}{*}{*} & 0.6860{*}{*}{*} & 0.7895{*}{*}{*} & 0.8157{*}{*}{*} & 0.6993{*}{*}{*} & 0.8043{*}{*}{*} \\
				& (0.0452) & (0.0403) & (0.0459) & (0.0455) & (0.0406) & (0.0462) \\
				Industry Growth & -0.2785{*}{*}{*} & -0.3620{*}{*}{*} & -0.2460{*}{*}{*} & -0.2880{*}{*}{*} & -0.3724{*}{*}{*} & -0.2550{*}{*}{*} \\
				& (0.0761) & (0.0773) & (0.0775) & (0.0762) & (0.0774) & (0.0776) \\
				\hline
				Fixed effects & Yes & Yes & Yes & Yes & Yes & Yes \\
				Time effects & Yes & Yes & Yes & Yes & Yes & Yes \\
				Industry linear trends & Yes & No & Yes & Yes & No & Yes \\
				Ind. dummy$\times$RGDP & No & Yes & Yes & No & Yes & Yes \\
				Observations & 84548 & 84548 & 84548 & 84548 & 84548 & 84548 \\
				$N$ & 3647 & 3647 & 3647 & 3647 & 3647 & 3647 \\
				$max(T_{i})$ & 44 & 44 & 44 & 44 & 44 & 44 \\
				$avg(T_{i})$ & 23.2 & 23.2 & 23.2 & 23.2 & 23.2 & 23.2 \\
				$med(T_{i})$ & 19 & 19 & 19 & 19 & 19 & 19 \\
				$min(T_{i})$ & 2 & 2 & 2 & 2 & 2 & 2 \\
				\hline
		\end{tabular}	}
\end{center}

%

\end{table}%


\subsubsection{FE-TE estimates based on the PanARDL(1) model
\label{Appx:ARDL1}}%

\begin{table}[H]
\caption{\textbf{FE--TE estimates of the net short-run effects of LSAPs on debt to asset ratios of non-financial firms based on the PanARDL(1) model}}\label{tab:SR_ARDL1}
\vspace{-0.2cm}
\footnotesize
Estimates of net short-run effects of LSAPs on firms' debt to asset ratios
(DA) as well as the effects of both firm- and industry-specific variables on
DA, for the PanARDL(1) model from equation (\ref{ARDL_model_lsap}). Net
short-run effects are defined as the sum of the estimated coefficients of
current and lagged values of the regressor under consideration. The first
three columns report results for the single-threshold panel regression model,
where $\gamma_{pre}=\gamma_{post}$. The last three columns report results for
the two-threshold panel regression, where $\gamma_{pre}\neq\gamma_{post}$. The
estimated quantile threshold parameters are shown in Table
\ref{tab: thresholds_LSAPs_dyn}. All regressions include both firm-specific
effects and time effects. Columns (1) and (4) include industry-specific linear
time trends, columns (2) and (5) include the interaction of industry dummies
and real GDP growth, while columns (3) and (6) include both. $LSAP$ is the
(scaled) amount of U.S. Treasuries and agency MBS purchased by the Fed;
$\pi_{-1}(\gamma)$ denotes the one-quarter lagged proportion of firms in an
industry with DA below the $\gamma$-th quantile. The sample consists of an
unbalanced panel of $3,647$ U.S. publicly traded non-financial firms observed
at a quarterly frequency over the period $2007$:Q1 - $2018$:Q3. Robust
standard errors (in parentheses) are computed using the delta method (***
$p<0.01$, ** $p<0.05$, * $p<0.1$).
\normalsize

\begin{center}
\scalebox{0.85}{
			\begin{tabular}{l|ccc|ccc}
				\multicolumn{1}{c}{} & \multicolumn{6}{c}{Dependent variable: debt to assets (DA)} \\
				\cline{2-7} \cline{3-7} \cline{4-7} \cline{5-7} \cline{6-7} \cline{7-7}
				\multicolumn{1}{c}{} & \multicolumn{3}{c|}{$\gamma_{pre}=\gamma_{post}=\gamma$} & \multicolumn{3}{c}{$\gamma_{pre}\neq\gamma_{post}$} \\
				\multicolumn{1}{c}{} & (1) & (2) & (3) & (4) & (5) & (6) \\
				\hline
				$\pi_{-1}(\hat{\gamma}_{pre})$ & 0.0136{*}{*}{*} & 0.0093{*}{*} & 0.0162{*}{*}{*} & 0.0148{*}{*}{*} & 0.0118{*}{*}{*} & 0.0153{*}{*}{*} \\
				& (0.0045) & (0.0043) & (0.0045) & (0.0045) & (0.0041) & (0.0046) \\
				$LSAP\times\pi_{-1}(\hat{\gamma}_{post})$ & 0.0059{*}{*}{*} & 0.0030{*}{*}{*} & 0.0040{*}{*}{*} & 0.0074{*}{*}{*} & 0.0055{*}{*}{*} & 0.0067{*}{*}{*} \\
				& (0.0016) & (0.0012) & (0.0014) & (0.0015) & (0.0015) & (0.0016) \\
				Lagged DA & 0.8337{*}{*}{*} & 0.8357{*}{*}{*} & 0.8337{*}{*}{*} & 0.8337{*}{*}{*} & 0.8359{*}{*}{*} & 0.8337{*}{*}{*} \\
				& (0.0052) & (0.0052) & (0.0052) & (0.0052) & (0.0052) & (0.0052) \\
				Cash to assets & -0.0380{*}{*}{*} & -0.0384{*}{*}{*} & -0.0381{*}{*}{*} & -0.0380{*}{*}{*} & -0.0384{*}{*}{*} & -0.0381{*}{*}{*} \\
				& (0.0030) & (0.0030) & (0.0030) & (0.0030) & (0.0030) & (0.0030) \\
				PPE to assets & 0.0236{*}{*}{*} & 0.0227{*}{*}{*} & 0.0236{*}{*}{*} & 0.0237{*}{*}{*} & 0.0227{*}{*}{*} & 0.0237{*}{*}{*} \\
				& (0.0047) & (0.0046) & (0.0047) & (0.0047) & (0.0046) & (0.0047) \\
				Size & 0.0030{*}{*}{*} & 0.0033{*}{*}{*} & 0.0031{*}{*}{*} & 0.0030{*}{*}{*} & 0.0033{*}{*}{*} & 0.0031{*}{*}{*} \\
				& (0.0008) & (0.0007) & (0.0008) & (0.0008) & (0.0007) & (0.0008) \\
				Industry Leverage & 0.0631{*}{*}{*} & 0.0554{*}{*}{*} & 0.0633{*}{*}{*} & 0.0715{*}{*}{*} & 0.0588{*}{*}{*} & 0.0694{*}{*}{*} \\
				& (0.0069) & (0.0072) & (0.0074) & (0.0082) & (0.0073) & (0.0084) \\
				Industry Growth & -0.1024{*}{*}{*} & -0.1245{*}{*}{*} & -0.0981{*}{*}{*} & -0.1056{*}{*}{*} & -0.1266{*}{*}{*} & -0.1041{*}{*}{*} \\
				& (0.0176) & (0.0176) & (0.0182) & (0.0176) & (0.0177) & (0.0183) \\
				\hline
				Fixed effects & Yes & Yes & Yes & Yes & Yes & Yes \\
				Time effects & Yes & Yes & Yes & Yes & Yes & Yes \\
				Industry linear trends & Yes & No & Yes & Yes & No & Yes \\
				Ind. dummy$\times$RGDP & No & Yes & Yes & No & Yes & Yes \\
				Observations & 84548 & 84548 & 84548 & 84548 & 84548 & 84548 \\
				$N$ & 3647 & 3647 & 3647 & 3647 & 3647 & 3647 \\
				$max(T_{i})$ & 44 & 44 & 44 & 44 & 44 & 44 \\
				$avg(T_{i})$ & 23.2 & 23.2 & 23.2 & 23.2 & 23.2 & 23.2 \\
				$med(T_{i})$ & 19 & 19 & 19 & 19 & 19 & 19 \\
				$min(T_{i})$ & 2 & 2 & 2 & 2 & 2 & 2 \\
				\hline
		\end{tabular}	}
\end{center}

%

\end{table}%
%

\begin{table}[H]
\caption{\textbf{FE--TE estimates of the long-run effects of LSAPs on debt to asset ratios of non-financial firms based on the PanARDL(1) model}}\label{tab:LR_ARDL1}
\vspace{-0.2cm}
\footnotesize
Estimates of long-run effects of LSAPs, defined in equation
(\ref{LR_definition}), on firms' debt to asset ratios (DA) as well as the
effects of both firm- and industry-specific variables on DA, for the
PanARDL(1) model from equation (\ref{ARDL_model_lsap}). The first three
columns report results for the single-threshold panel regression model, where
$\gamma_{pre}=\gamma_{post}$. The last three columns report results for the
two-threshold panel regression, where $\gamma_{pre}\neq\gamma_{post}$. The
estimated quantile threshold parameters are shown in Table
\ref{tab: thresholds_LSAPs_dyn}. All regressions include both firm-specific
effects and time effects. Columns (1) and (4) include industry-specific linear
time trends, columns (2) and (5) include the interaction of industry dummies
and real GDP growth, while columns (3) and (6) include both. $LSAP$ is the
(scaled) amount of U.S. Treasuries and agency MBS purchased by the Fed;
$\pi_{-1}(\gamma)$ denotes the one-quarter lagged proportion of firms in an
industry with DA below the $\gamma$-th quantile. The sample consists of an
unbalanced panel of $3,647$ U.S. publicly traded non-financial firms observed
at a quarterly frequency over the period $2007$:Q1 - $2018$:Q3. Robust
standard errors (in parentheses) are computed using the delta method (***
$p<0.01$, ** $p<0.05$, * $p<0.1$).
\normalsize

\begin{center}
\scalebox{0.85}{
			\begin{tabular}{l|ccc|ccc}
				\multicolumn{1}{c}{} & \multicolumn{6}{c}{Dependent variable: debt to assets (DA)} \\
				\cline{2-7} \cline{3-7} \cline{4-7} \cline{5-7} \cline{6-7} \cline{7-7}
				\multicolumn{1}{c}{} & \multicolumn{3}{c|}{$\gamma_{pre}=\gamma_{post}=\gamma$} & \multicolumn{3}{c}{$\gamma_{pre}\neq\gamma_{post}$} \\
				\multicolumn{1}{c}{} & (1) & (2) & (3) & (4) & (5) & (6) \\
				\hline
				$\pi_{-1}(\hat{\gamma}_{pre})$ & 0.0816{*}{*}{*} & 0.0566{*}{*} & 0.0974{*}{*}{*} & 0.0890{*}{*}{*} & 0.0720{*}{*}{*} & 0.0919{*}{*}{*} \\
				& (0.0271) & (0.0265) & (0.0270) & (0.0276) & (0.0255) & (0.0279) \\
				$LSAP\times\pi_{-1}(\hat{\gamma}_{post})$ & 0.0353{*}{*}{*} & 0.0185{*}{*}{*} & 0.0243{*}{*}{*} & 0.0446{*}{*}{*} & 0.0334{*}{*}{*} & 0.0405{*}{*}{*} \\
				& (0.0099) & (0.0071) & (0.0085) & (0.0093) & (0.0091) & (0.0094) \\
				Cash to assets & -0.2287{*}{*}{*} & -0.2338{*}{*}{*} & -0.2293{*}{*}{*} & -0.2284{*}{*}{*} & -0.2341{*}{*}{*} & -0.2288{*}{*}{*} \\
				& (0.0175) & (0.0175) & (0.0176) & (0.0175) & (0.0175) & (0.0175) \\
				PPE to assets & 0.1421{*}{*}{*} & 0.1385{*}{*}{*} & 0.1421{*}{*}{*} & 0.1424{*}{*}{*} & 0.1384{*}{*}{*} & 0.1423{*}{*}{*} \\
				& (0.0283) & (0.0280) & (0.0283) & (0.0283) & (0.0281) & (0.0283) \\
				Size & 0.0183{*}{*}{*} & 0.0200{*}{*}{*} & 0.0184{*}{*}{*} & 0.0183{*}{*}{*} & 0.0201{*}{*}{*} & 0.0184{*}{*}{*} \\
				& (0.0045) & (0.0044) & (0.0045) & (0.0045) & (0.0044) & (0.0045) \\
				Industry Leverage & 0.3792{*}{*}{*} & 0.3373{*}{*}{*} & 0.3806{*}{*}{*} & 0.4297{*}{*}{*} & 0.3580{*}{*}{*} & 0.4171{*}{*}{*} \\
				& (0.0414) & (0.0439) & (0.0445) & (0.0497) & (0.0442) & (0.0505) \\
				Industry Growth & -0.6156{*}{*}{*} & -0.7581{*}{*}{*} & -0.5897{*}{*}{*} & -0.6349{*}{*}{*} & -0.7711{*}{*}{*} & -0.6261{*}{*}{*} \\
				& (0.1074) & (0.1100) & (0.1112) & (0.1081) & (0.1105) & (0.1122) \\
				\hline
				Fixed effects & Yes & Yes & Yes & Yes & Yes & Yes \\
				Time effects & Yes & Yes & Yes & Yes & Yes & Yes \\
				Industry linear trends & Yes & No & Yes & Yes & No & Yes \\
				Ind. dummy$\times$RGDP & No & Yes & Yes & No & Yes & Yes \\
				Observations & 84548 & 84548 & 84548 & 84548 & 84548 & 84548 \\
				$N$ & 3647 & 3647 & 3647 & 3647 & 3647 & 3647 \\
				$max(T_{i})$ & 44 & 44 & 44 & 44 & 44 & 44 \\
				$avg(T_{i})$ & 23.2 & 23.2 & 23.2 & 23.2 & 23.2 & 23.2 \\
				$med(T_{i})$ & 19 & 19 & 19 & 19 & 19 & 19 \\
				$min(T_{i})$ & 2 & 2 & 2 & 2 & 2 & 2 \\
				\hline
		\end{tabular}	}
\end{center}

%

\end{table}%

\newpage

\section{Separating the effects of large-scale MBS and Treasury purchases
\label{sec:TY_vs_MBS}}

This section reports estimation results when separating the effects of MBS
from Treasury purchases. Thus, the panel regression model described in
equation (\ref{ARDL_model_lsap}) now contains two separate quantitative
measures of LSAPs interacted with one-quarter lags of $\pi_{s,t}(\gamma)$, our
industry-specific measure of debt capacity.

In Subsection \ref{Appx_MBS_TY_thresh}, we report the estimated threshold
parameters for the PanARDL(2) specification, distinguishing between the case
of single \textit{versus} the two-threshold model. The corresponding estimated
net short-run and long-run effects of both MBS and Treasury purchases are
shown in Subsection \ref{Appx:MBS_TY_SRLR}. In Subsection
\ref{Appx:MBS_TY_SmallT}, we examine the extent to which our estimation
results hold after correcting for the small-T bias.

\subsection{Quantile threshold parameter estimates\label{Appx_MBS_TY_thresh}}%

\begin{table}[H]
\caption{\textbf{Estimated quantile threshold parameters}}\label{tab: thresholds_TY_MBS}
\vspace{-0.2cm}
\footnotesize
Estimates of the quantile threshold parameters from a grid search procedure
for the PanARDL(2) model described in equation (\ref{ARDL_model_lsap}),
separating the effects of large-scale MBS and Treasury purchases. Panel A
shows the estimated threshold parameters for the single-threshold panel
regression model, where $\gamma_{pre}=\gamma_{post}$. Panel B displays results
for the two-threshold model, where $\gamma_{pre}\neq\gamma_{post}$. In column
(1) and (2), we use linear time trends or real GDP growth as a proxy for
$f_{t}$, respectively. Column (3) reports results when including both linear
trends and real GDP growth at the same time. The estimation sample consists of
an unbalanced panel of $3,647$ U.S. publicly traded non-financial firms
observed at a quarterly frequency over the period $2007$:Q1 - $2018$:Q3.
\normalsize

\begin{center}%
\begin{tabular}
[c]{cccc}\cline{2-4}\cline{3-4}\cline{4-4}
& (1) & (2) & (3)\\\hline
\textit{Panel A:} & \multicolumn{3}{c}{$\gamma_{pre}=\gamma_{post}=\gamma$}\\
$\hat{\gamma}$ & 0.76 & 0.56 & 0.56\\[0.2cm]%
\textit{Panel B:} & \multicolumn{3}{c}{$\gamma_{pre}\neq\gamma_{post}$}\\
$\hat{\gamma}_{pre}$ & 0.56 & 0.56 & 0.56\\
$\hat{\gamma}_{post}$ & 0.77 & 0.78 & 0.77\\\hline
linear trends & Yes & No & Yes\\
RGDP growth & No & Yes & Yes\\\hline
\end{tabular}

\end{center}

%

\end{table}%


\subsection{Short- and long-run effects of large-scale MBS and Treasury
purchases \label{Appx:MBS_TY_SRLR}}%

\begin{table}[H]
\caption{\textbf{FE--TE estimates of the net short-run and long-run effects of of large-scale MBS and Treasury purchases on debt to asset ratios of non-financial firms}}\label{tab:SRLR_MBS_TY}
\vspace{-0.2cm}
\footnotesize
Panel A (Panel B) reports the net short-run (long-run) effects of MBS and
Treasury purchases on firms' debt to asset ratios (DA), for the PanARDL(2)
model described in equation (\ref{ARDL_model_lsap}). Net short-run effects are
defined as the sum of the estimated coefficients of current and lagged values
of the regressor under consideration. The long-run effects are defined in
equation (\ref{LR_definition}). The first three columns report results for the
single-threshold panel regression model, where $\gamma_{pre}=\gamma_{post}$.
The last three columns report results for the two-threshold panel regression,
where $\gamma_{pre}\neq\gamma_{post}$. The estimated quantile threshold
parameters are shown in Table \ref{tab: thresholds_TY_MBS}. All regressions
include both firm-specific effects and time effects. Columns (1) and (4)
include industry-specific linear time trends, columns (2) and (5) include the
interaction of industry dummies and real GDP growth, while columns (3) and (6)
include both. $ty$ and $mbs$ denote the (scaled) amount of U.S. Treasuries and
agency MBS purchased by the Fed, respectively; $\pi_{-1}(\gamma)$ denotes the
one-quarter lagged proportion of firms in an industry with DA below the
$\gamma$-th quantile. The sample consists of an unbalanced panel of $3,647$
U.S. publicly traded non-financial firms observed at a quarterly frequency
over the period $2007$:Q1 - $2018$:Q3. Robust standard errors (in parentheses)
are computed using the delta method (***$p<0.01$, ** $p<0.05$, * $p<0.1$).
\normalsize

\begin{center}
\scalebox{0.85}{
			\begin{tabular}{l|ccc|ccc}
				\multicolumn{1}{c}{} & \multicolumn{6}{c}{Dependent variable: debt to assets (DA)} \\
				\cline{2-7} \cline{3-7} \cline{4-7} \cline{5-7} \cline{6-7} \cline{7-7}
				\multicolumn{1}{c}{} & \multicolumn{3}{c|}{$\gamma_{pre}=\gamma_{post}=\gamma$} & \multicolumn{3}{c}{$\gamma_{pre}\neq\gamma_{post}$} \\
				\multicolumn{1}{c}{} & (1) & (2) & (3) & (4) & (5) & (6) \\
				\hline
				\multicolumn{1}{c}{} & \multicolumn{6}{c}{\textit{Panel A: Short-run effects}} \\
				\hline
				$\pi_{-1}(\hat{\gamma}_{pre})$ & 0.0159{*}{*}{*} & 0.0108{*}{*} & 0.0174{*}{*}{*} & 0.0188{*}{*}{*} & 0.0124{*}{*}{*} & 0.0196{*}{*}{*} \\
				& (0.0050) & (0.0048) & (0.0055) & (0.0050) & (0.0046) & (0.0051) \\
				$ty\times\pi_{-1}(\hat{\gamma}_{post})$ & 0.0059{*} & 0.0018 & 0.0026 & 0.0078{*}{*} & 0.0053{*} & 0.0067{*}{*} \\
				& (0.0033) & (0.0025) & (0.0026) & (0.0031) & (0.0031) & (0.0032) \\
				$mbs\times\pi_{-1}(\hat{\gamma}_{post})$ & 0.0070{*}{*}{*} & 0.0041{*}{*} & 0.0045{*}{*}{*} & 0.0092{*}{*}{*} & 0.0060{*}{*}{*} & 0.0080{*}{*}{*} \\
				& (0.0022) & (0.0017) & (0.0017) & (0.0021) & (0.0021) & (0.0022) \\[0.1cm]
				\hline
				\multicolumn{1}{c}{} & \multicolumn{6}{c}{\textit{Panel B: Long-run effects}} \\
				\hline
				$\pi_{-1}(\hat{\gamma}_{pre})$ & 0.0988{*}{*}{*} & 0.0676{*}{*} & 0.1077{*}{*}{*} & 0.1166{*}{*}{*} & 0.0780{*}{*}{*} & 0.1216{*}{*}{*} \\
				& (0.0312) & (0.0302) & (0.0341) & (0.0317) & (0.0290) & (0.0322) \\
				$ty\times\pi_{-1}(\hat{\gamma}_{post})$ & 0.0368{*} & 0.0112 & 0.0162 & 0.0482{*}{*} & 0.0331{*} & 0.0418{*}{*} \\
				& (0.0202) & (0.0157) & (0.0164) & (0.0191) & (0.0193) & (0.0198) \\
				$mbs\times\pi_{-1}(\hat{\gamma}_{post})$ & 0.0433{*}{*}{*} & 0.0255{*}{*} & 0.0280{*}{*}{*} & 0.0569{*}{*}{*} & 0.0379{*}{*}{*} & 0.0493{*}{*}{*} \\
				& (0.0138) & (0.0105) & (0.0107) & (0.0134) & (0.0134) & (0.0137) \\[0.1cm]
				\hline
				Fixed effects & Yes & Yes & Yes & Yes & Yes & Yes \\
				Time effects & Yes & Yes & Yes & Yes & Yes & Yes \\
				Industry linear trends & Yes & No & Yes & Yes & No & Yes \\
				Ind. dummy$\times$RGDP & No & Yes & Yes & No & Yes & Yes \\
				Observations & 84548 & 84548 & 84548 & 84548 & 84548 & 84548 \\
				$N$ & 3647 & 3647 & 3647 & 3647 & 3647 & 3647 \\
				$max(T_{i})$ & 44 & 44 & 44 & 44 & 44 & 44 \\
				$avg(T_{i})$ & 23.2 & 23.2 & 23.2 & 23.2 & 23.2 & 23.2 \\
				$med(T_{i})$ & 19 & 19 & 19 & 19 & 19 & 19 \\
				$min(T_{i})$ & 2 & 2 & 2 & 2 & 2 & 2 \\
				\hline
		\end{tabular}	}
\end{center}

%

\end{table}%


\subsection{Robustness of the results to small-T bias
\label{Appx:MBS_TY_SmallT}}%

\begin{table}[H]
\caption{\textbf{FE--TE estimates of the net short-run and long-run effects of of large-scale MBS and Treasury purchases on debt to asset ratios of non-financial firms with at least $8$ observations}}\label{tab:SRLR_MBS_TY_mint8}
\vspace{-0.2cm}
\footnotesize
Panel A (Panel B) reports the net short-run (long-run) effects of MBS and
Treasury purchases on firms' debt to asset ratios (DA), for the PanARDL(2)
model described in equation (\ref{ARDL_model_lsap}). Net short-run effects are
defined as the sum of the estimated coefficients of current and lagged values
of the regressor under consideration. The long-run effects are defined in
equation (\ref{LR_definition}). The first three columns report results for the
single-threshold panel regression model, where $\gamma_{pre}=\gamma_{post}$.
The last three columns report results for the two-threshold panel regression,
where $\gamma_{pre}\neq\gamma_{post}$. The estimated quantile threshold
parameters are shown in Table \ref{tab: thresholds_TY_MBS}. All regressions
include both firm-specific effects and time effects. Columns (1) and (4)
include industry-specific linear time trends, columns (2) and (5) include the
interaction of industry dummies and real GDP growth, while columns (3) and (6)
include both. $ty$ and $mbs$ denote the (scaled) amount of U.S. Treasuries and
agency MBS purchased by the Fed, respectively; $\pi_{-1}(\gamma)$ denotes the
one-quarter lagged proportion of firms in an industry with DA below the
$\gamma$-th quantile. he sample only includes firms with at least $8$ time
observations, resulting in an unbalanced panel of $3,236$ U.S. publicly traded
non-financial firms observed at a quarterly frequency over the period
$2007$:Q1 - $2018$:Q3. Robust standard errors (in parentheses) are computed
using the delta method (***$p<0.01$, ** $p<0.05$, * $p<0.1$).
\normalsize

\begin{center}
\scalebox{0.85}{
			\begin{tabular}{l|ccc|ccc}
				\multicolumn{1}{c}{} & \multicolumn{6}{c}{Dependent variable: debt to assets (DA)} \\
				\cline{2-7} \cline{3-7} \cline{4-7} \cline{5-7} \cline{6-7} \cline{7-7}
				\multicolumn{1}{c}{} & \multicolumn{3}{c|}{$\gamma_{pre}=\gamma_{post}=\gamma$} & \multicolumn{3}{c}{$\gamma_{pre}\neq\gamma_{post}$} \\
				\multicolumn{1}{c}{} & (1) & (2) & (3) & (4) & (5) & (6) \\
				\hline
				\multicolumn{1}{c}{} & \multicolumn{6}{c}{\textit{Panel A: Short-run effects}} \\
				\hline
				$\pi_{-1}(\hat{\gamma}_{pre})$ & 0.0159{*}{*}{*} & 0.0107{*}{*} & 0.0172{*}{*}{*} & 0.0183{*}{*}{*} & 0.0123{*}{*}{*} & 0.0193{*}{*}{*} \\
				& (0.0050) & (0.0048) & (0.0054) & (0.0050) & (0.0046) & (0.0051) \\
				$ty\times\pi_{-1}(\hat{\gamma}_{post})$ & 0.0059{*} & 0.0016 & 0.0023 & 0.0078{*}{*} & 0.0051{*} & 0.0064{*}{*} \\
				& (0.0033) & (0.0025) & (0.0026) & (0.0031) & (0.0031) & (0.0032) \\
				$mbs\times\pi_{-1}(\hat{\gamma}_{post})$ & 0.0071{*}{*}{*} & 0.0041{*}{*} & 0.0045{*}{*}{*} & 0.0093{*}{*}{*} & 0.0061{*}{*}{*} & 0.0080{*}{*}{*} \\
				& (0.0022) & (0.0017) & (0.0017) & (0.0021) & (0.0021) & (0.0022) \\ [0.1cm]
				\hline
				\multicolumn{1}{c}{} & \multicolumn{6}{c}{\textit{Panel B: Long-run effects}} \\
				\hline
				$\pi_{-1}(\hat{\gamma}_{pre})$ & 0.1001{*}{*}{*} & 0.0686{*}{*} & 0.1083{*}{*}{*} & 0.1155{*}{*}{*} & 0.0788{*}{*}{*} & 0.1217{*}{*}{*} \\
				& (0.0317) & (0.0307) & (0.0347) & (0.0322) & (0.0295) & (0.0327) \\
				$ty\times\pi_{-1}(\hat{\gamma}_{post})$ & 0.0371{*} & 0.0100 & 0.0145 & 0.0490{*}{*} & 0.0324{*} & 0.0406{*}{*} \\
				& (0.0206) & (0.0160) & (0.0167) & (0.0194) & (0.0196) & (0.0202) \\
				$mbs\times\pi_{-1}(\hat{\gamma}_{post})$ & 0.0449{*}{*}{*} & 0.0263{*}{*} & 0.0286{*}{*}{*} & 0.0589{*}{*}{*} & 0.0390{*}{*}{*} & 0.0502{*}{*}{*} \\
				& (0.0140) & (0.0107) & (0.0109) & (0.0137) & (0.0136) & (0.0139) \\
				[0.1cm]
				\hline
				Fixed effects & Yes & Yes & Yes & Yes & Yes & Yes \\
				Time effects & Yes & Yes & Yes & Yes & Yes & Yes \\
				Industry linear trends & Yes & No & Yes & Yes & No & Yes \\
				Ind. dummy$\times$RGDP & No & Yes & Yes & No & Yes & Yes \\
				Observations & 83290 & 83290 & 83290 & 83290 & 83290 & 83290 \\
				$N$ & 3236 & 3236 & 3236 & 3236 & 3236 & 3236 \\
				$max(T_{i})$ & 44 & 44 & 44 & 44 & 44 & 44 \\
				$avg(T_{i})$ & 25.7 & 25.7 & 25.7 & 25.7 & 25.7 & 25.7 \\
				$med(T_{i})$ & 23 & 23 & 23 & 23 & 23 & 23 \\
				$min(T_{i})$ & 5 & 5 & 5 & 5 & 5 & 5 \\
				\hline
		\end{tabular}	}
\end{center}

%

\end{table}%

%

\begin{table}[H]
\caption{\textbf{FE--TE estimates of the net short-run and long-run effects of of large-scale MBS and Treasury purchases on debt to asset ratios of non-financial firms with at least $10$ observations}}\label{tab:SRLR_MBS_TY_mint10}
\vspace{-0.2cm}
\footnotesize
Panel A (Panel B) reports the net short-run (long-run) effects of MBS and
Treasury purchases on firms' debt to asset ratios (DA), for the PanARDL(2)
model described in equation (\ref{ARDL_model_lsap}). Net short-run effects are
defined as the sum of the estimated coefficients of current and lagged values
of the regressor under consideration. The long-run effects are defined in
equation (\ref{LR_definition}). The first three columns report results for the
single-threshold panel regression model, where $\gamma_{pre}=\gamma_{post}$.
The last three columns report results for the two-threshold panel regression,
where $\gamma_{pre}\neq\gamma_{post}$. The estimated quantile threshold
parameters are shown in Table \ref{tab: thresholds_TY_MBS}. All regressions
include both firm-specific effects and time effects. Columns (1) and (4)
include industry-specific linear time trends, columns (2) and (5) include the
interaction of industry dummies and real GDP growth, while columns (3) and (6)
include both. $ty$ and $mbs$ denote the (scaled) amount of U.S. Treasuries and
agency MBS purchased by the Fed, respectively; $\pi_{-1}(\gamma)$ denotes the
one-quarter lagged proportion of firms in an industry with DA below the
$\gamma$-th quantile. he sample only includes firms with at least $10$ time
observations, resulting in an unbalanced panel of $3,011$ U.S. publicly traded
non-financial firms observed at a quarterly frequency over the period
$2007$:Q1 - $2018$:Q3. Robust standard errors (in parentheses) are computed
using the delta method (***$p<0.01$, ** $p<0.05$, * $p<0.1$).
\normalsize

\begin{center}
\scalebox{0.85}{
			\begin{tabular}{l|ccc|ccc}
				\multicolumn{1}{c}{} & \multicolumn{6}{c}{Dependent variable: debt to assets (DA)} \\
				\cline{2-7} \cline{3-7} \cline{4-7} \cline{5-7} \cline{6-7} \cline{7-7}
				\multicolumn{1}{c}{} & \multicolumn{3}{c|}{$\gamma_{pre}=\gamma_{post}=\gamma$} & \multicolumn{3}{c}{$\gamma_{pre}\neq\gamma_{post}$} \\
				\multicolumn{1}{c}{} & (1) & (2) & (3) & (4) & (5) & (6) \\
				\hline
				\multicolumn{1}{c}{} & \multicolumn{6}{c}{\textit{Panel A: Short-run effects}} \\
				\hline
				$\pi_{-1}(\hat{\gamma}_{pre})$ & 0.0158{*}{*}{*} & 0.0105{*}{*} & 0.0169{*}{*}{*} & 0.0181{*}{*}{*} & 0.0122{*}{*}{*} & 0.0192{*}{*}{*} \\
				& (0.0050) & (0.0048) & (0.0054) & (0.0050) & (0.0046) & (0.0051) \\
				$ty\times\pi_{-1}(\hat{\gamma}_{post})$ & 0.0056{*} & 0.0015 & 0.0023 & 0.0076{*}{*} & 0.0049 & 0.0062{*} \\
				& (0.0032) & (0.0025) & (0.0026) & (0.0031) & (0.0031) & (0.0032) \\
				$mbs\times\pi_{-1}(\hat{\gamma}_{post})$ & 0.0074{*}{*}{*} & 0.0043{*}{*}{*} & 0.0048{*}{*}{*} & 0.0097{*}{*}{*} & 0.0065{*}{*}{*} & 0.0082{*}{*}{*} \\
				& (0.0022) & (0.0017) & (0.0017) & (0.0021) & (0.0021) & (0.0022) \\ [0.1cm]
				\hline
				\multicolumn{1}{c}{} & \multicolumn{6}{c}{\textit{Panel B: Long-run effects}} \\
				\hline
				$\pi_{-1}(\hat{\gamma}_{pre})$ & 0.1015{*}{*}{*} & 0.0686{*}{*} & 0.1087{*}{*}{*} & 0.1167{*}{*}{*} & 0.0800{*}{*}{*} & 0.1234{*}{*}{*} \\
				& (0.0324) & (0.0313) & (0.0354) & (0.0328) & (0.0301) & (0.0334) \\
				$ty\times\pi_{-1}(\hat{\gamma}_{post})$ & 0.0363{*} & 0.0100 & 0.0145 & 0.0487{*}{*} & 0.0318 & 0.0398{*} \\
				& (0.0210) & (0.0163) & (0.0170) & (0.0198) & (0.0200) & (0.0206) \\
				$mbs\times\pi_{-1}(\hat{\gamma}_{post})$ & 0.0476{*}{*}{*} & 0.0283{*}{*}{*} & 0.0306{*}{*}{*} & 0.0622{*}{*}{*} & 0.0422{*}{*}{*} & 0.0528{*}{*}{*} \\
				& (0.0142) & (0.0109) & (0.0111) & (0.0139) & (0.0138) & (0.0141) \\[0.1cm]
				\hline
				Fixed effects & Yes & Yes & Yes & Yes & Yes & Yes \\
				Time effects & Yes & Yes & Yes & Yes & Yes & Yes \\
				Industry linear trends & Yes & No & Yes & Yes & No & Yes \\
				Ind. dummy$\times$RGDP & No & Yes & Yes & No & Yes & Yes \\
				Observations & 82038 & 82038 & 82038 & 82038 & 82038 & 82038 \\
				$N$ & 3011 & 3011 & 3011 & 3011 & 3011 & 3011 \\
				$max(T_{i})$ & 44 & 44 & 44 & 44 & 44 & 44 \\
				$avg(T_{i})$ & 27.2 & 27.2 & 27.2 & 27.2 & 27.2 & 27.2 \\
				$med(T_{i})$ & 25 & 25 & 25 & 25 & 25 & 25 \\
				$min(T_{i})$ & 7 & 7 & 7 & 7 & 7 & 7 \\
				\hline
		\end{tabular}	}
\end{center}

%

\end{table}%

%

\begin{table}[H]
\caption{\textbf{Half-panel jackknife FE--TE estimates of the net short-run and long-run effects of of large-scale MBS and Treasury purchases on debt to asset ratios of non-financial firms}}\label{tab:SRLR_MBS_TY_jackk}
\vspace{-0.2cm}
\footnotesize
Panel A (Panel B) reports the net short-run (long-run) effects of MBS and
Treasury purchases on firms' debt to asset ratios (DA), for the PanARDL(2)
model described in equation (\ref{ARDL_model_lsap}). Net short-run effects are
defined as the sum of the estimated coefficients of current and lagged values
of the regressor under consideration. The long-run effects are defined in
equation (\ref{LR_definition}). The first three columns report results for the
single-threshold panel regression model, where $\gamma_{pre}=\gamma_{post}$.
The last three columns report results for the two-threshold panel regression,
where $\gamma_{pre}\neq\gamma_{post}$. The estimated quantile threshold
parameters are shown in Table \ref{tab: thresholds_TY_MBS}. All regressions
include both firm-specific effects and time effects. Columns (1) and (4)
include industry-specific linear time trends, columns (2) and (5) include the
interaction of industry dummies and real GDP growth, while columns (3) and (6)
include both. $ty$ and $mbs$ denote the (scaled) amount of U.S. Treasuries and
agency MBS purchased by the Fed, respectively; $\pi_{-1}(\gamma)$ denotes the
one-quarter lagged proportion of firms in an industry with DA below the
$\gamma$-th quantile. The sample only includes firms with at least $8$ time
observations, resulting in an unbalanced panel of $3,236$ U.S. publicly traded
non-financial firms observed at a quarterly frequency over the period
$2007$:Q1 - $2018$:Q3. Robust standard errors (in parentheses) are computed
using the delta method (***$p<0.01$, ** $p<0.05$, * $p<0.1$).
\normalsize

\begin{center}
\scalebox{0.85}{
			\begin{tabular}{l|ccc|ccc}
				\multicolumn{1}{c}{} & \multicolumn{6}{c}{Dependent variable: debt to assets (DA)} \\
				\cline{2-7} \cline{3-7} \cline{4-7} \cline{5-7} \cline{6-7} \cline{7-7}
				\multicolumn{1}{c}{} & \multicolumn{3}{c|}{$\gamma_{pre}=\gamma_{post}=\gamma$} & \multicolumn{3}{c}{$\gamma_{pre}\neq\gamma_{post}$} \\
				\multicolumn{1}{c}{} & (1) & (2) & (3) & (4) & (5) & (6) \\
				\hline
				\multicolumn{1}{c}{} & \multicolumn{6}{c}{\textit{Panel A: Short-run effects}} \\
				\hline
				$\pi_{-1}(\hat{\gamma}_{pre})$ & 0.0098 & 0.0067 & 0.0072 & 0.0074 & 0.0096 & 0.0105 \\
				& (0.0063) & (0.0064) & (0.0069) & (0.0064) & (0.0062) & (0.0065) \\
				$ty\times\pi_{-1}(\hat{\gamma}_{post})$ & -0.0007 & 0.0010 & 0.0014 & 0.0012 & 0.0035 & 0.0006 \\
				& (0.0042) & (0.0034) & (0.0034) & (0.0040) & (0.0041) & (0.0041) \\
				$mbs\times\pi_{-1}(\hat{\gamma}_{post})$ & 0.0099{*}{*}{*} & 0.0063{*}{*}{*} & 0.0061{*}{*}{*} & 0.0107{*}{*}{*} & 0.0066{*}{*} & 0.0090{*}{*}{*} \\
				& (0.0026) & (0.0020) & (0.0021) & (0.0025) & (0.0026) & (0.0026) \\ [0.1cm]
				\hline
				\multicolumn{1}{c}{} & \multicolumn{6}{c}{\textit{Panel B: Long-run effects}} \\
				\hline
				$\pi_{-1}(\hat{\gamma}_{pre})$ & 0.1950 & 0.1349 & 0.1416 & 0.1463 & 0.1930 & 0.2078 \\
				& (0.1280) & (0.1306) & (0.1386) & (0.1303) & (0.1286) & (0.1340) \\
				$ty\times\pi_{-1}(\hat{\gamma}_{post})$ & -0.0130 & 0.0196 & 0.0280 & 0.0232 & 0.0706 & 0.0113 \\
				& (0.0839) & (0.0677) & (0.0677) & (0.0792) & (0.0835) & (0.0804) \\
				$mbs\times\pi_{-1}(\hat{\gamma}_{post})$ & 0.1959{*}{*}{*} & 0.1260{*}{*}{*} & 0.1209{*}{*}{*} & 0.2124{*}{*}{*} & 0.1321{*}{*} & 0.1785{*}{*}{*} \\
				& (0.0581) & (0.0444) & (0.0448) & (0.0580) & (0.0548) & (0.0576) \\[0.1cm]
				\hline
				Fixed effects & Yes & Yes & Yes & Yes & Yes & Yes \\
				Time effects & Yes & Yes & Yes & Yes & Yes & Yes \\
				Industry linear trends & Yes & No & Yes & Yes & No & Yes \\
				Ind. dummy$\times$RGDP & No & Yes & Yes & No & Yes & Yes \\
				Observations & 82092 & 82092 & 82092 & 82092 & 82092 & 82092 \\
				$N$ & 3236 & 3236 & 3236 & 3236 & 3236 & 3236 \\
				$max(T_{i})$ & 44 & 44 & 44 & 44 & 44 & 44 \\
				$avg(T_{i})$ & 25.4 & 25.4 & 25.4 & 25.4 & 25.4 & 25.4 \\
				$med(T_{i})$ & 22 & 22 & 22 & 22 & 22 & 22 \\
				$min(T_{i})$ & 4 & 4 & 4 & 4 & 4 & 4 \\
				\hline
		\end{tabular}	}
\end{center}

%

\end{table}%

\section{Estimating the effects of four asset purchase programs using
qualitative measures of LSAPs \label{Apx_QE_sep}}

Here, we compare the effects of each Fed's asset purchase program by replacing
the two aforementioned quantitative measures of LSAPs with four qualitative
variables which take the value of one during policy on periods and zero
otherwise. Following the literature, we label these policy indicators as QE1
(covering the period 2008Q4 to 2010Q1), QE2 (2010Q4 - 2011Q2), MEP (the
maturity extension program of 2011Q3 - 2012Q4), and QE3 (2012Q3 - 2012Q4).
Further information on these programs can be found in Table \ref{Table_LSAP}
of Subsection \ref{Appx: Var_Construct} in this online supplement.

In Subsection \ref{Apx_QE_sep_thresh}, we report the estimated threshold
parameters for the PanARDL(2) panel regression model. The estimated net
short-run and long-run effects of the various Fed's programs are shown in
Subsection \ref{Apx_QE_sep_SRLR}. In Subsection \ref{Apx_QE_sep_smallT}, we
show the estimation results after correcting for potential small-sample bias.

\subsection{Quantile threshold parameter estimates \label{Apx_QE_sep_thresh}}%

\begin{table}[H]
\caption{\textbf{Estimated quantile threshold parameters}}\label{tab: thresholds_QE_sep}
\vspace{-0.2cm}
\footnotesize
Estimates of the quantile threshold parameters from a grid search procedure
for the PanARDL(2) model described in equation (\ref{ARDL_model_lsap}), using
qualitative measures of LSAPs. Panel A shows the estimated threshold
parameters for the single-threshold panel regression model, where
$\gamma_{pre}=\gamma_{post}$. Panel B displays results for the two-threshold
model, where $\gamma_{pre}\neq\gamma_{post}$. In column (1) and (2), we use
linear time trends or real GDP growth as a proxy for $f_{t}$, respectively.
Column (3) reports results when including both linear trends and real GDP
growth at the same time. The estimation sample consists of an unbalanced panel
of $3,647$ U.S. publicly traded non-financial firms observed at a quarterly
frequency over the period $2007$:Q1 - $2018$:Q3.
\normalsize

\begin{center}%
\begin{tabular}
[c]{cccc}\cline{2-4}\cline{3-4}\cline{4-4}
& (1) & (2) & (3)\\\hline
\textit{Panel A:} & \multicolumn{3}{c}{$\gamma_{pre}=\gamma_{post}=\gamma$}\\
$\hat{\gamma}$ & 0.72 & 0.55 & 0.56\\[0.2cm]%
\textit{Panel B:} & \multicolumn{3}{c}{$\gamma_{pre}\neq\gamma_{post}$}\\
$\hat{\gamma}_{pre}$ & 0.56 & 0.56 & 0.56\\
$\hat{\gamma}_{post}$ & 0.73 & 0.72 & 0.72\\\hline
linear trends & Yes & No & Yes\\
RGDP growth & No & Yes & Yes\\\hline
\end{tabular}

\end{center}

%

\end{table}%

\subsection{Short- and long-run effects of four episodes of LSAPs
\label{Apx_QE_sep_SRLR}}%

\begin{table}[H]
\caption{\textbf{FE--TE estimates of the net short-run effects of four episodes of LSAPs on debt to asset ratios of non-financial firms}}\label{tab:SR_QE_sep}
\vspace{-0.2cm}
\footnotesize
Estimates of net short-run effects of the first four asset purchase programs
on firms' debt to asset ratios (DA) for the PanARDL(2) model described in
equation (\ref{ARDL_model_lsap}). Net short-run effects are defined as the sum
of the estimated coefficients of current and lagged values of the regressor
under consideration. The first three columns report results for the
single-threshold panel regression model, where $\gamma_{pre}=\gamma_{post}$.
The last three columns report results for the two-threshold panel regression,
where $\gamma_{pre}\neq\gamma_{post}$. The estimated quantile threshold
parameters are shown in Table \ref{tab: thresholds_QE_sep}. All regressions
include both firm-specific effects and time effects. Columns (1) and (4)
include industry-specific linear time trends, columns (2) and (5) include the
interaction of industry dummies and real GDP growth, while columns (3) and (6)
include both. $QE1$ and $QE2$ are two indicator variables equal to one during
the period 2008Q4 - 2010Q1, and 2010Q4 - 2011Q2 and zero otherwise,
respectively. $MEP$ denotes the maturity extension program of 2011Q3 - 2012Q4,
while $QE3$ is equal to one between 2012Q3 - 2012Q4; $\pi(\gamma)$ denotes the
proportion of firms in an industry with DA below the $\gamma$-th quantile. The
sample consists of an unbalanced panel of $3,647$ U.S. publicly traded
non-financial firms observed at a quarterly frequency over the period
$2007$:Q1 - $2018$:Q3. Robust standard errors (in parentheses) are computed
using the delta method (*** $p<0.01$, ** $p<0.05$, * $p<0.1$).
\normalsize

\begin{center}
\scalebox{0.85}{
			\begin{tabular}{l|ccc|ccc}
				\multicolumn{1}{c}{} & \multicolumn{6}{c}{Dependent variable: debt to assets (DA)} \\
				\cline{2-7} \cline{3-7} \cline{4-7} \cline{5-7} \cline{6-7} \cline{7-7}
				\multicolumn{1}{c}{} & \multicolumn{3}{c|}{$\gamma_{pre}=\gamma_{post}=\gamma$} & \multicolumn{3}{c}{$\gamma_{pre}\neq\gamma_{post}$} \\
				\multicolumn{1}{c}{} & (1) & (2) & (3) & (4) & (5) & (6) \\
				\hline
				$\pi_{-1}(\hat{\gamma}_{pre})$ & 0.0173{*}{*}{*} & 0.0135{*}{*}{*} & 0.0215{*}{*}{*} & 0.0199{*}{*}{*} & 0.0136{*}{*}{*} & 0.0212{*}{*}{*} \\
				& (0.0053) & (0.0048) & (0.0056) & (0.0051) & (0.0046) & (0.0052) \\
				$QE_{1}\times\pi_{-1}(\hat{\gamma}_{post})$ & 0.0140{*}{*}{*} & 0.0066{*}{*} & 0.0105{*}{*}{*} & 0.0189{*}{*}{*} & 0.0092{*}{*} & 0.0156{*}{*}{*} \\
				& (0.0045) & (0.0032) & (0.0039) & (0.0044) & (0.0038) & (0.0044) \\
				$QE_{2}\times\pi_{-1}(\hat{\gamma}_{post})$ & 0.0081 & 0.0021 & 0.0048 & 0.0114{*}{*} & 0.0043 & 0.0089{*} \\
				& (0.0054) & (0.0046) & (0.0050) & (0.0053) & (0.0050) & (0.0053) \\
				$MEP\times\pi_{-1}(\hat{\gamma}_{post})$ & -0.0018 & -0.0015 & -0.0007 & 0.0019 & -0.0036 & -0.0012 \\
				& (0.0039) & (0.0033) & (0.0034) & (0.0038) & (0.0037) & (0.0038) \\
				$QE_{3}\times\pi_{-1}(\hat{\gamma}_{post})$ & 0.0045 & -0.0009 & 0.0000 & 0.0069{*}{*} & 0.0029 & 0.0040 \\
				& (0.0035) & (0.0029) & (0.0030) & (0.0033) & (0.0033) & (0.0034) \\
				\hline
				Fixed effects & Yes & Yes & Yes & Yes & Yes & Yes \\
				Time effects & Yes & Yes & Yes & Yes & Yes & Yes \\
				Industry linear trends & Yes & No & Yes & Yes & No & Yes \\
				Ind. dummy$\times$RGDP & No & Yes & Yes & No & Yes & Yes \\
				Observations & 84548 & 84548 & 84548 & 84548 & 84548 & 84548 \\
				$N$ & 3647 & 3647 & 3647 & 3647 & 3647 & 3647 \\
				$max(T_{i})$ & 44 & 44 & 44 & 44 & 44 & 44 \\
				$avg(T_{i})$ & 23.2 & 23.2 & 23.2 & 23.2 & 23.2 & 23.2 \\
				$med(T_{i})$ & 19 & 19 & 19 & 19 & 19 & 19 \\
				$min(T_{i})$ & 2 & 2 & 2 & 2 & 2 & 2 \\
				\hline
		\end{tabular}	}
\end{center}

%

\end{table}%
%

\begin{table}[H]
\caption{\textbf{FE--TE estimates of the long-run effects of four episodes of LSAPs on debt to asset ratios of non-financial firms}}\label{tab:LR_QE_sep}
\vspace{-0.2cm}
\footnotesize
Estimates of long-run effects, defined in equation (\ref{LR_definition}), of
the first four asset purchase programs on firms' debt to asset ratios (DA) for
the PanARDL(2) model described in equation (\ref{ARDL_model_lsap}). The first
three columns report results for the single-threshold panel regression model,
where $\gamma_{pre}=\gamma_{post}$. The last three columns report results for
the two-threshold panel regression, where $\gamma_{pre}\neq\gamma_{post}$. The
estimated quantile threshold parameters are shown in Table
\ref{tab: thresholds_QE_sep}. All regressions include both firm-specific
effects and time effects. Columns (1) and (4) include industry-specific linear
time trends, columns (2) and (5) include the interaction of industry dummies
and real GDP growth, while columns (3) and (6) include both. $QE1$ and $QE2$
are two indicator variables equal to one during the period 2008Q4 - 2010Q1,
and 2010Q4 - 2011Q2 and zero otherwise, respectively. $MEP$ denotes the
maturity extension program of 2011Q3 - 2012Q4, while $QE3$ is equal to one
between 2012Q3 - 2012Q4; $\pi(\gamma)$ denotes the proportion of firms in an
industry with DA below the $\gamma$-th quantile. The sample consists of an
unbalanced panel of $3,647$ U.S. publicly traded non-financial firms observed
at a quarterly frequency over the period $2007$:Q1 - $2018$:Q3. Robust
standard errors (in parentheses) are computed using the delta method (***
$p<0.01$, ** $p<0.05$, * $p<0.1$).
\normalsize

\begin{center}
\scalebox{0.85}{
			\begin{tabular}{l|ccc|ccc}
				\multicolumn{1}{c}{} & \multicolumn{6}{c}{Dependent variable: debt to assets (DA)} \\
				\cline{2-7} \cline{3-7} \cline{4-7} \cline{5-7} \cline{6-7} \cline{7-7}
				\multicolumn{1}{c}{} & \multicolumn{3}{c|}{$\gamma_{pre}=\gamma_{post}=\gamma$} & \multicolumn{3}{c}{$\gamma_{pre}\neq\gamma_{post}$} \\
				\multicolumn{1}{c}{} & (1) & (2) & (3) & (4) & (5) & (6) \\
				\hline
				$\pi_{-1}(\hat{\gamma}_{pre})$ & 0.1074{*}{*}{*} & 0.0847{*}{*}{*} & 0.1331{*}{*}{*} & 0.1233{*}{*}{*} & 0.0855{*}{*}{*} & 0.1315{*}{*}{*} \\
				& (0.0328) & (0.0304) & (0.0351) & (0.0321) & (0.0294) & (0.0328) \\
				$QE_{1}\times\pi_{-1}(\hat{\gamma}_{post})$ & 0.0867{*}{*}{*} & 0.0413{*}{*} & 0.0652{*}{*}{*} & 0.1169{*}{*}{*} & 0.0577{*}{*} & 0.0967{*}{*}{*} \\
				& (0.0279) & (0.0205) & (0.0242) & (0.0278) & (0.0238) & (0.0276) \\
				$QE_{2}\times\pi_{-1}(\hat{\gamma}_{post})$ & 0.0502 & 0.0135 & 0.0297 & 0.0705{*}{*} & 0.0267 & 0.0551{*} \\
				& (0.0334) & (0.0291) & (0.0308) & (0.0332) & (0.0314) & (0.0331) \\
				$MEP\times\pi_{-1}(\hat{\gamma}_{post})$ & -0.0109 & -0.0095 & -0.0046 & 0.0120 & -0.0228 & -0.0072 \\
				& (0.0244) & (0.0205) & (0.0211) & (0.0234) & (0.0232) & (0.0238) \\
				$QE_{3}\times\pi_{-1}(\hat{\gamma}_{post})$ & 0.0279 & -0.0055 & 0.0000 & 0.0426{*}{*} & 0.0183 & 0.0249 \\
				& (0.0215) & (0.0182) & (0.0186) & (0.0204) & (0.0209) & (0.0208) \\
				\hline
				Fixed effects & Yes & Yes & Yes & Yes & Yes & Yes \\
				Time effects & Yes & Yes & Yes & Yes & Yes & Yes \\
				Industry linear trends & Yes & No & Yes & Yes & No & Yes \\
				Ind. dummy$\times$RGDP & No & Yes & Yes & No & Yes & Yes \\
				Observations & 84548 & 84548 & 84548 & 84548 & 84548 & 84548 \\
				$N$ & 3647 & 3647 & 3647 & 3647 & 3647 & 3647 \\
				$max(T_{i})$ & 44 & 44 & 44 & 44 & 44 & 44 \\
				$avg(T_{i})$ & 23.2 & 23.2 & 23.2 & 23.2 & 23.2 & 23.2 \\
				$med(T_{i})$ & 19 & 19 & 19 & 19 & 19 & 19 \\
				$min(T_{i})$ & 2 & 2 & 2 & 2 & 2 & 2 \\
				\hline
		\end{tabular}	}
\end{center}

%

\end{table}%


\subsection{Robustness of the results to small-T bias
\label{Apx_QE_sep_smallT}}

\subsubsection{Estimates when firms have at least $8$ or $10$ observations}%

\begin{table}[H]
\caption{\textbf{FE--TE estimates of the net short-run effects of four episodes of LSAPs on debt to asset ratios of non-financial firms with at least $8$ or $10$ observations}}\label{tab:SR_QE_sep_minT}
\vspace{-0.2cm}
\footnotesize
Estimates of net short-run effects of the first four asset purchase programs
on firms' debt to asset ratios (DA) for the PanARDL(2) model described in
equation (\ref{ARDL_model_lsap}), focusing on the two-threshold panel
regression, where $\gamma_{pre}\neq\gamma_{post}$. The estimated quantile
threshold parameters are shown in Table \ref{tab: thresholds_QE_sep}. Net
short-run effects are defined as the sum of the estimated coefficients of
current and lagged values of the regressor under consideration. All
regressions include both firm-specific effects and time effects. Columns (1)
and (4) include industry-specific linear time trends, columns (2) and (5)
include the interaction of industry dummies and real GDP growth, while columns
(3) and (6) include both. In the first (last) three columns, the sample only
includes firms with at least 8 (10) observations. $QE1$ and $QE2$ are two
indicator variables equal to one during the period 2008Q4 - 2010Q1, and 2010Q4
- 2011Q2 and zero otherwise, respectively. $MEP$ denotes the maturity
extension program of 2011Q3 - 2012Q4, while $QE3$ is equal to one between
2012Q3 - 2012Q4; $\pi_{-1}(\gamma)$ denotes the one-quarter lagged proportion
of firms in an industry with DA below the $\gamma$-th quantile. Robust
standard errors (in parentheses) are computed using the delta method (***
$p<0.01$, ** $p<0.05$, * $p<0.1$).
\normalsize

\begin{center}
\scalebox{0.85}{
			\begin{tabular}{l|ccc|ccc}
				\multicolumn{1}{c}{} & \multicolumn{6}{c}{Dependent variable: debt to assets (DA)} \\
				\cline{2-7} \cline{3-7} \cline{4-7} \cline{5-7} \cline{6-7} \cline{7-7}
				\multicolumn{1}{c}{} & \multicolumn{3}{c|}{\textit{At least 8 observations}} & \multicolumn{3}{c}{\textit{At least 10 observations}} \\
				\multicolumn{1}{c}{} & (1) & (2) & (3) & (4) & (5) & (6) \\
				\hline
				$\pi(\hat{\gamma}_{pre})$ & 0.0194{*}{*}{*} & 0.0137{*}{*}{*} & 0.0210{*}{*}{*} & 0.0193{*}{*}{*} & 0.0136{*}{*}{*} & 0.0209{*}{*}{*} \\
				& (0.0051) & (0.0046) & (0.0052) & (0.0051) & (0.0046) & (0.0052) \\
				$QE_{1}\times\pi(\hat{\gamma}_{post})$ & 0.0189{*}{*}{*} & 0.0095{*}{*} & 0.0154{*}{*}{*} & 0.0182{*}{*}{*} & 0.0091{*}{*} & 0.0148{*}{*}{*} \\
				& (0.0044) & (0.0038) & (0.0044) & (0.0044) & (0.0037) & (0.0044) \\
				$QE_{2}\times\pi(\hat{\gamma}_{post})$ & 0.0113{*}{*} & 0.0040 & 0.0082 & 0.0109{*}{*} & 0.0039 & 0.0079 \\
				& (0.0053) & (0.0050) & (0.0053) & (0.0053) & (0.0050) & (0.0053) \\
				$MEP\times\pi(\hat{\gamma}_{post})$ & 0.0018 & -0.0039 & -0.0016 & 0.0016 & -0.0040 & -0.0018 \\
				& (0.0038) & (0.0037) & (0.0038) & (0.0038) & (0.0037) & (0.0038) \\
				$QE_{3}\times\pi(\hat{\gamma}_{post})$ & 0.0069{*}{*} & 0.0025 & 0.0035 & 0.0065{*}{*} & 0.0023 & 0.0033 \\
				& (0.0033) & (0.0033) & (0.0034) & (0.0033) & (0.0033) & (0.0034) \\
				\hline
				Fixed effects & Yes & Yes & Yes & Yes & Yes & Yes \\
				Time effects & Yes & Yes & Yes & Yes & Yes & Yes \\
				Industry linear trends & Yes & No & Yes & Yes & No & Yes \\
				Ind. dummy$\times$RGDP & No & Yes & Yes & No & Yes & Yes \\
				Observations & 83290 & 83290 & 83290 & 82038 & 82038 & 82038 \\
				$N$ & 3236 & 3236 & 3236 & 3011 & 3011 & 3011 \\
				$max(T_{i})$ & 44 & 44 & 44 & 44 & 44 & 44 \\
				$avg(T_{i})$ & 25.7 & 25.7 & 25.7 & 27.2 & 27.2 & 27.2 \\
				$med(T_{i})$ & 23 & 23 & 23 & 25 & 25 & 25 \\
				$min(T_{i})$ & 5 & 5 & 5 & 7 & 7 & 7 \\
				\hline
		\end{tabular}	}
\end{center}

%

\end{table}%

\subsubsection{Half-panel jackknife FE--TE estimates of the short- and
long-run effects of four episodes of LSAPs}%

\begin{table}[H]
\caption{\textbf{Half-panel jackknife FE--TE estimates of the net short-run effects of four episodes of LSAPs on debt to asset ratios of non-financial firms}}\label{tab:SR_QE_sep_Jackk}
\vspace{-0.2cm}
\footnotesize
Estimates of net short-run effects of LSAPs on firms' debt to asset ratios
(DA), for the PanARDL(2) model described in equation (\ref{ARDL_model_lsap}).
Net short-run effects are defined as the sum of the estimated coefficients of
current and lagged values of the regressor under consideration. The first
three columns report results for the single-threshold panel regression model,
where $\gamma_{pre}=\gamma_{post}$. The last three columns report results for
the two-threshold panel regression, where $\gamma_{pre}\neq\gamma_{post}$. The
estimated quantile threshold parameters are shown in Table
\ref{tab: thresholds_QE_sep}. All regressions include both firm-specific
effects and time effects. Columns (1) and (4) include industry-specific linear
time trends, columns (2) and (5) include the interaction of industry dummies
and real GDP growth, while columns (3) and (6) include both. $QE1$ and $QE2$
are two indicator variables equal to one during the period 2008Q4 - 2010Q1,
and 2010Q4 - 2011Q2 and zero otherwise, respectively. $MEP$ denotes the
maturity extension program of 2011Q3 - 2012Q4, while $QE3$ is equal to one
between 2012Q3 - 2012Q4; $\pi_{-1}(\gamma)$ denotes the one-quarter lagged
proportion of firms in an industry with DA below the $\gamma$-th quantile. The
sample only includes firms with at least $8$ time observations, resulting in
an unbalanced panel of $3,236$ U.S. publicly traded non-financial firms
observed at a quarterly frequency over the period $2007$:Q1 - $2018$:Q3.
Robust standard errors (in parentheses) are computed using the delta method
(*** $p<0.01$, ** $p<0.05$, * $p<0.1$).
\normalsize

\begin{center}
\scalebox{0.85}{
			\begin{tabular}{l|ccc|ccc}
				\multicolumn{1}{c}{} & \multicolumn{6}{c}{Dependent variable: debt to assets (DA)} \\
				\cline{2-7} \cline{3-7} \cline{4-7} \cline{5-7} \cline{6-7} \cline{7-7}
				\multicolumn{1}{c}{} & \multicolumn{3}{c|}{$\gamma_{pre}=\gamma_{post}=\gamma$} & \multicolumn{3}{c}{$\gamma_{pre}\neq\gamma_{post}$} \\
				\multicolumn{1}{c}{} & (1) & (2) & (3) & (4) & (5) & (6) \\
				\hline
				$\pi_{-1}(\hat{\gamma}_{pre})$ & 0.0147{*}{*} & 0.0090 & 0.0100 & 0.0082 & 0.0105{*} & 0.0115{*} \\
				& (0.0066) & (0.0063) & (0.0070) & (0.0065) & (0.0062) & (0.0066) \\
				$QE_{1}\times\pi_{-1}(\hat{\gamma}_{post})$ & 0.0112{*}{*} & 0.0031 & 0.0121{*}{*} & 0.0111{*}{*} & 0.0023 & 0.0112{*}{*} \\
				& (0.0056) & (0.0045) & (0.0050) & (0.0055) & (0.0052) & (0.0055) \\
				$QE_{2}\times\pi_{-1}(\hat{\gamma}_{post})$ & 0.0092 & 0.0004 & 0.0102{*} & 0.0036 & -0.0002 & 0.0072 \\
				& (0.0065) & (0.0059) & (0.0060) & (0.0064) & (0.0063) & (0.0063) \\
				$MEP\times\pi_{-1}(\hat{\gamma}_{post})$ & 0.0032 & 0.0015 & 0.0072{*} & 0.0021 & -0.0024 & 0.0022 \\
				& (0.0048) & (0.0042) & (0.0042) & (0.0046) & (0.0047) & (0.0046) \\
				$QE_{3}\times\pi_{-1}(\hat{\gamma}_{post})$ & 0.0110{*}{*} & 0.0007 & 0.0056 & 0.0104{*}{*} & 0.0068 & 0.0092{*}{*} \\
				& (0.0045) & (0.0037) & (0.0038) & (0.0043) & (0.0043) & (0.0043) \\
				\hline
				Fixed effects & Yes & Yes & Yes & Yes & Yes & Yes \\
				Time effects & Yes & Yes & Yes & Yes & Yes & Yes \\
				Industry linear trends & Yes & No & Yes & Yes & No & Yes \\
				Ind. dummy$\times$RGDP & No & Yes & Yes & No & Yes & Yes \\
				Observations & 82092 & 82092 & 82092 & 82092 & 82092 & 82092 \\
				$N$ & 3236 & 3236 & 3236 & 3236 & 3236 & 3236 \\
				$max(T_{i})$ & 44 & 44 & 44 & 44 & 44 & 44 \\
				$avg(T_{i})$ & 25.4 & 25.4 & 25.4 & 25.4 & 25.4 & 25.4 \\
				$med(T_{i})$ & 22 & 22 & 22 & 22 & 22 & 22 \\
				$min(T_{i})$ & 4 & 4 & 4 & 4 & 4 & 4 \\
				\hline
		\end{tabular}	}
\end{center}

%

\end{table}%
%

\begin{table}[H]
\caption{\textbf{Half-panel jackknife FE--TE estimates of the long-run effects of four episodes of LSAPs on debt to asset ratios of non-financial firms}}\label{tab:LR_QE_sep_Jackk}
\vspace{-0.2cm}
\footnotesize
Estimates of long-run effects of LSAPs, defined in equation
(\ref{LR_definition}), on firms' debt to asset ratios (DA), for the PanARDL(2)
model described in equation (\ref{ARDL_model_lsap}). The first three columns
report results for the single-threshold panel regression model, where
$\gamma_{pre}=\gamma_{post}$. The last three columns report results for the
two-threshold panel regression, where $\gamma_{pre}\neq\gamma_{post}$. The
estimated quantile threshold parameters are shown in Table
\ref{tab: thresholds_QE_sep}. All regressions include both firm-specific
effects and time effects. Columns (1) and (4) include industry-specific linear
time trends, columns (2) and (5) include the interaction of industry dummies
and real GDP growth, while columns (3) and (6) include both. $QE1$ and $QE2$
are two indicator variables equal to one during the period 2008Q4 - 2010Q1,
and 2010Q4 - 2011Q2 and zero otherwise, respectively. $MEP$ denotes the
maturity extension program of 2011Q3 - 2012Q4, while $QE3$ is equal to one
between 2012Q3 - 2012Q4;; $\pi(\gamma)$ denotes the proportion of firms in an
industry with DA below the $\gamma$-th quantile. The sample only includes
firms with at least $8$ time observations, resulting in an unbalanced panel of
$3,236$ U.S. publicly traded non-financial firms observed at a quarterly
frequency over the period $2007$:Q1 - $2018$:Q3. Robust standard errors (in
parentheses) are computed using the delta method (*** $p<0.01$, ** $p<0.05$, *
$p<0.1$).
\normalsize

\begin{center}
\scalebox{0.85}{
			\begin{tabular}{l|ccc|ccc}
				\multicolumn{1}{c}{} & \multicolumn{6}{c}{Dependent variable: debt to assets (DA)} \\
				\cline{2-7} \cline{3-7} \cline{4-7} \cline{5-7} \cline{6-7} \cline{7-7}
				\multicolumn{1}{c}{} & \multicolumn{3}{c|}{$\gamma_{pre}=\gamma_{post}=\gamma$} & \multicolumn{3}{c}{$\gamma_{pre}\neq\gamma_{post}$} \\
				\multicolumn{1}{c}{} & (1) & (2) & (3) & (4) & (5) & (6) \\
				\hline
				$\pi_{-1}(\hat{\gamma}_{pre})$ & 0.2936{*}{*} & 0.1802 & 0.1986 & 0.1634 & 0.2104 & 0.2285{*} \\
				& (0.1381) & (0.1312) & (0.1436) & (0.1322) & (0.1299) & (0.1364) \\
				$QE_{1}\times\pi_{-1}(\hat{\gamma}_{post})$ & 0.2237{*} & 0.0626 & 0.2397{*}{*} & 0.2212{*} & 0.0452 & 0.2218{*} \\
				& (0.1162) & (0.0914) & (0.1052) & (0.1156) & (0.1048) & (0.1159) \\
				$QE_{2}\times\pi_{-1}(\hat{\gamma}_{post})$ & 0.1828 & 0.0086 & 0.2021{*} & 0.0717 & -0.0039 & 0.1433 \\
				& (0.1323) & (0.1174) & (0.1222) & (0.1281) & (0.1272) & (0.1281) \\
				$MEP\times\pi_{-1}(\hat{\gamma}_{post})$ & 0.064 & 0.0294 & 0.1419{*} & 0.0419 & -0.0481 & 0.0444 \\
				& (0.0964) & (0.0840) & (0.0860) & (0.0915) & (0.0941) & (0.0926) \\
				$QE_{3}\times\pi_{-1}(\hat{\gamma}_{post})$ & 0.2185{*}{*} & 0.0131 & 0.1112 & 0.2072{*}{*} & 0.1358 & 0.1820{*}{*} \\
				& (0.0932) & (0.0749) & (0.0776) & (0.0884) & (0.0876) & (0.0887) \\
				\hline
				Fixed effects & Yes & Yes & Yes & Yes & Yes & Yes \\
				Time effects & Yes & Yes & Yes & Yes & Yes & Yes \\
				Industry linear trends & Yes & No & Yes & Yes & No & Yes \\
				Ind. dummy$\times$RGDP & No & Yes & Yes & No & Yes & Yes \\
				Observations & 82092 & 82092 & 82092 & 82092 & 82092 & 82092 \\
				$N$ & 3236 & 3236 & 3236 & 3236 & 3236 & 3236 \\
				$max(T_{i})$ & 44 & 44 & 44 & 44 & 44 & 44 \\
				$avg(T_{i})$ & 25.4 & 25.4 & 25.4 & 25.4 & 25.4 & 25.4 \\
				$med(T_{i})$ & 22 & 22 & 22 & 22 & 22 & 22 \\
				$min(T_{i})$ & 4 & 4 & 4 & 4 & 4 & 4 \\
				\hline
		\end{tabular}	}
\end{center}

%

\end{table}%

\newpage

\section{Estimation results using firm-specific debt capacity
indicators\label{Appx:firmdebtcpcty}}

In this section, we report estimates of the effects of LSAPs on firm capital
structure using an identification strategy which exploits variation in debt
capacity across firms within each industry. To this end, we interact our
measures of LSAPs with one-quarter lag of a firm-specific debt capacity
indicator, $d_{is,t}(\gamma)$, defined as a dummy variable equal to one if
firm $i$'s debt to assets ratio (DA) is below the $\gamma^{th}$ quantile of
the cross-sectional distribution of DA across all firms in industry $s$ at
time $t$. Specifically,
\begin{equation}
d_{is,t}(\gamma)=\mathcal{I}\left[  y_{is,t}<g_{st}(\gamma)\right]  ,
\label{Firm_DebtCap_apx}%
\end{equation}
where $y_{is,t}$ is the ratio of debt to assets of firm $i$ in industry $s$
for quarter $t$, and $\mathcal{I}\left(  A\right)  $ is an indicator variable
that takes the value of one if $A$ is true and zero otherwise.

Because $d_{is,t}(\gamma)$ varies across firms, we can now include
industry-time fixed effects, $\phi_{st}$, in the panel regression model
without the need of imposing restrictions of the type described in equation
(\ref{phi_st}) in Subsection \ref{subsec: Ident_Strategy} of the paper. In
this case, because of industry-time fixed effects, we do not include
industry-specific variables (such as industry leverage and industry growth)
among the regressors and consider only firm-specific variables.

For comparison, we also consider the case where the regression model includes
both firm- and industry-specific variables, replacing the industry-time fixed
effects, $\phi_{st}$, with time effects and the interaction of industry
dummies and selected macro-variables, namely $\phi_{st}=\delta_{t}%
+\boldsymbol{\phi}_{s}^{\prime}\mathbf{f}_{t}$.

\subsection{Quantile threshold parameter estimates \label{Apx_FDC_thresh}}

As before, for a given choice of the lag order, $p$, for the panel ARDL
specification, we estimate the quantile threshold parameter, $\gamma$, by grid
search over the values of $\gamma$ in the range $0.25\leq\gamma_{pre}%
,\gamma_{post}\leq0.9$ in increments of $0.01$. For $p=2$, the estimates of
$\gamma$ are reported in Table \ref{tab: thresholds_firmdebtcap}. Panel A
shows the estimated thresholds when our measure of LSAPs, $q_{t}$, is the
(scaled) amount of U.S. Treasuries and agency MBS purchased by the
Fed.\footnote{See Subsection \ref{Sec:qt} in the main paper, and Section \ref{Appendix_DataAnalysis} of this online supplement, for further
details on our measures of LSAPs.} Panel B reports the estimates when the PanARDL(2) model includes two
separate measures of LSAPs, namely MBS and Treasury purchases. Panel C
displays results when including four qualitative measures of LSAPs, namely a
set of dummy variables which take the value of one during policy on periods
and zero otherwise.

When using the firm-specific debt capacity indicator, we find that the
estimated threshold parameters are either identical or extremely close before
and after the introduction of LSAPs. Therefore, here we focus on the
single-threshold parameter case, where $\gamma_{pre}=\gamma_{post}=\gamma$.

The estimated quantile threshold parameter, $\hat{\gamma}$, is around $0.69$
in all PanARDL(2) regressions when using the quantitative measures of LSAPs,
regardless of whether the model includes industry-time fixed effects (as in
column (1)) or time effects and macro-variables interacted with industry
dummies (columns (2) to (4)). We estimate a similar threshold when using the
qualitative measures of LSAPs. In this case the estimated threshold parameter
varies between $0.65$ and $0.69$, as shown in Panel C.%

\begin{table}[H]
\caption{\textbf{Estimated quantile threshold parameters}}\label{tab: thresholds_firmdebtcap}
\vspace{-0.2cm}
\footnotesize
Estimates of the quantile threshold parameters from a grid search procedure
for the single-threshold PanARDL(2) model. Panel A shows results for the case
where our measure of LSAPs is the (scaled) amount of U.S. Treasuries and
agency MBS purchased by the Fed. Panel B displays results when separating the
effects of MBS from Treasury purchases. Panel C displays results for the
qualitative measures of LSAPs, a set of dummy variables which take the value
of one during policy on periods and zero otherwise. In column (1) the
regression model does not include industry-specific regressors, noting that
$\phi_{st}$ is unconstrained. In the remaining columns, the regression model
includes both firm- and industry-specific regressors. Column (2), (3), and (4)
report results when including industry linear trends, real GDP growth, or both
as a proxy for $f_{t}$, respectively. The estimation sample consists of an
unbalanced panel of $3,647$ U.S. publicly traded non-financial firms observed
at a quarterly frequency over the period $2007$:Q1 - $2018$:Q3.
\normalsize

\begin{center}%
\begin{tabular}
[c]{ccccc}\cline{2-5}\cline{3-5}\cline{4-5}\cline{5-5}
& (1) & (2) & (3) & (4)\\\hline
\textit{Panel A:} & \multicolumn{4}{c}{\textit{Tot. MBS and TY}}\\
$\hat{\gamma}$ & 0.69 & 0.69 & 0.69 & 0.69\\[0.2cm]%
\textit{Panel B:} & \multicolumn{4}{c}{\textit{ MBS versus TY}}\\
$\hat{\gamma}$ & 0.69 & 0.69 & 0.69 & 0.69\\[0.2cm]%
\textit{Panel C:} & \multicolumn{4}{c}{\textit{4 QE episodes}}\\
$\hat{\gamma}$ & 0.69 & 0.65 & 0.65 & 0.65\\\hline
Time effects & No & Yes & Yes & Yes\\
Ind. $\times$ quarter & Yes & No & No & No\\
Ind. $\times$ lin. trend & No & Yes & No & Yes\\
Ind. $\times$ RGDP gr. & No & No & Yes & Yes\\\hline
\end{tabular}

\end{center}

%

\end{table}%

Given the estimated threshold parameters, in the next sections we present the
estimates of the policy parameters of interest. In particular, in Section
\ref{Apx_FDC_bench}, we report results when $q_{t}$ is the (scaled) total
amount of MBS and Treasuries purchased by the Fed. In Section
\ref{Apx_FDC_MBSvsTY}, we separate the effects of MBS from Treasury purchases.
In Section \ref{Apx_FDC_4QE}, we evaluate the first four large-scale asset
purchases by the Fed, using qualitative measures of LSAPs.

\pagebreak

\subsection{Estimation results when $q_{t}$ measures the total size of
LSAPs\label{Apx_FDC_bench}}

Table \ref{tab:SR_benchmark_FirmDC} displays the estimates of the net
short-run (SR) effects defined as the sum of estimated coefficients of current
and lagged values of the regressor under consideration. In column (1), the
PanARDL(2) model includes firm-specific fixed effects and industry-time fixed
effects. The estimates under columns (2) to (4) are based on PanARDL(2)
regressions that include both firm-specific fixed effects and time effects as
well as the interaction of selected macro-variables with industry dummies.

The estimates of the policy SR effects ($LSAP\times d_{-1}(\hat{\gamma})$ in
Table \ref{tab:SR_benchmark_FirmDC}) are positive and highly statistically
significant under all specifications, although the magnitude is rather small.
The estimates for the other regressors are very much in line with the results
obtained using the industry-specific debt capacity indicators ($\pi
_{st}(\gamma)$), shown in Table \ref{tab:SR_benchmark} of the paper.
Interestingly, the estimated SR effects are very close regardless of whether
we use industry-time fixed effects, $\phi_{st}$, or its restricted version
with time effects and the interaction of industry dummies and selected
macro-variables. This further corroborates the identification strategy
described in Subsection \ref{subsec: Ident_Strategy}.

Table \ref{tab:LR_benchmark_FirmDC} reports the estimates of the long-run (LR)
effects of LSAPs and other determinants of firms' debt to asset ratios. The
policy long-run effects are defined by equation (\ref{LR_definition}), in
Section \ref{ARDL}. Also in this case, we find that the effects of LSAPs on
firm capital structure are long-lasting.%

\begin{table}[H]
\caption{\textbf{Estimates of the net short-run effects of LSAPs on debt to asset ratios of non-financial firms}}\label{tab:SR_benchmark_FirmDC}
\vspace{-0.2cm}
\footnotesize
Estimates of net short-run effects of LSAPs on firms' debt to asset ratios
(DA) as well as the effects of both firm- and industry-specific variables on
DA, for the single-threshold PanARDL(2) model, using the firm-specific debt
capacity indicator, $d_{is,t}(\gamma)$, defined by (\ref{Firm_DebtCap_apx}).
Net short-run effects are defined as the sum of the estimated coefficients of
current and lagged values of the regressor under consideration. The estimated
quantile threshold parameters are shown in Panel A of Table
\ref{tab: thresholds_firmdebtcap}. All regressions include firm-specific fixed
effects. Column (1) includes industry-time fixed effects. Columns (2) to (4)
include time effects and the interaction of industry dummies with either
linear trend, real GDP growth, or both. $LSAP$ is the (scaled) amount of U.S.
Treasuries and agency MBS purchased by the Fed; $d_{-1}(\gamma)$ denotes the
one-quarter lagged firm-specific debt capacity indicator. The sample consists
of an unbalanced panel of $3,647$ U.S. publicly traded non-financial firms
observed at a quarterly frequency over the period $2007$:Q1 - $2018$:Q3.
Robust standard errors (in parentheses) are computed using the delta method
(*** $p<0.01$, ** $p<0.05$, * $p<0.1$).
\normalsize

\begin{center}
\scalebox{0.9}{
			\begin{tabular}{l|cccc}
				\multicolumn{1}{c}{} & \multicolumn{4}{c}{Dependent variable: debt to assets (DA)} \\
				\cline{2-5} \cline{3-5} \cline{4-5} \cline{5-5}
				\multicolumn{1}{c}{} & (1) & (2) & (3) & (4) \\
				\hline
				$d_{-1}(\hat{\gamma})$ & -0.0118{*}{*}{*} & -0.0122{*}{*}{*} & -0.0121{*}{*}{*} & -0.0122{*}{*}{*} \\
				& (0.0014) & (0.0014) & (0.0014) & (0.0014) \\
				$LSAP\times d_{-1}(\hat{\gamma})$ & 0.0041{*}{*}{*} & 0.0040{*}{*}{*} & 0.0040{*}{*}{*} & 0.0040{*}{*}{*} \\
				& (0.0006) & (0.0007) & (0.0007) & (0.0007) \\
				Lagged DA & 0.8228{*}{*}{*} & 0.8205{*}{*}{*} & 0.8231{*}{*}{*} & 0.8205{*}{*}{*} \\
				& (0.0063) & (0.0063) & (0.0063) & (0.0063) \\
				Cash to assets & -0.0365{*}{*}{*} & -0.0365{*}{*}{*} & -0.0366{*}{*}{*} & -0.0365{*}{*}{*} \\
				& (0.0030) & (0.0030) & (0.0029) & (0.0030) \\
				PPE to assets & 0.0221{*}{*}{*} & 0.0229{*}{*}{*} & 0.0221{*}{*}{*} & 0.0228{*}{*}{*} \\
				& (0.0046) & (0.0047) & (0.0046) & (0.0047) \\
				Size & 0.0033{*}{*}{*} & 0.0033{*}{*}{*} & 0.0035{*}{*}{*} & 0.0033{*}{*}{*} \\
				& (0.0008) & (0.0008) & (0.0007) & (0.0008) \\
				Industry leverage &  & 0.0558{*}{*}{*} & 0.0482{*}{*}{*} & 0.0525{*}{*}{*} \\
				&  & (0.0071) & (0.0060) & (0.0073) \\
				Industry growth &  & -0.0860{*}{*}{*} & -0.1202{*}{*}{*} & -0.0921{*}{*}{*} \\
				&  & (0.0207) & (0.0203) & (0.0216) \\
				\hline
				Fixed effects & Yes & Yes & Yes & Yes \\
				Time effects & No & Yes & Yes & Yes \\
				Industry $\times$ quarter & Yes & No & No & No \\
				Industry $\times$ linear trend & No & Yes & No & Yes \\
				Industry $\times$ RGDP & No & No & Yes & Yes \\
				Observations & 84548 & 84548 & 84548 & 84548 \\
				$N$ & 3647 & 3647 & 3647 & 3647 \\
				$max(T_{i})$ & 44 & 44 & 44 & 44 \\
				$avg(T_{i})$ & 23.2 & 23.2 & 23.2 & 23.2 \\
				$med(T_{i})$ & 19 & 19 & 19 & 19 \\
				$min(T_{i})$ & 2 & 2 & 2 & 2 \\
				\hline
		\end{tabular}	}
\end{center}

%

\end{table}%

%

\begin{table}[H]
\caption{\textbf{Estimates of the long-run effects of LSAPs on debt to asset ratios of non-financial firms}}\label{tab:LR_benchmark_FirmDC}
\vspace{-0.2cm}
\footnotesize
Estimates of long-run effects of LSAPs, defined in equation
(\ref{LR_definition}), on firms' debt to asset ratios (DA) as well as the
effects of both firm- and industry-specific variables on DA, for the
single-threshold PanARDL(2) model, using the firm-specific debt capacity
indicator, $d_{is,t}(\gamma)$, defined by (\ref{Firm_DebtCap_apx}). The
estimated quantile threshold parameters are shown in Panel A of Table
\ref{tab: thresholds_firmdebtcap}. All regressions include firm-specific fixed
effects. Column (1) includes industry-time fixed effects. Columns (2) to (4)
include time effects and the interaction of industry dummies with either
linear trend, real GDP growth, or both. $LSAP$ is the (scaled) amount of U.S.
Treasuries and agency MBS purchased by the Fed; $d_{-1}(\gamma)$ denotes the
one-quarter lagged firm-specific debt capacity indicator. The sample consists
of an unbalanced panel of $3,647$ U.S. publicly traded non-financial firms
observed at a quarterly frequency over the period $2007$:Q1 - $2018$:Q3.
Robust standard errors (in parentheses) are computed using the delta method
(*** $p<0.01$, ** $p<0.05$, * $p<0.1$).
\normalsize

\begin{center}
\scalebox{0.9}{
			\begin{tabular}{l|cccc}
				\multicolumn{1}{c}{} & \multicolumn{4}{c}{Dependent variable: debt to assets (DA)} \\
				\cline{2-5} \cline{3-5} \cline{4-5} \cline{5-5}
				\multicolumn{1}{c}{} & (1) & (2) & (3) & (4) \\
				\hline
				$d_{-1}(\hat{\gamma})$ & -0.0665{*}{*}{*} & -0.0680{*}{*}{*} & -0.0684{*}{*}{*} & -0.0682{*}{*}{*} \\
				& (0.0065) & (0.0065) & (0.0066) & (0.0065) \\
				$LSAP\times d_{-1}(\hat{\gamma})$ & 0.0230{*}{*}{*} & 0.0224{*}{*}{*} & 0.0229{*}{*}{*} & 0.0225{*}{*}{*} \\
				& (0.0036) & (0.0036) & (0.0037) & (0.0036) \\
				Cash to assets & -0.2062{*}{*}{*} & -0.2032{*}{*}{*} & -0.2067{*}{*}{*} & -0.2036{*}{*}{*} \\
				& (0.0165) & (0.0163) & (0.0163) & (0.0163) \\
				PPE to assets & 0.1248{*}{*}{*} & 0.1274{*}{*}{*} & 0.1249{*}{*}{*} & 0.1272{*}{*}{*} \\
				& (0.0263) & (0.0261) & (0.0259) & (0.0261) \\
				Size & 0.0188{*}{*}{*} & 0.0183{*}{*}{*} & 0.0200{*}{*}{*} & 0.0184{*}{*}{*} \\
				& (0.0042) & (0.0042) & (0.0041) & (0.0042) \\
				Industry leverage &  & 0.3110{*}{*}{*} & 0.2724{*}{*}{*} & 0.2928{*}{*}{*} \\
				&  & (0.0378) & (0.0321) & (0.0389) \\
				Industry growth &  & -0.4793{*}{*}{*} & -0.6794{*}{*}{*} & -0.5130{*}{*}{*} \\
				&  & (0.1161) & (0.1168) & (0.1216) \\
				\hline
				Fixed effects & Yes & Yes & Yes & Yes \\
				Time effects & No & Yes & Yes & Yes \\
				Industry $\times$quarter & Yes & No & No & No \\
				Industry $\times$ linear trend & No & Yes & No & Yes \\
				Industry $\times$ RGDP & No & No & Yes & Yes \\
				Observations & 84548 & 84548 & 84548 & 84548 \\
				$N$ & 3647 & 3647 & 3647 & 3647 \\
				$max(T_{i})$ & 44 & 44 & 44 & 44 \\
				$avg(T_{i})$ & 23.2 & 23.2 & 23.2 & 23.2 \\
				$med(T_{i})$ & 19 & 19 & 19 & 19 \\
				$min(T_{i})$ & 2 & 2 & 2 & 2 \\
				\hline
			\end{tabular}
			}
\end{center}

%

\end{table}%

\pagebreak

\subsection{Separating the effects of MBS and Treasury purchases
\label{Apx_FDC_MBSvsTY}}

In this subsection, we separate the effects of Treasury and MBS purchases.
Panel A and B of Table \ref{tab:SRLR_MBS_TY_FDC} report the policy SR and LR
effects, respectively.

When using $d_{is,t}(\gamma)$, we find that both Treasury and MBS purchases
have significant impacts on firm leverage but the effects are now stronger for
Treasuries. The opposite holds when using $\pi_{st}(\gamma)$. Part of this
difference can be explained by the different nature of the two indicators. The
identification strategy based on $d_{is,t}(\gamma)$ exploits cross-firm
variation within an industry, implying that firms which are not over-leveraged
should benefit more from LSAPs relative to peers in the same industry.
Instead, estimation based on $\pi_{st}(\gamma)$ exploits cross-industry
variation, suggesting that firms in less leveraged industries should benefit
more, thus allowing for spillover effects within an industry. Taken together,
these results suggest that both MBS and Treasury purchases can facilitate
firms' access to external financing, although the magnitude of the effects is
rather small.%

\begin{table}[H]
\caption{\textbf{Net short-run and long-run effects of large-scale MBS and Treasury purchases on debt to asset ratios of non-financial firms}}\label{tab:SRLR_MBS_TY_FDC}
\vspace{-0.2cm}
\footnotesize
Panel A (Panel B) reports the net short-run (long-run) effects of MBS and
Treasury purchases on firms' debt to asset ratios (DA), for the
single-threshold PanARDL(2) model, using the firm-specific debt capacity
indicator, $d_{is,t}(\gamma)$, defined by (\ref{Firm_DebtCap_apx}). Net
short-run effects are defined as the sum of the estimated coefficients of
current and lagged values of the regressor under consideration. The long-run
effects are defined in equation (\ref{LR_definition}). The
estimated quantile threshold parameters are shown in Panel B of Table
\ref{tab: thresholds_firmdebtcap}. All panel regressions include firm-specific
effects. Column (1) includes industry-time fixed effects. Columns (2) to (4)
include time effects and the interaction of industry dummies with linear
trend, real GDP growth, or both. $ty$ and $mbs$ denote the (scaled) amount of
U.S. Treasuries and agency MBS purchased by the Fed, respectively;
$d_{-1}(\gamma)$ denotes the one-quarter lagged firm-specific debt capacity
indicator. The sample consists
of an unbalanced panel of $3,647$ U.S. publicly traded non-financial firms
observed at a quarterly frequency over the period $2007$:Q1 - $2018$:Q3.
Robust standard errors (in parentheses) are computed using the delta method
(*** $p<0.01$, ** $p<0.05$, * $p<0.1$).
\normalsize

\begin{center}
\scalebox{0.9}{
			\begin{tabular}{l|cccc}
				\multicolumn{1}{c}{} & \multicolumn{4}{c}{Dependent variable: debt to assets (DA)} \\
				\cline{2-5} \cline{3-5} \cline{4-5} \cline{5-5}
				\multicolumn{1}{c}{} & (1) & (2) & (3) & (4) \\
				\hline
				\multicolumn{1}{c}{} & \multicolumn{4}{c}{\textit{Panel A: Short-run effects}} \\
				\hline
				$d_{-1}(\hat{\gamma})$ & -0.0122{*}{*}{*} & -0.0126{*}{*}{*} & -0.0125{*}{*}{*} & -0.0128{*}{*}{*} \\
				& (0.0014) & (0.0014) & (0.0014) & (0.0023) \\
				$ty\times d_{-1}(\hat{\gamma})$ & 0.0062{*}{*}{*} & 0.0062{*}{*}{*} & 0.0062{*}{*}{*} & 0.0064{*}{*}{*} \\
				& (0.0012) & (0.0012) & (0.0012) & (0.0022) \\
				$mbs\times d_{-1}(\hat{\gamma})$ & 0.0034{*}{*}{*} & 0.0034{*}{*}{*} & 0.0034{*}{*}{*} & 0.0033{*}{*} \\
				& (0.0008) & (0.0008) & (0.0008) & (0.0015) \\
				\hline
				\multicolumn{1}{c}{} & \multicolumn{4}{c}{\textit{Panel B: Long-run effects}} \\
				\hline
				$d_{-1}(\hat{\gamma})$ & -0.0687{*}{*}{*} & -0.0701{*}{*}{*} & -0.0705{*}{*}{*} & -0.0714{*}{*}{*} \\
				& (0.0066) & (0.0065) & (0.0066) & (0.0110) \\
				$ty\times d_{-1}(\hat{\gamma})$ & 0.0351{*}{*}{*} & 0.0342{*}{*}{*} & 0.0351{*}{*}{*} & 0.0358{*}{*}{*} \\
				& (0.0066) & (0.0066) & (0.0068) & (0.0122) \\
				$mbs\times d_{-1}(\hat{\gamma})$ & 0.0191{*}{*}{*} & 0.0186{*}{*}{*} & 0.0189{*}{*}{*} & 0.0186{*}{*} \\
				& (0.0044) & (0.0044) & (0.0044) & (0.0083) \\
				\hline
				Fixed effects & Yes & Yes & Yes & Yes \\
				Time effects & No & Yes & Yes & Yes \\
				Industry $\times$ quarter & Yes & No & No & No \\
				Industry $\times$ linear trend & No & Yes & No & Yes \\
				Industry $\times$ RGDP & No & No & Yes & Yes \\
				Observations & 84548 & 84548 & 84548 & 84548 \\
				$N$ & 3647 & 3647 & 3647 & 3647 \\
				$max(T_{i})$ & 44 & 44 & 44 & 44 \\
				$avg(T_{i})$ & 23.2 & 23.2 & 23.2 & 23.2 \\
				$med(T_{i})$ & 19 & 19 & 19 & 19 \\
				$min(T_{i})$ & 2 & 2 & 2 & 2 \\
				\hline
			\end{tabular}
				}
\end{center}

%

\end{table}%

\pagebreak

\subsection{Estimating the effects of the first four asset purchase programs
\label{Apx_FDC_4QE}}

We now compare the effects of each Fed's program separately by replacing the
two aforementioned quantitative measures of LSAPs with four qualitative
variables which take the value of one during policy on periods and zero
otherwise. More details on each program can be found in Table \ref{Table_LSAP}.

The estimates of policy SR and LR effects can be found in Panel A and B of
Table \ref{tab:SRLR_4QE_FDC}, respectively. We find that QE1, QE2, and QE3 had
positive and statistically significant effects on firm leverage, both in the
short- and the long-term. When using $d_{is,t}(\gamma)$, differences in
magnitudes across these programs are less marked. MEP continues to have the
lowest impact, and is generally non significant at the 5 per cent level.%

\begin{table}[H]
\caption{\textbf{Net short-run and long-run effects of of large-scale MBS and Treasury purchases on debt to asset ratios of non-financial firms}}\label{tab:SRLR_4QE_FDC}
\vspace{-0.2cm}
\footnotesize
Panel A (Panel B) reports net short-run (long-run) effects of the first four
asset purchase programs by Fed on firms' debt to asset ratios (DA), for the
single-threshold PanARDL(2) model using the firm-specific debt capacity
indicator, $d_{is,t}(\gamma)$, defined by (\ref{Firm_DebtCap_apx}). Net
short-run effects are defined as the sum of the estimated coefficients of
current and lagged values of the regressor under consideration. The long-run
effects are defined in equation (\ref{LR_definition}). The estimated quantile
threshold parameters are shown in Panel C of Table
\ref{tab: thresholds_firmdebtcap}. All regressions include firm-specific fixed
effects. Column (1) includes industry-time fixed effects. Columns (2) to (4)
include time effects and the interaction of industry dummies with either
linear trend, real GDP growth, or both. $QE1$ and $QE2$ are two indicator
variables equal to one during the period 2008Q4 - 2010Q1, and 2010Q4 - 2011Q2
and zero otherwise, respectively. $MEP$ denotes the maturity extension program
of 2011Q3 - 2012Q4, while $QE3$ is equal to one between 2012Q3 - 2012Q4;
$d_{-1}(\gamma)$ denotes the one-quarter lagged firm-specific debt capacity
indicator. The sample consists of an unbalanced panel of $3,647$ U.S. publicly
traded non-financial firms observed at a quarterly frequency over the period
$2007$:Q1 - $2018$:Q3. Robust standard errors (in parentheses) are computed
using the delta method (*** $p<0.01$, ** $p<0.05$, * $p<0.1$).
\normalsize

\begin{center}
\scalebox{0.85}{
			\begin{tabular}{l|cccc}
				\multicolumn{1}{c}{} & \multicolumn{4}{c}{Dependent variable: debt to assets (DA)} \\
				\cline{2-5} \cline{3-5} \cline{4-5} \cline{5-5}
				\multicolumn{1}{c}{} & (1) & (2) & (3) & (4) \\
				\hline
				\multicolumn{1}{c}{} & \multicolumn{4}{c}{\textit{Panel A: Short-run effects}} \\
				\hline
				$d_{-1}(\hat{\gamma})$ & -0.0115{*}{*}{*} & -0.0110{*}{*}{*} & -0.0109{*}{*}{*} & -0.0110{*}{*}{*} \\
				& (0.0014) & (0.0014) & (0.0014) & (0.0014) \\
				$QE_{1}\times d_{-1}(\hat{\gamma})$ & 0.0072{*}{*}{*} & 0.0067{*}{*}{*} & 0.0067{*}{*}{*} & 0.0068{*}{*}{*} \\
				& (0.0016) & (0.0015) & (0.0015) & (0.0015) \\
				$QE_{2}\times d_{-1}(\hat{\gamma})$ & 0.0092{*}{*}{*} & 0.0086{*}{*}{*} & 0.0087{*}{*}{*} & 0.0086{*}{*}{*} \\
				& (0.0021) & (0.0020) & (0.0020) & (0.0020) \\
				$MEP\times d_{-1}(\hat{\gamma})$ & 0.0018 & 0.0024{*} & 0.0025{*} & 0.0024{*} \\
				& (0.0015) & (0.0015) & (0.0015) & (0.0015) \\
				$QE_{3}\times d_{-1}(\hat{\gamma})$ & 0.0056{*}{*}{*} & 0.0043{*}{*}{*} & 0.0044{*}{*}{*} & 0.0043{*}{*}{*} \\
				& (0.0013) & (0.0013) & (0.0013) & (0.0013) \\
				\hline
				\multicolumn{1}{c}{} & \multicolumn{4}{c}{\textit{Panel B: Long-run effects}} \\
				\hline
				$d_{-1}(\hat{\gamma})$ & -0.0646{*}{*}{*} & -0.0615{*}{*}{*} & -0.0620{*}{*}{*} & -0.0616{*}{*}{*} \\
				& (0.0066) & (0.0066) & (0.0066) & (0.0066) \\
				$QE_{1}\times d_{-1}(\hat{\gamma})$ & 0.0402{*}{*}{*} & 0.0376{*}{*}{*} & 0.0381{*}{*}{*} & 0.0378{*}{*}{*} \\
				& (0.0090) & (0.0085) & (0.0086) & (0.0085) \\
				$QE_{2}\times d_{-1}(\hat{\gamma})$ & 0.0515{*}{*}{*} & 0.0482{*}{*}{*} & 0.0491{*}{*}{*} & 0.0481{*}{*}{*} \\
				& (0.0119) & (0.0112) & (0.0114) & (0.0112) \\
				$MEP\times d_{-1}(\hat{\gamma})$ & 0.0102 & 0.0137{*} & 0.0140{*} & 0.0137{*} \\
				& (0.0086) & (0.0083) & (0.0084) & (0.0083) \\
				$QE_{3}\times d_{-1}(\hat{\gamma})$ & 0.0316{*}{*}{*} & 0.0240{*}{*}{*} & 0.0251{*}{*}{*} & 0.0242{*}{*}{*} \\
				& (0.0074) & (0.0071) & (0.0073) & (0.0071) \\
				\hline
				Fixed effects & Yes & Yes & Yes & Yes \\
				Time effects & No & Yes & Yes & Yes \\
				Industry $\times$quarter & Yes & No & No & No \\
				Industry $\times$ linear trend & No & Yes & No & Yes \\
				Industry $\times$ RGDP & No & No & Yes & Yes \\
				Observations & 84548 & 84548 & 84548 & 84548 \\
				$N$ & 3647 & 3647 & 3647 & 3647 \\
				$max(T_{i})$ & 44 & 44 & 44 & 44 \\
				$avg(T_{i})$ & 23.2 & 23.2 & 23.2 & 23.2 \\
				$med(T_{i})$ & 19 & 19 & 19 & 19 \\
				$min(T_{i})$ & 2 & 2 & 2 & 2 \\
				\hline
			\end{tabular}
		}
\end{center}

%

\end{table}%

\end{document}